\documentclass[12pt]{article} 
\usepackage[T1]{fontenc}
\usepackage{biolinum}
\usepackage[left=2cm,right=2cm,top=2cm,bottom=2cm]{geometry}
\usepackage[pdftex]{graphicx} 
\usepackage{amssymb} 
\usepackage{amsmath} 
\usepackage{amsthm}
\usepackage{wrapfig} 
\usepackage{calc} 
\usepackage{footmisc} 
\usepackage[table, svgnames, dvipsnames]{xcolor}
\usepackage{makecell, cellspace, caption}
\usepackage{subcaption}
\usepackage{mathtools}
\usepackage{url}
\usepackage{hyperref}
\hypersetup{colorlinks,linkcolor={},citecolor={blue},urlcolor={red}}

\usepackage[version=4]{mhchem}

\usepackage{booktabs}
\usepackage[square,sort,comma,numbers]{natbib}

\usepackage{epigraph}
\usepackage[T1]{fontenc}

\usepackage{etoolbox}

\makeatletter

\patchcmd{\maketitle}{\@fnsymbol}{\@alph}{}{}
\makeatother

\title{ Epidemic response to physical distancing policies and their impact on the outbreak risk}
\author{Fabio Vanni\thanks{Sciences Po, OFCE , France} \thanks{Institute of Economics, Sant'Anna School of Advanced Studies, Pisa, Italy} \thanks{Department of Physics, University of North Texas, USA }
\and	
	David Lambert\thanks{Department of Mathematics, University of North Texas, USA }\, \footnotemark[3]
	\and
	Luigi Palatella\footnote{ Liceo Scientifico Statale ``C. De Giorgi'', Lecce, Italy} 
}		

\date{}

\begin{document}
\maketitle
		\begin{abstract}
			 We introduce a theoretical framework that highlights the impact of physical distancing variables such as human mobility and physical proximity on the evolution of epidemics and, crucially, on the reproduction number.  In particular, in response to the coronavirus disease (CoViD-19) pandemic, countries have introduced various levels of 'lockdown' to reduce the number of new infections. Specifically we use a collisional approach to an infection-age structured model described by a renewal equation for the time homogeneous evolution of epidemics. As a result, we show how various contributions of the lockdown policies, namely physical proximity and human mobility, reduce the impact of SARS-CoV-2 and mitigate the risk of disease resurgence. We check our theoretical framework using real-world data on physical distancing with two different data repositories, obtaining consistent results. Finally, we propose an equation for the effective reproduction number which takes into account types of interactions among people, which may help policy makers to improve remote-working organizational structure.
		\end{abstract}
		
		\section{Introduction}
		As  the  coronavirus  disease  (CoViD-19)  epidemic  worsens, understanding  the  effectiveness  of  public  messaging  and  large-scale  physical distancing  interventions  is  critical in order to manage the acute and the long-term phases of the spread of the epidemic.	The CoViD-19 epidemic has forced many countries to react by imposing strategies primarily based on mobility and physical lockdowns together with intranational and international border limitations.  Data regarding these interventions can  help  refine future efforts  by  providing  near  real-time   information about changes in patterns of human movement.  In particular, estimates of aggregate flows of people are incredibly valuable. 	
		Infectiousness depends on the frequency of contacts and on the level of infection within each individual. In airborne infections, the former can be decomposed as a product of mobility and physical proximity, interpreted broadly as an effective distance measure which also includes the degree of physical protection used by individuals aware of the risk of infection. The latter involves an internal micro-scale competition between the virus and the immune system which depends on environmental factors like pollution levels and repeated viral exposure, which can modify the viral load shed by infectious individuals. 
		Moreover, one should distinguish between random movements (as supposed in our collision approach) and structured contacts (as in crowds or workplaces) so that interactions can be shaped by packing density of individuals.  In the example we analyze below, we assume these factors to be homogeneous in space (within each region of interest), which represents only an approximation of the real infection transmission.  In reality, these factors are spatially heterogeneous and the interactions involve decision processes which change the direction and speed of the individual so changing the kinematical dynamics of the epidemic.

		The economic impact of the CoViD-19 pandemic is mainly due to indirect effects related to policies of physical distancing, see \citet{bellomo2020multi}. So it would be desirable to show how the risk of contagion is related to heterogeneous exposure to infection which is more concentrated in certain worker categories such as those in the health sectors and less concentrated for workers able to work remotely.
		Indirect effects of the mitigatory response to the CoViD-19 are primarily due to the mobility side of the lockdown.  This is because reducing personal mobility implies closing many businesses and services. However, many jobs have been converted into remote-work from home.  This conversion is in a proportion which is higher than what is commonly supposed to be affordable by the job system without decreasing the national economic productivity.
		
		Mechanistic models of disease transmission are often used to forecast disease trajectories and likely disease burden but are currently hampered by substantial uncertainty in disease epidemiology.  Models of disease transmission dynamics are hindered by uncertainty in the role of asymptomatic transmission, the length of the incubation period, the generation interval, and the contribution of different modes of transmission.
		Phenomenological models provide a starting point for estimation of key transmission parameters, such as the reproduction number, and forecasts of epidemic impact.  They represent promising tools to generate early forecasts of epidemic impact particularly in the context of substantial uncertainty in epidemiological parameters, \citet{yan2019quantitative, chowell2009mathematical,breda2012formulation,nishiura2010correcting, flaxman2020report,metz1978epidemic}.
		From a practical point of view, it is fundamental to understand
		which approach best permits one to forecast epidemic dynamics
		in the presence of incomplete data that, due to overload in the healthcare
		system, very often are only available during the early phases of disease
		spread. 
	For CoViD-19 this is especially true due to the number of undetected cases,
	since the total daily number of tests that can be performed is limited.
\begin{figure}[!h]
	\centering
	\begin{subfigure}[c]{0.8\textwidth}
		\centering
		\includegraphics[angle=0,origin=c,width=0.9\linewidth]{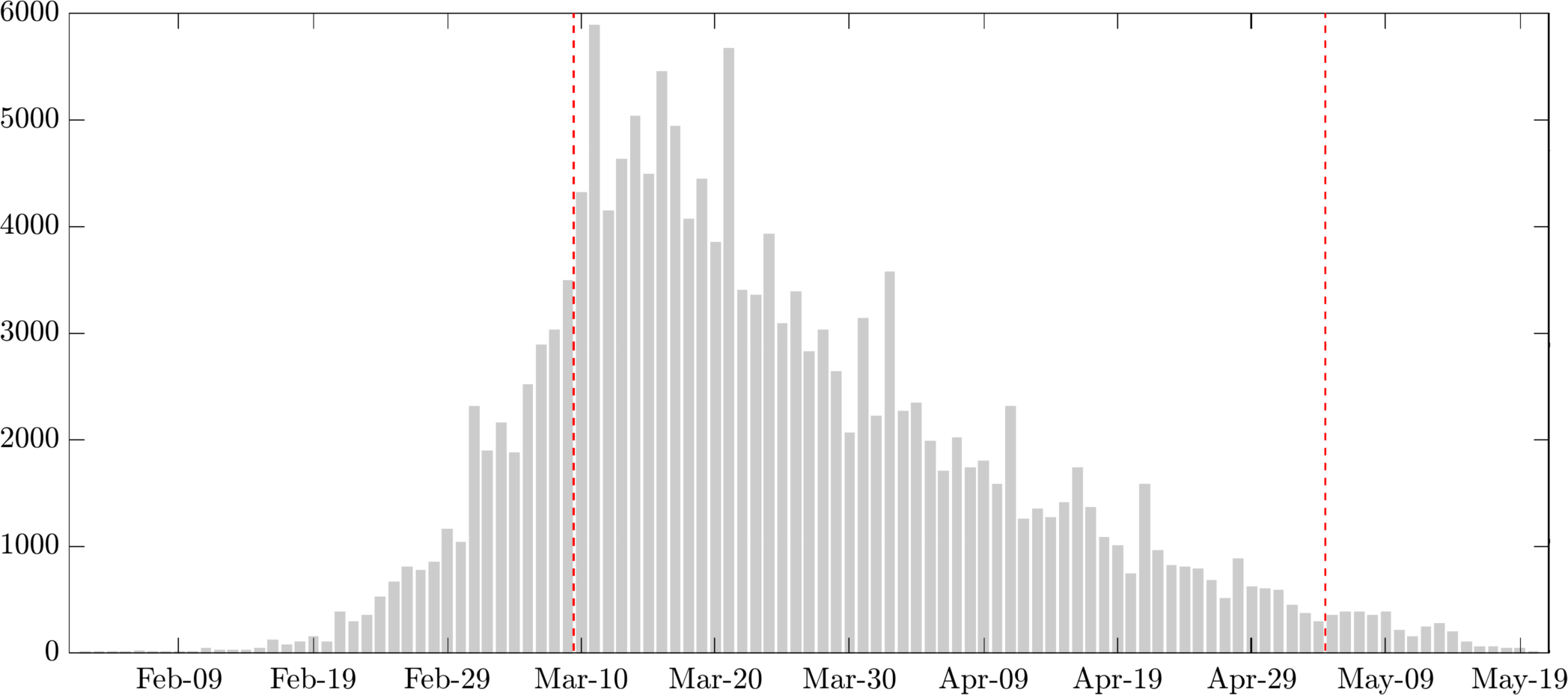}
		\caption{CoViD-19 infection incidence by onset of symptoms. Source: \citet{epicentro}.  }
		\label{fig_epi}  
	\end{subfigure}\\
	\vspace{0.5cm}
	\begin{subfigure}[c]{0.8\textwidth}
		\centering
		\includegraphics[angle=0,origin=c,width=0.9\linewidth]{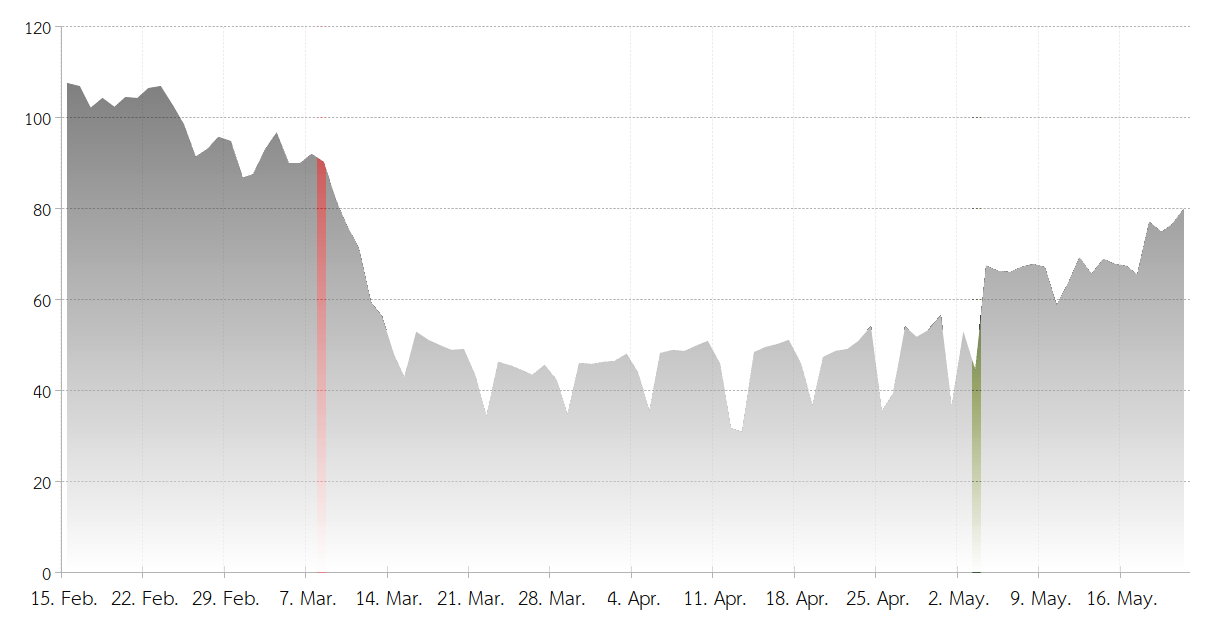}
		\caption{Mobility trend. Source: \citet{Google19} .  }
		\label{fig_mob}  
	\end{subfigure}\\
	\vspace{0.5cm}
	\begin{subfigure}[c]{0.8\textwidth}
		\centering
		\includegraphics[angle=0,origin=c,width=0.9\linewidth]{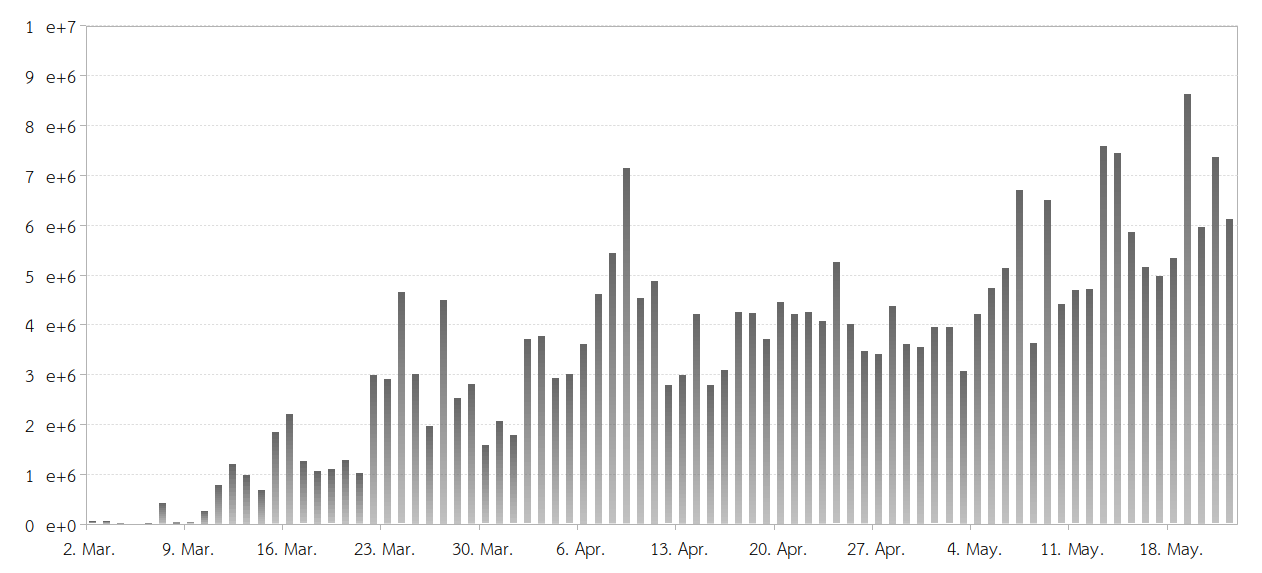}
		\caption{ Individual protection, source: \citet{ProtezCivileADI}. }
		\label{fig_masks}  
	\end{subfigure}%
	\caption{Social distancing data in Italy: (a) CoViD-19 incidence . (b) percent change in visits to different places with respect to baseline. (c) Number of respiratory protectors (face masks) as a proxy measure of increased distancing precautions. Vertical lines represent the beginning and the end of the lockdown policy in Italy.}
	\label{fig_socialdist}  
\end{figure}
In our study, we focus our attention on the contribution of asymptomatic or undiagnosed individuals to the propagation of the contagion, as well as the impact of physical distancing policies in response to the epidemic outbreak of CoViD-19 in Italy and the US.  First, we introduce the renewal equation approach to the evolution of epidemics to estimate several crucial variables from data.  We use it in the evaluation of the reproduction number of SARS-CoV-2 {under an active containment strategy}, and investigate implications for future risk. 
Later, we focus on the dynamic response of the epidemic to mobility lockdown and make an estimate of the hazard in relaxing this policy alongside other factors. 
In conclusion, the approach we follow concentrates on the fraction of people which are infectious but have not been detected, i.e., not reported as infected. We assume these hidden infectious agents have the ability to spread the disease in an environment where susceptible agents are present and all the individuals have certain mobility and physical distance parameters.  We interpret this approach in terms of a macroscopic collision theory of infected individuals in a region with a given susceptible population, taking into account the mobility of individuals as well as their radii of interaction as a proxy of physical distancing.

In the next section, we introduce the renewal theory of epidemic dynamics and embed within it a kinetic theory of collisional contacts among individuals with the contribution of two effects of physical distancing: mobility and interpersonal proximity.
In section 3, we depict and discuss the data we used and the results we obtain using the collisional theory of epidemics so as to express the effective reproduction number in terms of physical distancing variables. In Section 4, we introduce a reproduction number as combinations of various categories which contribute with different weights to the overall epidemic trend. We show how this definition could be used for policy purposes in managing mobility of workers in a smart-working perspective. Finally, we discuss the  possible future lines of research on top of our approach to epidemics.

\section{Collisional dynamics of epidemics via physical distancing}

The most important assumptions in our use of phenomenological models are (1) Short time scale of the epidemic (much shorter than the characteristic birth and death time scales of the population) (2) Well mixed population (force of infection homogeneously the same for all ages, sexes, etc.) (3) closed population (no immigration or emigration) (4) initial small shock (the initial infected group extremely small with respect to the size of the susceptible population).	
The renewal equation was introduced in the context of population dynamics studies.  Later it was reinterpreted along the lines of stochastic processes, as in \citet{fraser2007estimating}, where transmission occurred via a Poisson infection process.  This process is such that the probability that, between time $t$ and $t+\delta t$, someone infected a time $\tau$ ago successfully infects someone else is $A(t, \tau )\delta t$ , where $\delta t$ is a very small time interval.
As a consequence, the predicted  mean infectious incidence at time $t$ follows the so-called renewal equation:
\begin{align}
j(t)&=\int_{0}^{\infty}A(t,\tau)j(t-\tau)d\tau\\
j(t)&=-\frac{d}{dt}n_s(t),
\end{align}
where $\tau$ is the generation time, which describes the duration from the onset of infectiousness in the primary case to the onset of infectiousness in a secondary case (infected by the primary case), and $j(t)$ is the rate of production of infectious individuals.   The kernel $A(t,\tau)$ is the average rate at which an individual infected $\tau$ time units earlier generates secondary cases. In other words $A(t,\tau)$ is the expected infectivity of an individual with infection-age $\tau$, it can be interpreted as the reproduction function for new infections at time $t$.
{A practical issue concerns the extrinsic dynamics (e.g., public health interventions) of time inhomogeneities highlighting the depletion of susceptible individuals when contact tracing, quarantine, and isolation are implemented during the course of an epidemic.  The kernel $A$ }can be factorized as $A(t,\tau)=n_s(t)\beta(t,\tau)\Gamma(t,\tau)$, where $\beta(t,\tau)$ is the product of the contact rate and the risk of infection (i.e., the effective contact rate), and $\Gamma(t,\tau)$ is the probability of being infectious at infection age $\tau$.  So, reduction in contact frequency with calendar time $t$ affects $\beta(t, \tau)$ while early removal of infectious individuals at calendar time $t$ changes the form of $\Gamma(t, \tau )$.  An earlier average infection age at first transmission of the disease will result from contact tracing and isolation.
However, the classic approach to renewal equations for epidemics assumes, as in \citet{nishiura2010time,champredon2018equivalence,breda2012formulation},
that the non-linearity of an epidemic is characterized by the depletion of susceptible individuals alone, so that $A(t,\tau)=n_s(t)\beta(\tau)\Gamma(\tau)$.
Finally, the number of infected individuals is called prevalence, which indicates the proportion of persons who have the ability to infect at a given calendar time.  It can be written as:
\begin{equation}\label{eq_persistence}
p(t)=\int_{0}^{\infty} \Gamma(\tau) j(t-\tau) d\tau.
\end{equation}
Notice that $p(t)$ is not the number of active infected individuals generally reported
in epidemic data published by different national health services. This is because the
officially detected cases are actively confined (in hospitals or at home) and so their 
contribution to epidemic spreading is not so relevant.  On the contrary $p(t)$ represents
the infected people that are still conducting their lives as usual, possibly infecting other people.

An important variable is the incidence-persistence ratio $\text{IPR}_t$ at time $t$:
\begin{equation}\label{eq_ratio}
\text{IPR}_t =\frac{j(t)}{p(t)}.
\end{equation}
This is important, as it indicates the propensity of currently infected individuals to infect susceptibles.
Let $D:=\int_{0}^{\infty}\Gamma(\tau)d\tau$ be the average infectious period (mean generation time).  Taking $\beta$ independent of the calendar time $t$, the \textit{actual (or effective) reproduction number} can be written as the incidence-prevalence ratio:
\begin{equation}\label{eq_Rdefined}
R(t)= \text{IPR}_t\cdot D.
\end{equation}
The effective reproduction number represents the average number of secondary infections generated by each new infectious case (assuming $n_s$ and other environmental variables retain their current values forever).   It has been employed in interpreting the course of an epidemic\footnote{Diseases with long generation times usually exhibit strong dependency of infectiousness on infection-age, indicating that the effective reproduction number might not be as useful as the instantaneous reproduction number. Although it appears that the effective reproduction number may not be useful for a disease with a long generation time (e.g. HIV/AIDS), it might be extremely useful for a disease with acute course of illness, especially when we have both prevalence and incidence in hand, see \citet{yan2019quantitative}.}. 

		Let us assume that during an outbreak, only a certain fraction of infected persons are observed through direct testing, other infectious individuals are not observed, e.g., because of lack of symptoms or the mildness of their illness.
		In particular, asymptomatic secondary transmissions, caused by those who have been infected and have not developed symptoms yet, and also by those who have been infected and will not become symptomatic throughout the course of infection, must be considered.
		At a given calendar time, $t$, we imagine that the important new cases are not the observed newly infected (which are put in a position so as not to infect, via safety protocols), but rather the fraction of newly infected that are not observed.  These unobserved infected (or at least some of them) spread the disease around.  
		Therefore, we split the incidence into two parts:
		\begin{align*}
		\,\, j_o(t)&=\lambda_t \,j(t)\\
		j(t)&=j_o(t)+j_\text{x}(t),
		\end{align*}
		where $\lambda_t$ is the rate of detection which can change over time depending on the details of and degree of adherence to testing protocols and medical screenings.\footnote{The subscript $o$ stands for ``observed''.}
		Thus, the relation between non-detected infected and detected infected individuals is:
		\begin{equation}
		j_\text{x}(t)= \dfrac{1-\lambda_t}{\lambda _t}\; j_o(t).
		\end{equation} 
		If the population screening procedure is effective, 
		we have $\lambda =1$. This could happen, for example, if the infected group is made up of only symptomatic persons which are infectious only after the onset of symptoms.  This is the perfect situation for stopping the outbreak, as all of the infected individuals are detected and (ideally) contained.
We relate the variable $\lambda$ to contact tracing technologies which can be used to make $\lambda$ closer and closer to the ideal value of $1$, as shown by \citet{ferretti2020quantifying}.  		
		Now, let us imagine that some infectious individuals have not been detected and isolated, so that some of them are free to move and have contacts with the susceptible population. We wish to evaluate a measure of risk of exposure for a given susceptible individual.  We assume a kinetic approach to the evaluation of this  risk, where unobserved spreaders are free to infect other individuals. We imagine that the contagion acts within a certain radius of an infected individual.  In a gas, this radius would correspond to the interaction crossection of a gas particle.  We imagine an environment in which two types of individuals are present at a calendar time $t$: $N_s$ is the number of susceptible individuals in a region along with another $J_x$ individuals, which are infectious but have not been detected and are free to move in that region.
		
		\begin{figure}[!ht]
			\centering
			\begin{subfigure}[c]{0.5\textwidth}
				\centering
				\framebox{\includegraphics[angle=0,origin=c,width=0.9\linewidth]{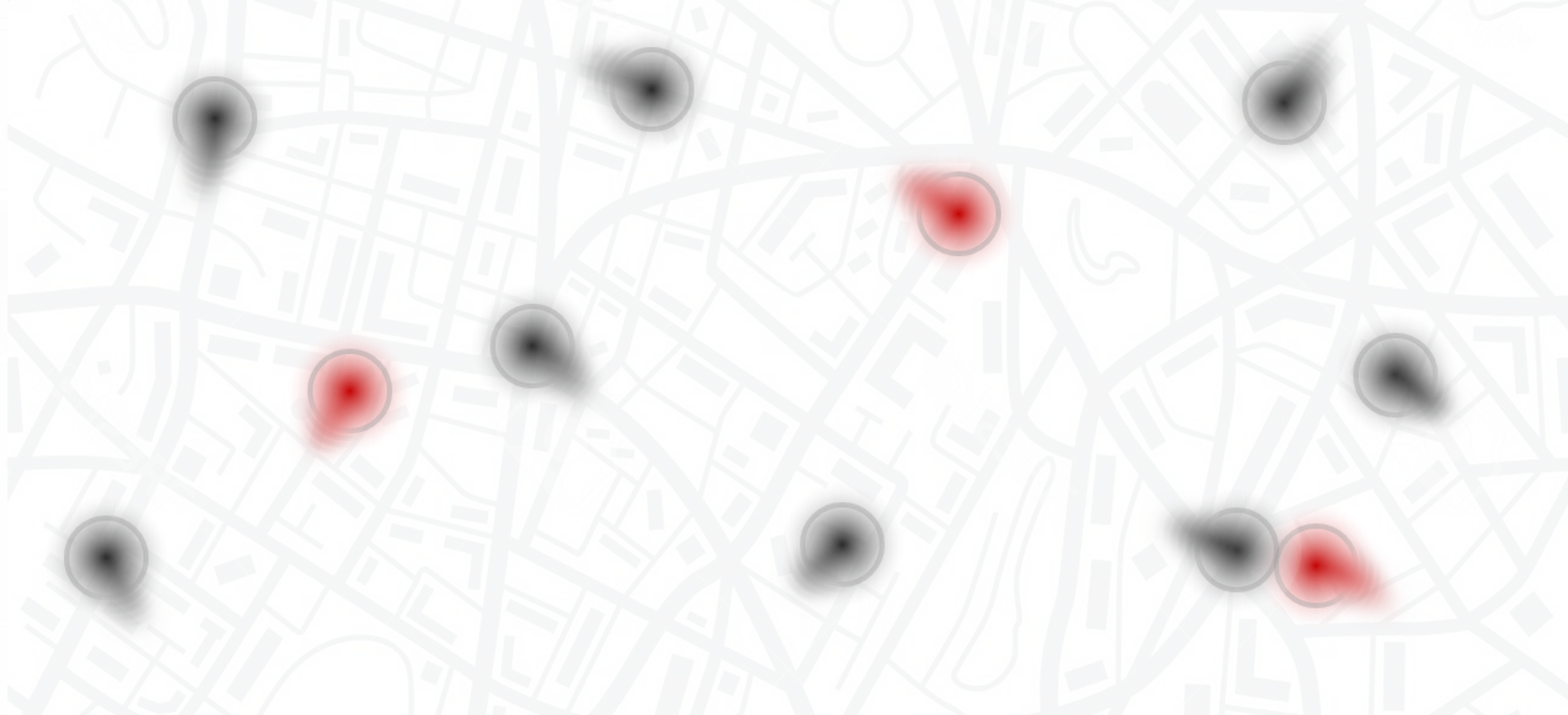}}
				\caption{For a given mobility, red circles are the undetected persons that could interact with susceptible individuals (black circles) which are at risk of infection. }
			\end{subfigure}\quad 
			\begin{subfigure}[c]{0.4\textwidth}
				\centering
				\includegraphics[angle=0,origin=c,width=0.6\linewidth]{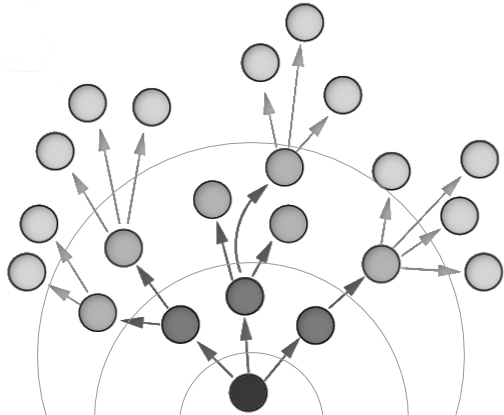}
				\caption{Temporal intervals between consecutive generations of infectious individuals.}
				\label{fig_collisionGT}  
			\end{subfigure}%
			\caption{Frequency of contacts of infectious agents with susceptible individuals in a given region during time intervals of size equal to the infection-generation timescale, $\tau_g$}
			\label{fig_collision}  
		\end{figure}
		
		We consider the regional mobility, $\mu$, to be the average distance explored by each individual during the time interval, $t$, (usually daily).  We define the distance, $r$, to be the maximum distance that an infected person can be from a susceptible person (in the model) and still cause them to become infected.  This distance depends, for example, on the virus' infectiousness as a function of distance and on the viral load. Physical distancing regulations and hygienic norms (such as mask wearing) will result in a decrease in $r$. 
		Using the collision theory for chemical reactions in solution with two types of molecules, we can write down the rate of contacts between the two types in a given volume, per unit time:
		\begin{equation}\label{eq_Bcrosssection}
		z= n_s j_\text{x} 2\pi r \mu.
		\end{equation}  
		Where we have assumed that all agents are ideal point particles that do not interact directly, and travel through space in straight lines.  We further assume that the collisions are instantaneous and elastic. 
		However, not all contacts will result in secondary infectious, rather only those contacts that have sufficient viral load so as to surmount a certain threshold for triggering the infection. Such transmission efficacy should depend inversely on the physical distance between individuals.  Moreover, the collision rate, in reality, depends on time and, in general, on the epidemic's evolution.  This is because the total number of agents changes over time. 
		As an approximation, we embed all of these complexities in the choice of $r$, so to maintain the simplest form of eq.\eqref{eq_Bcrosssection}.
		
		As discussed in \citet{gielen1997epidemics,champredon2018equivalence}, a phenomenological approach to epidemics based on a renewal equation for the incidence (number of newly infected individuals), is defined using the rate of secondary transmissions at a time, $t$, and infection-age, $\tau$, for a population concentrated in the unit of area $d{A}$:
		\begin{equation}\label{eq_incidence}
		j(t,\boldsymbol{x}_0)=\frac{j_o(t,\boldsymbol{x}_0)}{\lambda (t)}=\int_{0}^{t}\Gamma(\tau)\int_{\mathcal{A}} Z(t,\tau,\boldsymbol{x},\boldsymbol{x}_0) \,j_o(t-\tau,\boldsymbol{x}) d{A} d\tau +i(t).
		\end{equation}
		Where $\Gamma(\tau)$  is the survival function of infectiousness, i.e., the probability of being infectious through at least an infection-age of $\tau$. The non-linearity of an epidemic is characterized by the secondary transmission rate  $Z(t,\tau,\boldsymbol{x})$, that is the integral kernel informing the rate of secondary transmissions per single primary case at infection-age $\tau$, at the position $\boldsymbol{x}$ of the region of area $\mathcal{A}$ .  The quantity $j_o(t-\tau)$ is the rate of infection of new cases (incidence) at a time $\tau$ before the present. Finally, $i(t)$ is a function that describes the effects of an external source of infected persons.  For the special case $i(t) = A\delta(t)$, it encodes the number of initially imported infected individuals.\footnote{Note that one could completely disregard the external source of infectious individuals, by modelling an infinitely old epidemic where $\tau \in [0,\infty)$ in the renewal integral: 
			\begin{equation}\label{eq_incidence1}
			j(t,\boldsymbol{x}_0)=\int_{0}^{\infty}\Gamma(\tau)\int_{\mathcal{A}} Z(t,\tau,\boldsymbol{x},\boldsymbol{x}_0) \,j_o(t-\tau,\boldsymbol{x}) d{A} d\tau.
			\end{equation}}

		The propagator kernel $Z$ can be expressed in terms of the collision theory of non-interacting spheres of radius $R$ in eq.\eqref{eq_incidence} as:
		\begin{equation}\label{eq_full}
		j_o(t,\boldsymbol{x}_0)= n_s(t,\boldsymbol{x}_0)\lambda(t)\int_\mathcal{A} \Theta\left[r(t,\boldsymbol{x})-\lvert\boldsymbol{x}-\boldsymbol{x}_0\rvert\right]\mu(t,\boldsymbol{x})\int_0^t \tfrac{1-\lambda(t-\tau,\boldsymbol{x})}{\lambda(t-\tau,\boldsymbol{x})}\,j_o(t-\tau,\boldsymbol{x})\eta(t,\tau)\Gamma(\tau -\tau_g)d\tau d\boldsymbol{x}.
		\end{equation}
		Here, $\eta(t)$ captures the probability that a contact will transmit the infection, and $r$ is the infectious radius.  The latter is inversely related to the minimum mandated physical distance between individuals, $\rho$, so that the greater the physical distance between individuals is the smaller the infectious zone $r=(2\pi\rho)^{-1}$ is.
		Note that if it were possible to detect, track, and isolate every newly infected individual, an epidemic could be stopped within a time $\tau_A$.  It is useful to note that the detection rate, $\lambda$, is essentially a scaling factor for the survival probability $\Gamma(\tau)$.  The value of $\lambda$ could change with infection age as well as $t$ during the disease outbreak.  These changes might depend, for example, on the ability to detect and isolate individuals, or the efficiency of contact tracing during the epidemic.  The contact tracing efficiency varies with the characteristics of the infection and the speed and coverage of the tracing process.  At the beginning of a large outbreak, testing and manual tracing quickly becomes an unmanageable strategy and a lockdown to reduce physical contact may then become a more efficient and effective means of controlling the epidemic.  However, lockdowns aren't sustainable in the long term because of their social, economic, physical, and mental health effects.  Lockdown policies have reduced the spread of CoViD-19, but as restrictions are relaxed transmission may go up again.  Hopefully, with a testing, tracking, and tracing strategy, and additional hygienic precautions in place it will still be possible to keep the epidemic under control. 
		In table \ref{tab_collision}, we make a summary of the typical factors which contribute to the transmission of a disease.  The biological and environmental properties are accounted for in the Transmissivity variable $\eta$.  Physical proximity, viral load, and environmental conditions determine the infectious dose necessary to trigger the infection in a new host. For example, closed environments such as workplaces and schools correspond to higher $\eta$ values in the model as compared to an outdoor space.  Another important component is the temporal duration of a contact that in our model is considered negligible.  This approximation is then collapsed into $\eta$ as an average exposure to the viral particles so as to determine an infection after the existence of a contact has been established.  
		\begin{table}[!ht]
	\centering\captionsetup{justification = centering}
	\caption{Parameters of kinetic approach to infectious contacts. Mobility times proximity gives the rate of interactions. }
	\label{tab_collision}
	{
		\setlength\arrayrulewidth{.001pt}
		\begin{tabular}{Sc|Sc}
			\makecell{\textbf{Collision variable}} & \makecell{ \textbf{Description}} \\ \toprule 
			\rowcolor{Gainsboro!20}
			\makecell{\textsl{Mobility}\\ $\mu$}     & {\small  \makecell{ \textit{Movement trends over time}} }   \\ 
			\makecell{{Social Movements} }     & {\footnotesize  \makecell{ Average velocity and the path length\\ of individual trajectories} }   \\ \midrule
			\rowcolor{Gainsboro!20}
			\makecell{\textsl{ Infectious Zone }\\ $r$}      & {\small  \makecell{\textit{ The area in which contact with an infectious individual} \\ \textit{can trigger a secondary infection in airborne diseases}} }   \\ 
			{ \makecell{{Physical Proximity} } }    & {\footnotesize  \makecell{ Average distance between persons  \\ to be considered for airborne diseases  \\ (interpersonal distance, protection devices and hygienic procedures affect it)}} \\ \midrule
			\rowcolor{Gainsboro!20}
			\makecell{\textsl{Transmissibility}\\ $\eta$}      & {\small  \makecell{ \textit{The chance that a contact results in an infection}} }  \\ 
			\makecell{ Viral Load}      & {\footnotesize   \makecell{ Concentration of viral particles \\ in the material being shed by an infected patient} }  \\ 
			{\makecell{Contact duration}    } &  {\footnotesize \makecell{ Period of time of a collision}}  \\ 
			{\makecell{ Environment}    } &  {\footnotesize \makecell{ Air flow, UV exposure, climate factors such as\\ temperature and humidity that influence infectiousness }}  \\ \bottomrule		
		\end{tabular} 
	}
\end{table}
The availability of reliable data over time is key to lifting containment measures.  In particular, there needs to be sufficient monitoring of the progression of the coronavirus pandemic, including through large-scale testing. There are two main types of CoViD-19 tests.  Swab tests, which usually take a sample from the throat or nose, to detect viral RNA. These determine if one currently has CoViD-19.  The other type is blood tests, which detect antibodies.  This type of test can provide evidence to determine whether one has had CoViD-19, and is now immune.  Although tests can perform well in ideal laboratory conditions, in practice lots of other factors affect accuracy including: the timing of the test, how the swab was taken, and the handling of the specimen.  The meaning of a test result for a given person depends not only on the accuracy of the test, but also on the estimated risk of disease before testing.

		{Making reasonable assumptions, we can write a mean-field approximation of the renewal equation for incidence evolution over time.  This equation is useful for evaluating the growth rate of epidemics directly from the reported data of the disease.}
		Firstly, we assume the spatial homogeneity of every variable.  In particular, we consider the average distance, $\rho(t)$, between individuals, their average mobility, $\mu(t)$, and the fraction of missed cases, $\lambda$, to be spatially homogeneous and constant with respect to infection age.{ Moreover, we define a typical time interval necessary to detect an individual to be infected.  We conservatively assume it to be equal to its maximum possible value, the serial interval which (we assume for simplcity) equals the generation time, $\tau_g$.  The detection time can, in general, depend on calendar time (as screening procedures improve over time).  We consider the detection-age to be our time-scale for the evolution of the observed infected individuals.}  The detection time is always at least equal to the latent period and can be thought of as equivalent to the incubation time plus the time needed to screen for the infection and isolate the infected individual.
		This is why we take the mean detection time to have the same value as the generation time and the serial interval of the contagion.  Thus, we can assume a window of infectiousness which takes into account the fact that the secondary infected person has to be infected at least $\tau_g$ days after contact with the primary infected person to be infectious.  Consequently, we consider the survival probability to be windowed between $\tau_g$ and $\tau_A$.  Therefore, eq.\eqref{eq_full} can be rewritten in terms of the expected number of new cases $J_o(t)=j_o(t)N$:
		\begin{equation}
		J_o(t)\approx \eta(t) N_s(t-\tau_g)\,r(t-\tau_g)\,\mu(t-\tau_g) \sum_{\tau=\tau_g}^{\tau_A}\lambda(t)\dfrac{1-\lambda(t-\tau)}{\lambda(t-\tau)}J_o(t-\tau).
		\end{equation}
		We assume that the mobility (and infectious zone size) of an infector in the past doesn't affect their ability to infect someone, except on the day the person is actually infected.
		Furthermore, the sum over $\tau$ is there to account for the fact that people with a range of infection ages can infect someone.\footnote{If we were taking into account variations in the time between infection and detection, we would have to integrate over changes in mobility during that time period.  However, that would not be the same sum as the sum over $\tau$.}

The actual reproduction number can  be  used  as  a  predictive  tool  to track the epidemic's evolution.  It is also a measure of epidemic risk, in the sense that if it is above one for an extended time period, then an outbreak is possible.  Thus, by linking a dynamical model with time-series data, one obtains a measure of epidemic risk.  This risk can be derived from eq.\eqref{eq_Rdefined}, leading to the effective reproduction number
\begin{align}\label{eq_Rtmob}
{R(t)}&\sim \frac{n_s(t-\tau_g)}{n_s(t_0-\tau_g)}\frac{\eta(t-\tau_g)}{\eta(t_0-\tau_g)} \frac{\rho (t_0-\tau_g)}{\rho(t-\tau_g)} \frac{\mu(t-\tau_g)}{\mu(t_0-\tau_g)} R(t_0),
\end{align}
where $t_0$ is an initial (or calibration) time. The above equation represents the change in the average number of secondary cases caused by a single primary case throughout the course of infection at calendar time $t$ with respect to an initial value (for example, before the lockdown).  All variables should be interpreted as average values.   Note that the expression for $R(t)$ does not depend explicitly on the distribution of infection survivial times, $\Gamma$.  It only depends on the typical time between infection and detection.  So, the $\Gamma$ distribution can be any distribution as long as that timescale does not change perceptibly.

 From eq.\eqref{eq_Rtmob} it is evident that as the epidemic evolves, the force of infection is reduced for various reasons, primarily due to physical distancing policies adopted by most countries in the form of a lockdown of the population's mobility.  Since it is not practical to reduce physical distancing beyond a certain socially and economically acceptable level, the only possible ultimate reasons for the end of an epidemic are the depletion of susceptible population (immunization), a change in the intrinsic infectiousness of the virus, a sustained change in public hygiene habits (mask wearing, physical distancing, etc.), or innovation in contact tracing, testing, and isolation, see \citet{nepomuceno2020besides} for a discussion.

\section{Data  and Results}	
In the present section, we apply our theoretical approach to real-world data.  The data repositories used to obtain our results are listed in Table \ref{tab_data}.  This table includes both the epidemiological and social distancing data sources.  We follow two approachs in evaluating proxies for physical proximity. The first approach is in regards to the Italian evolution of the CoViD-19 epidemic along the lines of social distancing trends. The rest of the studies address the course of the epidemic in some US states and three European countries: the UK, Ireland, and the Czech Republic, for which data on physical proximity are available to us.

\begin{table}[!ht]
	\centering\captionsetup{justification = centering}
	\caption{Repositories used for epidemiological and social distancing data. }
	\label{tab_data}
	{
		\setlength\arrayrulewidth{.001pt}
		\begin{tabular}{Sl}
			\makecell{\textbf{Data Repositories}}  \\ \toprule 
			\rowcolor{Gainsboro!30}
			\makecell{\textsl{Epidemiological}}      \\ 
			\makecell{{Johns Hopkins University \cite{JohnHopk}}} \\ {CoViD Tracking Project \cite{CovidTracking}} \\ Dip. Protezione Civile \cite{ProtezCivileCov}\\ Epicentro ISS \cite{epicentro}  \\ 
			Gov.UK CoViD-19 \cite{UKgov,GreatLondon}			\\ \hline
			\rowcolor{Gainsboro!30}\hline
			\makecell{\textsl{$R_t$ estimations }}      \\ 
			Rt live \cite{rtlive}  \\ Epiforecast \cite{epiforecasts} \\ CoViD-19 Projections  \cite{covidprojections} \\  \midrule
			\rowcolor{Gainsboro!30}
			\makecell{\textsl{Social Distancing}}       \\ 
			Google Mobility \cite{Google19} \\ Voxel51 Physical Distancing Index\cite{voxel} \\ Analisi Distribuzione Aiuti \cite{ProtezCivileADI} \\ 
			\hline Unacast Social Distancing Dataset \cite{unacast}\\
			 \bottomrule		
		\end{tabular} 
	}
\end{table}

Physical distancing (also known as social distancing) is a practice recommended by public health officials to stop or slow down the spread of contagious diseases. It requires the creation of physical space between individuals who may spread certain infectious diseases. The use of cloth face coverings should reduce the transmission of CoViD-19 by individuals who do not have symptoms and may reinforce physical distancing. Public health officials also caution that face coverings may increase risk if users reduce their use of strong defenses such as physical distancing and frequent hand washing. 
We split physical distancing into two components: movement (mobility) and distance (proximity, interpreted as radius of contact).

As regards to the former, we essentially have relied on \citet{Google19} mobility open source data.  We interpret that data as movement trends, reflected in changes in numbers of visits to various categories of places (parks, public transportation hubs, residences, etc.).  Google reports these changes relative to a average baseline on the same day of the week evaluated before the pandemic outbreak. We have used the the average mobility across all the places (residential included).
The mobility is not intended to be the average speed of a given active particle.  Since we are interested in the rate of collisions, the speed of a particle approaching the particle of interest is just as important as the speed of the particle of interest. So, we take mobility to be the average relative speed of the group.  If the velocities of all individuals are uncorrelated, the mobility, i.e., the relative velocity in a certain society is proportionally related to the mobility of each individual\footnote{  There are many definitions of mobility and there are multitudinous data repositories. Mobility trends are subdivided into movements, distances traveled, and shelter-in-place trends. The first accounts for  changes in movement fluxes, as in the data from Google and \citet{Apple19}.  The second type is the change in average distance traveled by individuals.  This type of mobility data is available from Safegraph \citet{safegraph}, or other repositories such as \citet{unacast,cuebiq} for the US.   These repositories also include shelter-in-place analysis, and contact/encounter density. As for international data, other repositories provide similar data.  For example, Facebook Data for Good Mobility Dashboard \cite{FacebookMob}, which gives the change in frequency of travel and the percentage of people staying put.  When kilometers traveled per day is used as a mobility proxy, one can use the principle of maximum entropy, under the assumptions that the individuals are independent particles and that there exists an average daily trip length in the population, see \citet{bazzani2010statistical}. Thus, the distribution of distances travelled is exponential, $p(L)=L_0e^{-L/L_0}$.  Here $L_0$ is the characteristic daily path length reported in mobility data. We checked that distance traveled per day and movement requests show almost exactly the same trend.}. 

As regards the physical proximity we use two different proxies according to the availability of data in the aforementioned repositories.  For the case of Italy, we use data on the number of facial coverings and masks distributed among the population which we interpret as inversely proportional to physical proximity in the model.  This can be considered reasonable at least during the period of lockdown. In the work of \citet{macintyre2020physical,chu2020physical}, the authors show that the risk of getting infected drops by half for every additional meter of distancing\footnote{In particular, by keeping a physical distance from another person of 1 meter, or 3.3 feet, the chance of transmission falls to 12.8 percent, and a distance over 3.3 feet reduces the chance to 2.6 percent}.
 Essentially, we assume that the use of face coverings in this population corresponds proportionally to an awareness of the importance of increasing interpersonal distance. In particular, we assume that the number of people wearing a face mask is consistent with an increase of human distance. If this tendency is constant over time, the ratio between the sizes of infectious zones at two different periods of time will cancel out the proportionality factors, thus accounting for the share of face masks used.  
Regarding the US and certain other European countries (Ireland and the Czech Republic), we use an inferred measure of average proximity among individuals, as explained in the following paragraphs.
Deep learning models are able to detect and identify pedestrians, vehicles, and other human-centric objects in the frames of each live street cam video stream in real-time. Using images sampled from each video stream every 15 minutes, \citet{voxel} computes the Physical Distancing Index or PDI, an aggregate statistical measure that captures the average density of human activity within view of the camera over time. The PDI value for a particular day will not exactly correspond to the number of people in that exact frame. Rather, the PDI value at time, $t$, is an average measure of peak activity during a window of time (a few days) around that time. So, a large PDI value means that there were a lot of people out and about around that time, at some point. Voxel51's PDI website presents measures of activity at a single location in the city. While these measures are likely correlated with overall trends in activity in the city, and are thus an interesting proxy for public behavior, correlation is not guaranteed. This project uses Artificial Intelligence (AI) and camera feeds to compute a Physical Distancing Index to track physical distancing behaviors in real-time in cities around the world. 
We know from data the concentration of people in a certain area, and so we have an estimation of the density of individuals in a two-dimensional plane. From that information, one can infer the mean distance among individuals, assuming random positions.
The maximum entropy distribution would be a uniform probability density of particle positions.  This corresponds to an exponential distribution of inter-particle distances.  The characteristic length scale (equal to the mean distance between particles) is the inverse of the square root of the particle number density (up to a factor of order $\pi$).

Consequently the average time between collisions is $\tau_{\mu}=\ell /\mu$, where $\ell$ is the mean free path, defined to be the average distance traveled by the particle between each collision $\ell=1/4\pi r d$, where $r$ is the radius of the particle and $d$ is the particle density ($d=N/A$) over a region of area $A$.
Suppose that the probability that a molecule undergoes a collision between a time, $t$, and a time, $t+dt$, is given by $\gamma dt$,for some constant $\gamma$ (the collision rate).  If we assume $\gamma$ is a constant, this implies that no memory of previous collisions remains at the time of any later collision. Calling $P(t)$ the probability that the particle has not undergone to a contact from time $0$ up until time $t$, then the probability that it further makes it to a time $t+dt$ without collision is $P(t+dt)=P(t)(1-\gamma dt)$ so that $P(t)=\gamma e^{-\gamma t}$.  Yielding the result that the collision rate is the inverse of the collision time $\gamma =1/\tau_{\mu}$.

In the general analysis of epidemic data we refer to reported infected persons by their dates of diagnosis via laboratory test. However, some countries also report infections by the date of first symptoms reported by patients. In particular, we have used the latter type of data when possible (Italy) and inferred it in the case of the USA and the UK via an analysis of the effective reproduction number assessed by \citet{epiforecasts}, \citet{covidprojections} and \citet{rtlive}.   This is significant, since up to three weeks can pass between the day of infection and that of diagnosis with CoViD-19 \footnote{Furthermore, reports by first date of symptoms do not suffer from the strong weekly periodicity due to lower rates of processing of tests (and other data) on weekends (and holidays).  The delay between infection and a positive test result comes from the incubation time of the disease (up to fourteen days) and from the time periods between the beginning of symptoms, seeking of medical assistance, and the completion of diagnostic laboratory tests. As a consequence, tracking the evolution of epidemics from data can suffer from a delay of about fifteen days.}. We use epidemiological data at the level of states, meanwhile we use mobility data at level of cities for US locations and at the level of state for EU countries as available from Google's mobility reports \cite{Google19}. Finally, the proximity index has been obtained at the level of cities. 

\begin{figure}[!h]
	\centering
	\begin{subfigure}[c]{0.8\textwidth}
		\centering
		\includegraphics[angle=0,origin=c,width=0.8\linewidth]{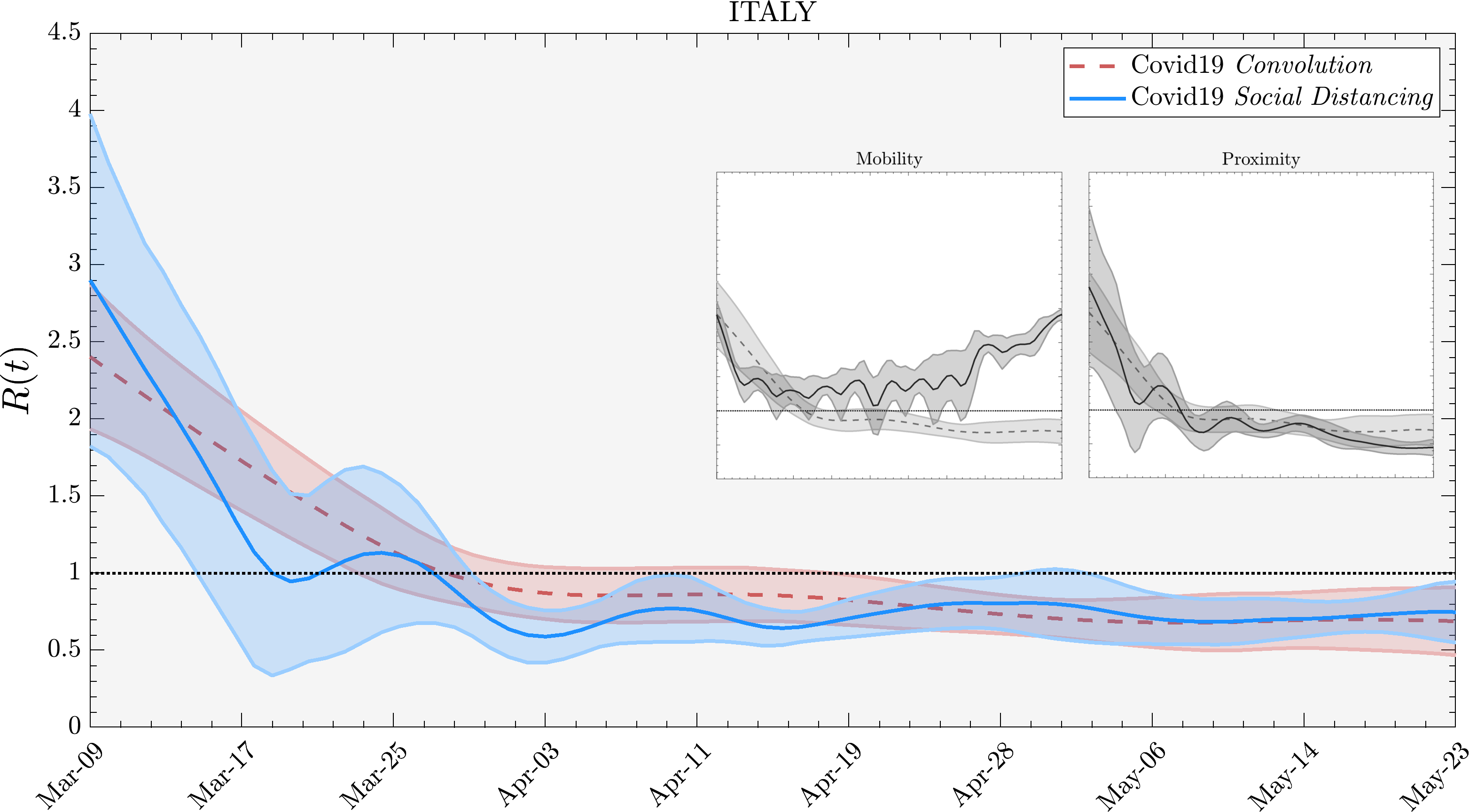}
	\end{subfigure} 
	\caption{Actual reproduction number for Italy during the lockdown period (March 9th to May 18th). We compare the traditional derivation with the analytical derivation from the collisional approach of this paper.  We have taken into account data on the depletion of the susceptible population and physical distancing policies made up of social mobility and physical proximity. In the inset we show a lockdown only on mobility without any physical proximity change. We see that without maintaining physical distance or  any protections such as facial masks, the epidemic growth increases with the disinhibition of social movements. In reality this has not happened.}
	\label{fig_Rtlockdown}  
\end{figure}

\begin{figure}[!ht]
	\centering
	\begin{subfigure}[c]{0.8\textwidth}
		\centering
		\includegraphics[angle=0,origin=c,width=0.8\linewidth]{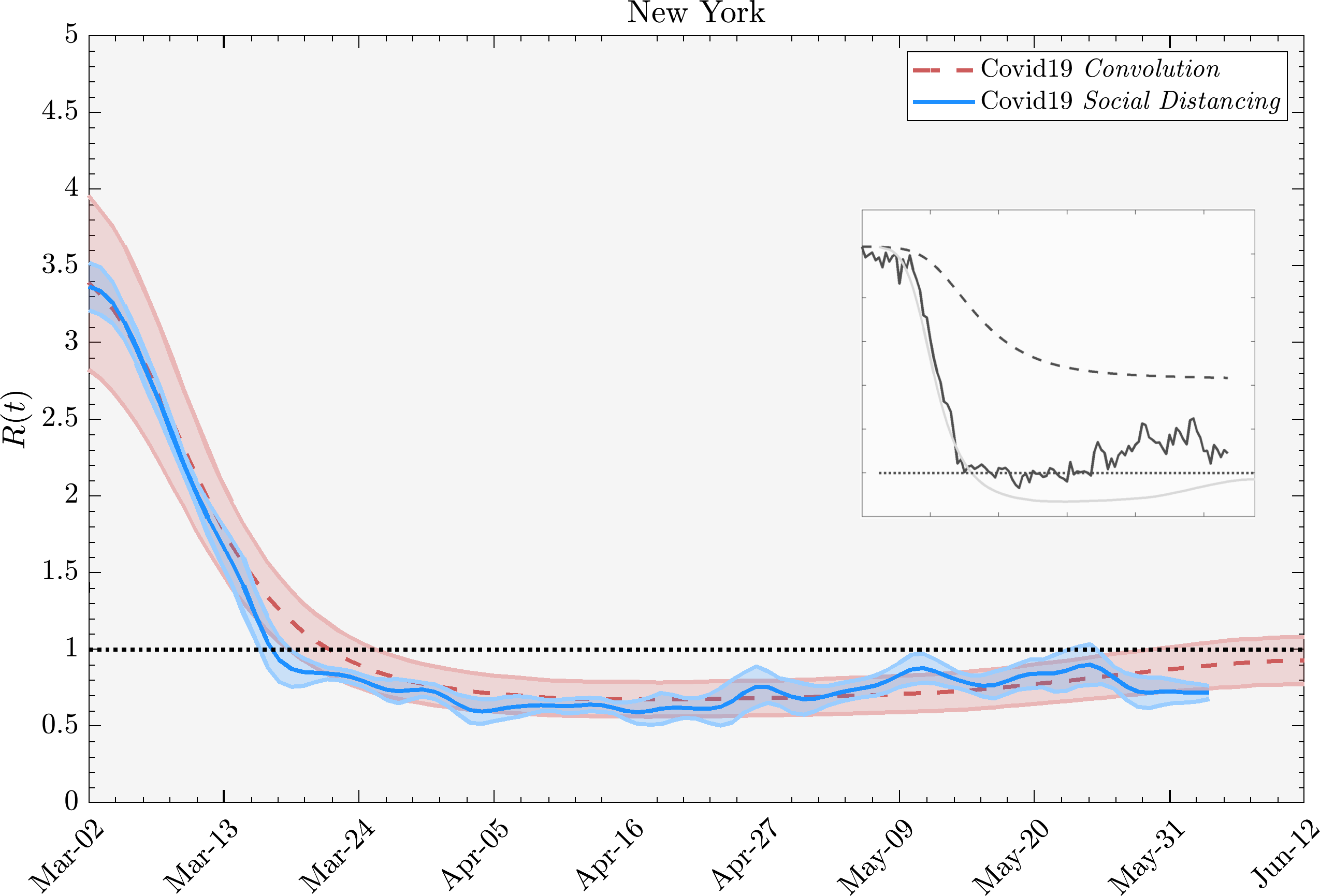}
		\caption{}
	\end{subfigure}%
	\\ \vspace{1cm}
	\begin{subfigure}[c]{0.8\textwidth}
		\centering
		\includegraphics[angle=0,origin=c,width=0.8\linewidth]{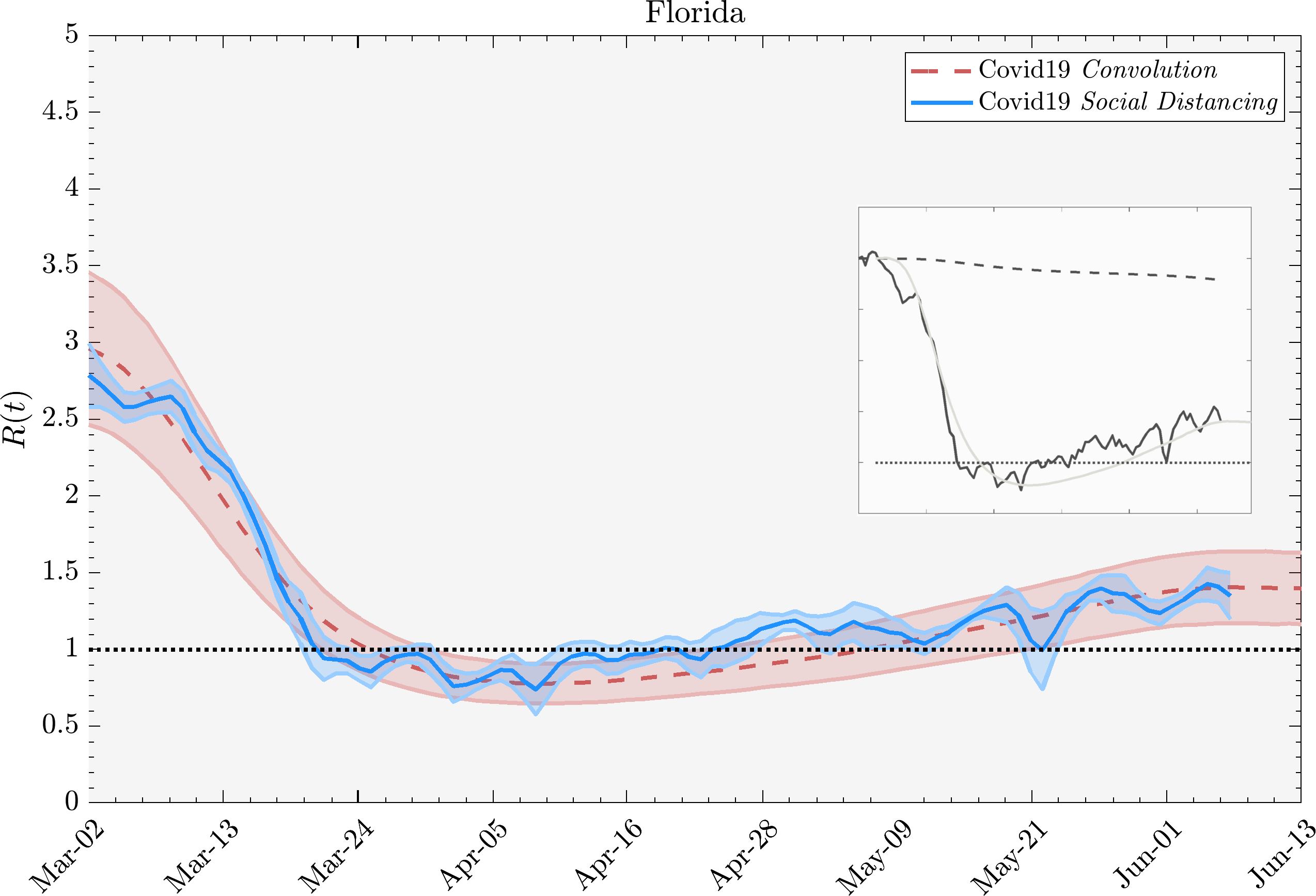}
		\caption{}
	\end{subfigure}%
	\caption{Reproduction number estimates for two US states. Comparison between the reproduction number calculated from symptom onset data as in literature \citet{rtlive} (red line) and the reproduction number computed according our kinetic approach, using data from \citet{Google19} for mobility, \citet{voxel} for the social proximity and \citet{CovidTracking} for epidemic data.   Ribbons are the $90\%$ credible interval obtained via bootstrapping. In the insets, the black solid line is $R(t)$ using physical distancing variables only, meanwhile the dashed black line is $R(t)$ due to the depletion of susceptibles only. Epidemiological data and $R(t)$ estimation are from the date of onset of symptoms (indirectly calculated). }	
	\label{fig_USArt}  
\end{figure}

We have studied and analyzed the regions reported in Table \ref{table_cities}, with the indicated dates of effect of lockdown policies taken from \citet{acaps}.
\begin{table}[!ht]
	\centering
	\begin{tabular}{l|ccc} 
		\toprule
		& Lockdown Started & {Reopening}     & Lockdown Ended  \\ \midrule
				{\scshape Italy}           & March 09  & May 18  & June 03  \\
		New York (US)   & March 22  & May 15  & May 28   \\
		Florida (US)    & April 03  &               & May 04   \\
		California (US) & March 18  & May 24  &                \\
		Nevada (US)     & April 01  &               & May 09   \\
		London (UK)     & March 23  & May 11    &          June 23      \\
		Dublin (IR)     & March 27  & May 18 &  June 08               \\
		Prague (CZ)  & March 16  &               & July 01  \\
		\bottomrule
	\end{tabular}
	\caption{Lockdown policy dates in states and regions in our study. Reopening stands for easing shelter policies before the very end of a lockdown. We use data from the cities as representative of  the physical distancing behavior for the entire regions they are in, for which estimations of reproduction number are available. Since the lockdown policies are made up of progressive enforcement or easing of restrictions, we consider these dates as reference time-stamps for government actions to combat the spreading of the disease. }
	\label{table_cities}
\end{table}
For the case of Italy as a nation, in figure \ref{fig_Rtlockdown} we show the two derivations of the effective reproduction number $R(t)$.  The first is found using an ensemble estimation with the convolutional renewal approach described in Appendix \ref{sec_appendixR}) and Epiforecast \cite{epiforecasts}.  The other derivation is assessed using the physical distancing approach via the above described kinetic theory of collisional infections. We see how the two behaviors are compatible in describing the behavior of the epidemic. We also call attention to the fact that mobility alone is not sufficient to explain the dynamics of epidemics, as discussed in \citet{cintia2020relationship}. We see how physical proximity is crucial in determining why, despite an increase in mobility after the end of the lockdown period, a relatively stable $R_t$  below $1$ persists. Recall that, in the case of Italy, the number of face masks distributed to the populations have been considered as proxy of physical proximity. On the other hand, one should subtract from the susceptible population the number of asymptomatic or undetected  infected individuals, which are not counted in official reports. We provide an estimate for this number in the next section. 
For the other regions under study, we use physical proximity in terms of population density able to move. In particular, for US states we use \cite{rtlive} as estimation of the reproduction number as well as the estimation of susceptible population considered. This is the context for figure Fig.\ref{fig_USArt}. When analyzing the UK, Ireland, and the Czech republic, we use an ensemble of $R_t$ estimates, averaging Epiforecast \cite{epiforecasts},  Covid19 projections \cite{covidprojections}, and our convolution estimation.

\begin{figure}[!ht]
	\centering
	\begin{subfigure}[c]{0.48\textwidth}
		\centering
		\includegraphics[angle=0,origin=c,width=0.9\linewidth]{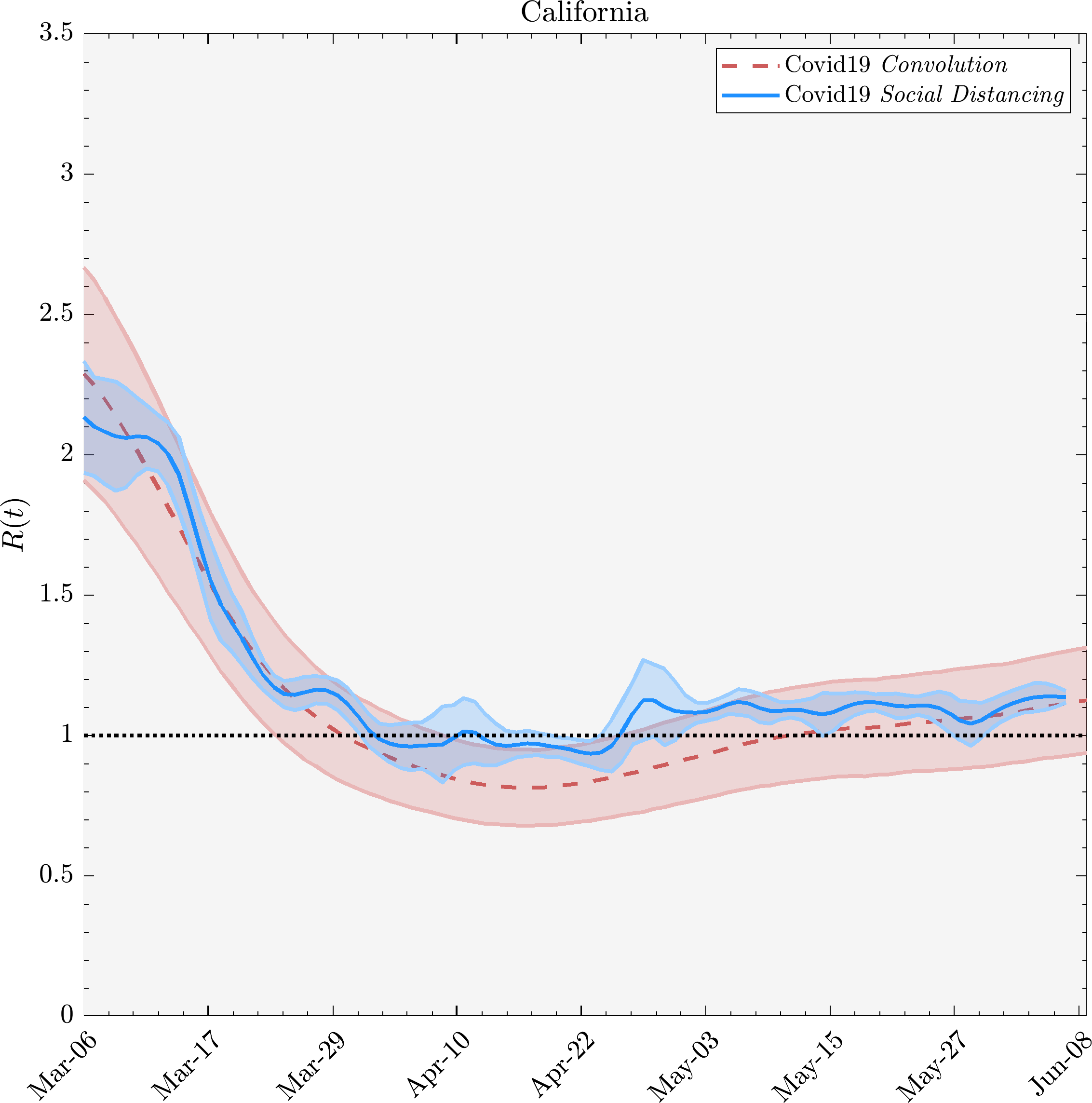}
		\caption{}
	\end{subfigure}%
	\begin{subfigure}[c]{0.48\textwidth}
		\centering
		\includegraphics[angle=0,origin=c,width=0.9\linewidth]{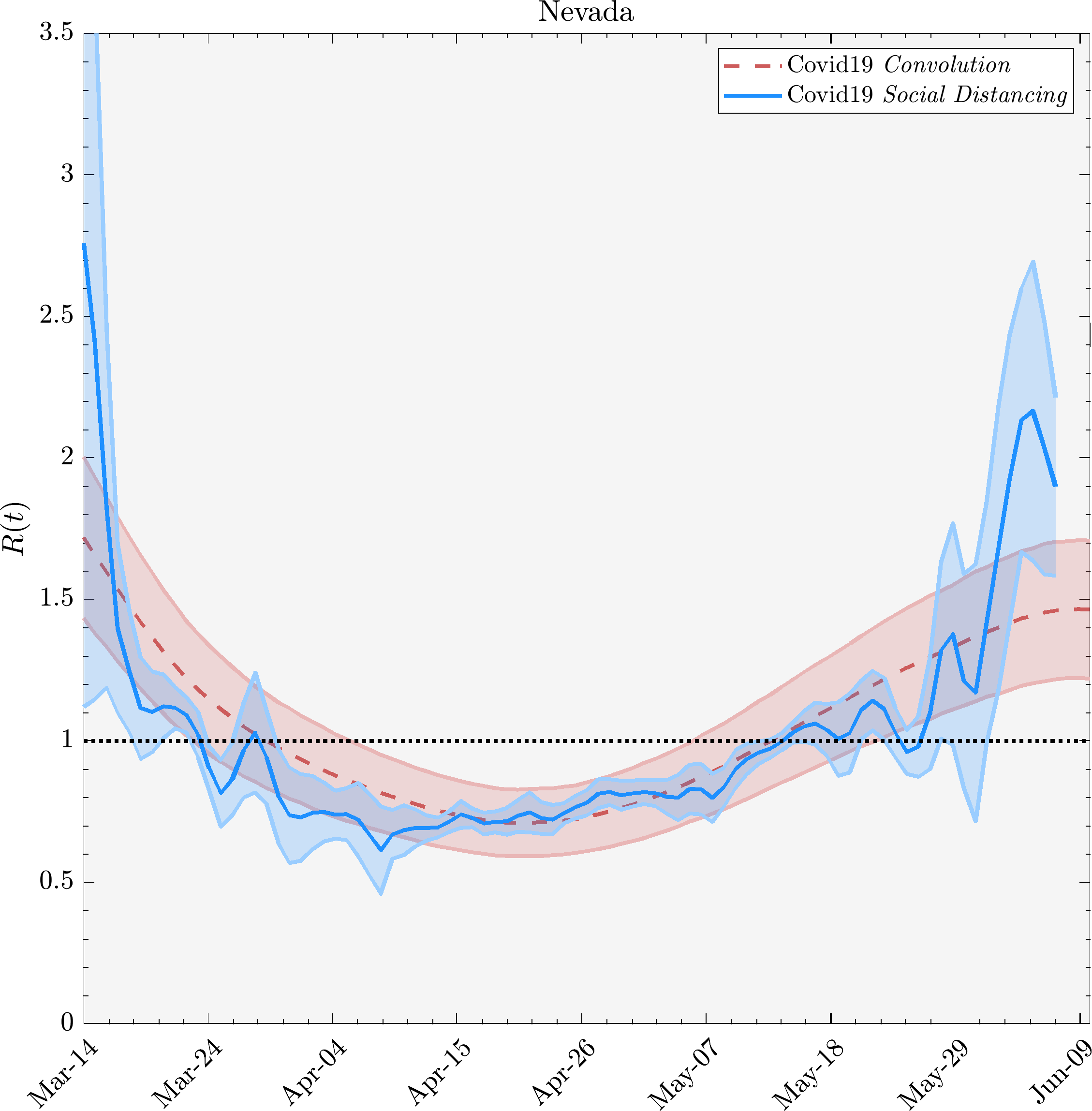}
		\caption{}
	\end{subfigure}%
	\caption{Reproduction number estimates for two US states: California (evaluated in Los Angeles) (a) and Nevada (evaluated in Las Vegas) (b). In the latter case the camera was in a location devoted to leisure activities, this yields a biased estimate of the mean interpersonal distance as compared with rest of the state. }	
	\label{fig_USmore}  
\end{figure}

In figure \ref{fig_USArt}, we show the hardest hit states in the US at the date of our work, New York and Florida.  The degree to which this figure show that one can reconstruct the reproduction number using the social mobility approach is remarkable.  Note that for New York state an important cause for the reduction $R_t$ is due to the depletion of the susceptible population, while physical distancing has a smaller impact.  Meanwhile, in Florida, the behavior is mainly due to physical distancing restrictions taken up at the end of the shelter-at-home policy. We perform the same analysis the US states of California and Nevada, in Fig.\ref{fig_USmore} obtaining similarly accurate results.  

\begin{figure}[!ht]
	\centering
	\begin{subfigure}[c]{0.48\textwidth}
		\centering
		\includegraphics[angle=0,origin=c,width=0.9\linewidth]{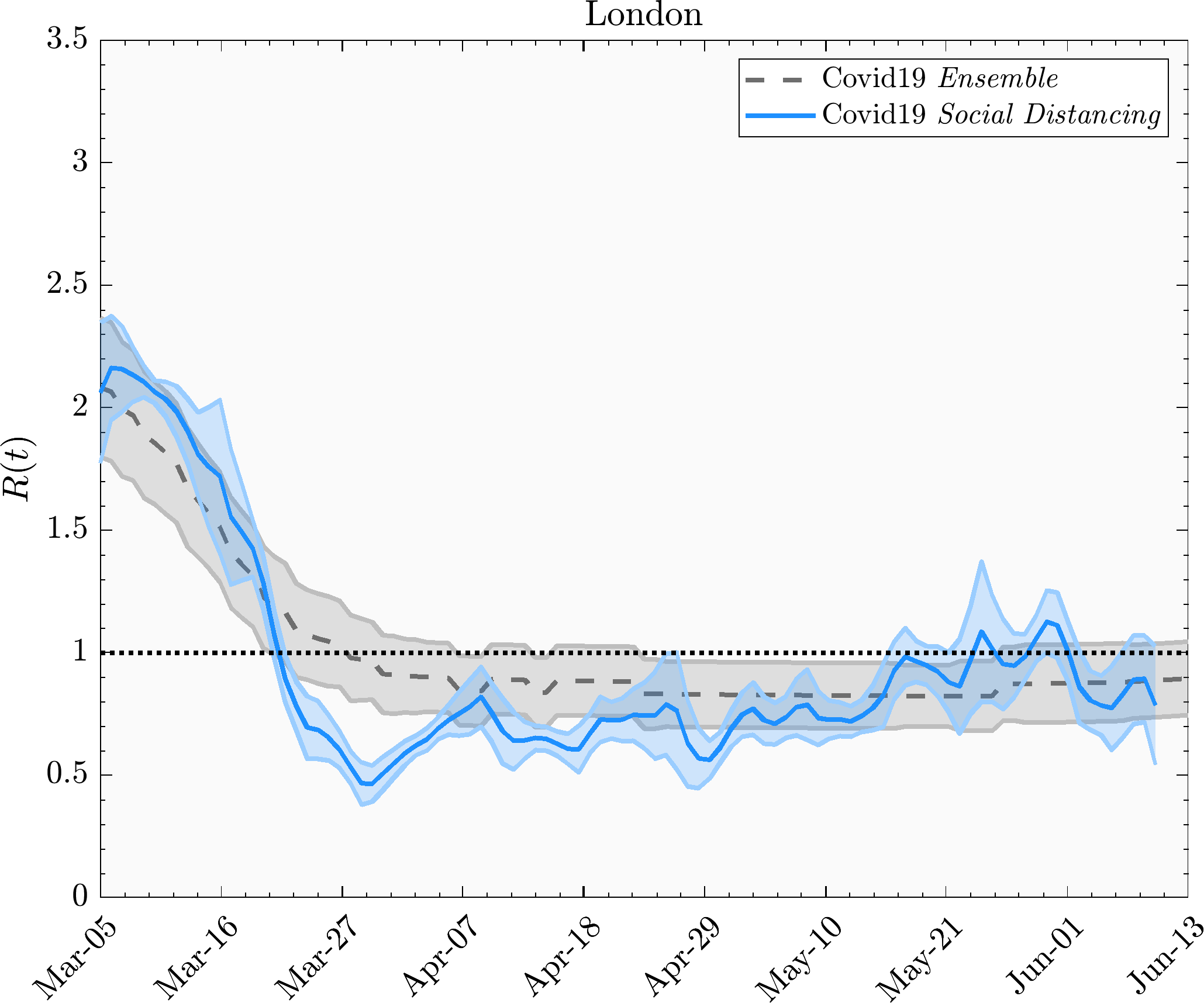}
		\caption{}
	\end{subfigure}%
	\begin{subfigure}[c]{0.48\textwidth}
		\centering
		\includegraphics[angle=0,origin=c,width=0.9\linewidth]{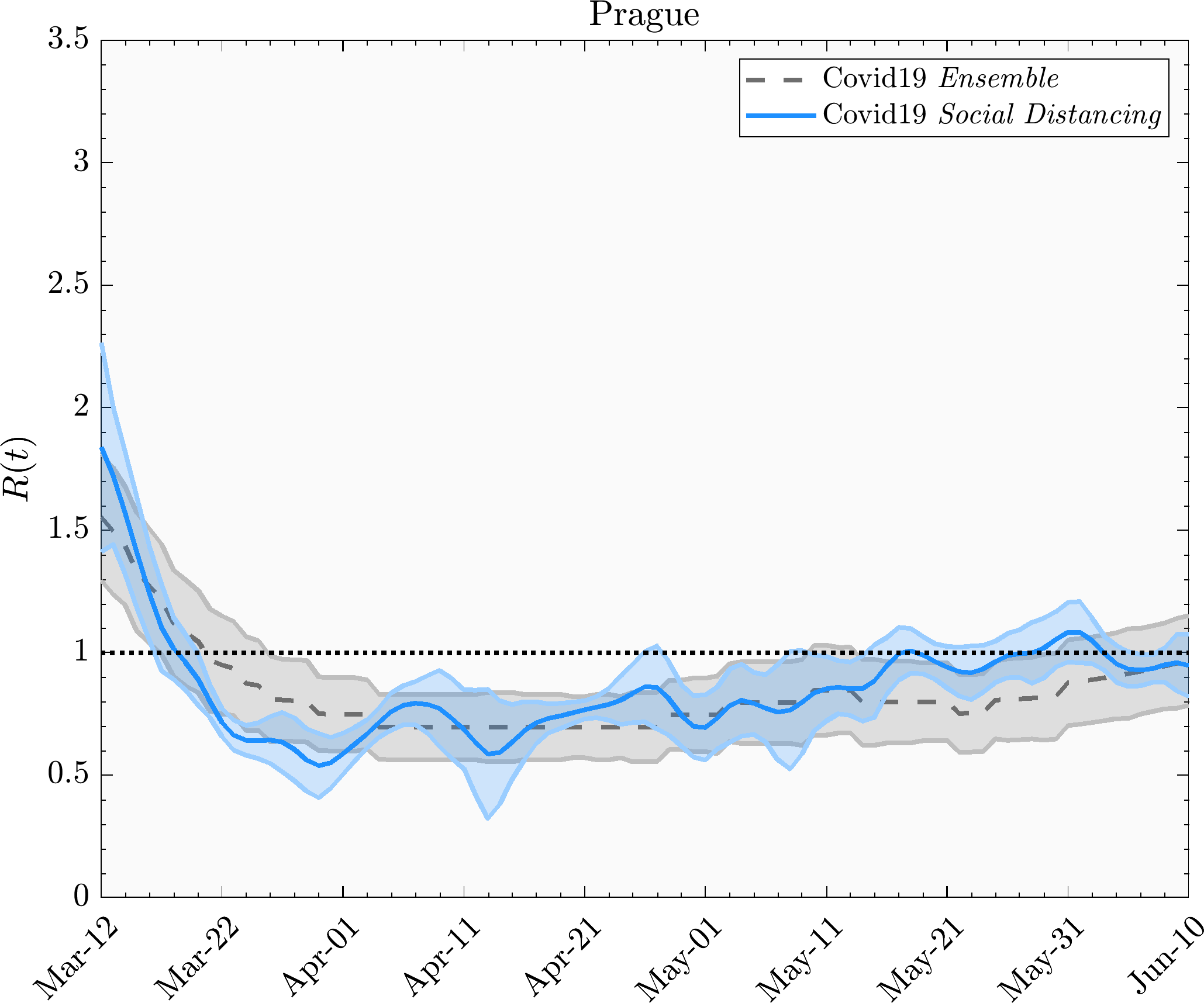}
		\caption{}
	\end{subfigure}%
	\\\vspace{0.5cm}
	\begin{subfigure}[c]{0.48\textwidth}
		\centering
		\includegraphics[angle=0,origin=c,width=0.9\linewidth]{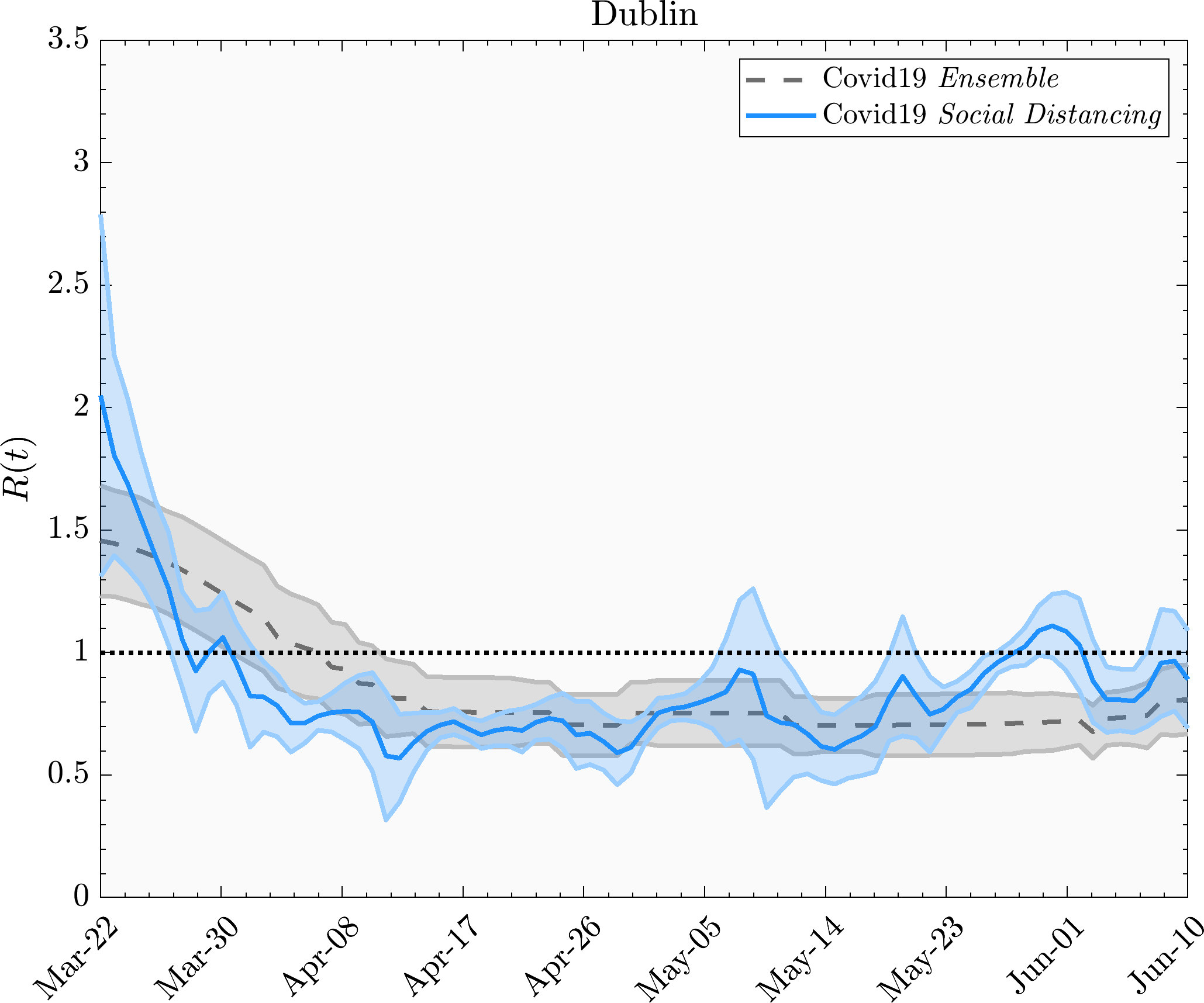}
		\caption{}
	\end{subfigure}%
	\caption{Reproduction number estimates for London, UK (a), Prague, Czech Republic (b), and Dublin, Ireland (c). Comparison between the reproduction number ensemble average of \citet{epiforecasts} and \citet{covidprojections} (red line) and the reproduction number computed according our kinetic approach, using data from \citet{Google19}  mobility, \citet{voxel} for  social proximity and \citet{CovidTracking} for epidemic data. Ribbons are the $80\%$ credible interval obtained via bootstrap. Epidemiological data and $R(t)$ estimation are from data referring to the date of lab diagnosis.}	
	\label{fig_EUrt}  
\end{figure}
If one does not have reliable data to estimate the epidemic risk through the reproduction number, one can look for a proxy of this variable that is less sensitive to a lack of physical distancing data.	
				In fact, in the absence of physical distancing data, the crude reproduction number can be very noisy.  This leads us to consider a proxy of the reproduction number that can predict a generation time period ahead. So we introduce the growth factor which is far more stable with respect to a lack of physical distancing data as long as the epidemiological data is still good, as explained in depth in Appendix \ref{app_growth}.

In the analysis presented in this work, we need to calibrate the measure of the reproduction obtained via physical distancing  and the one obtained by the traditional methods (the renewal approach, SIR model, curve fitting, and machine learning) reported in literature. So we evaluate the pre-multiplicative scaling factor in the reproduction number eq.\eqref{eq_Rtmob},   using linear regression with an intercept coefficient fixed to be zero.
Moreover, when plotting the reproduction number and growth rate, to visualize the trend, we use non-parametric regression analysis with LoWeSS (Locally Weighted Scatterplot Smoothing) surrounded by a $90\%$ confidence interval obtained through bootstrapping.

\begin{figure}
	\centering
	\begin{subfigure}[c]{0.45\textwidth}
		\centering
		\includegraphics[width=1\linewidth]{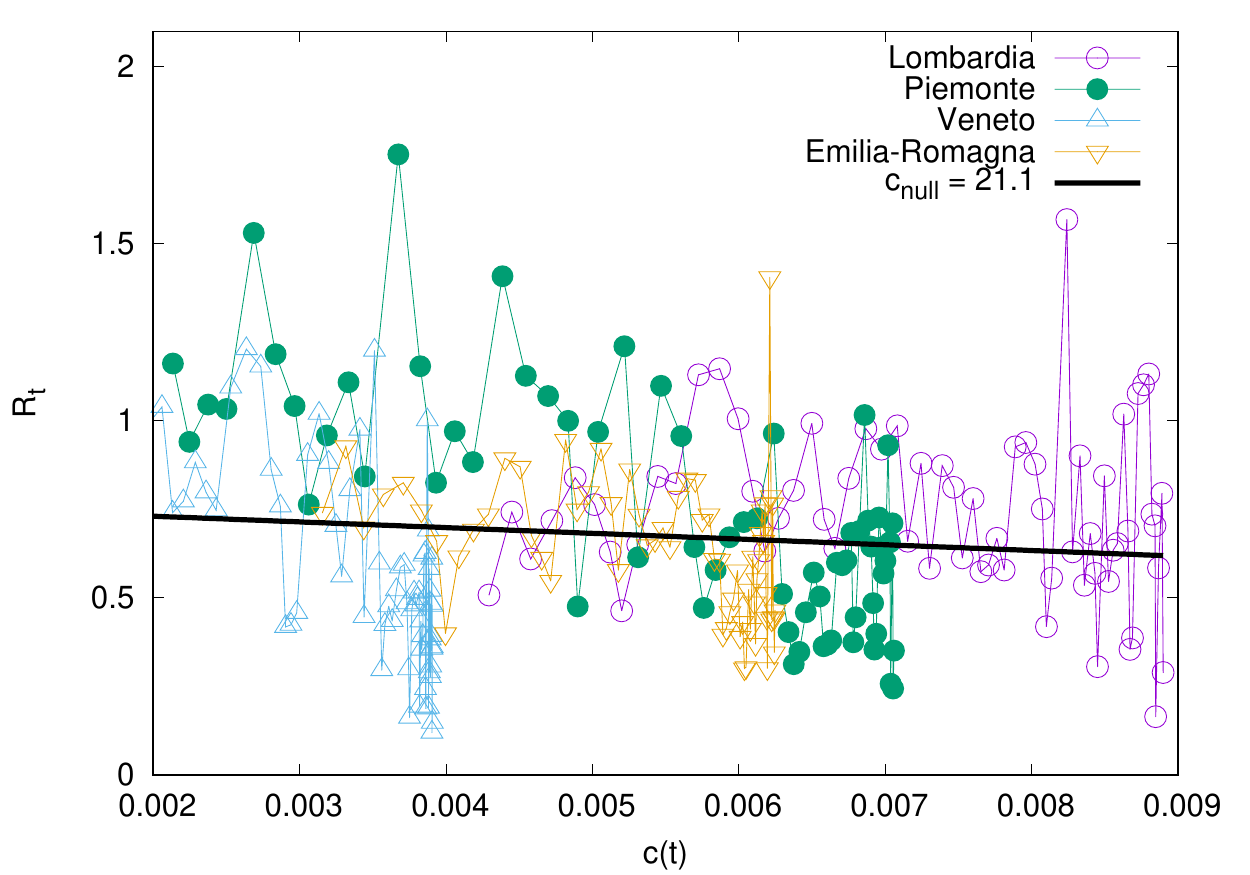}
		\caption{\label{lamia} Reproduction number, $R$, calculated via convolution as a function of the fraction of officially infected population.}
	\end{subfigure}%
	\qquad 
	\begin{subfigure}[c]{0.45\textwidth}
		\centering
		\includegraphics[width=1\linewidth]{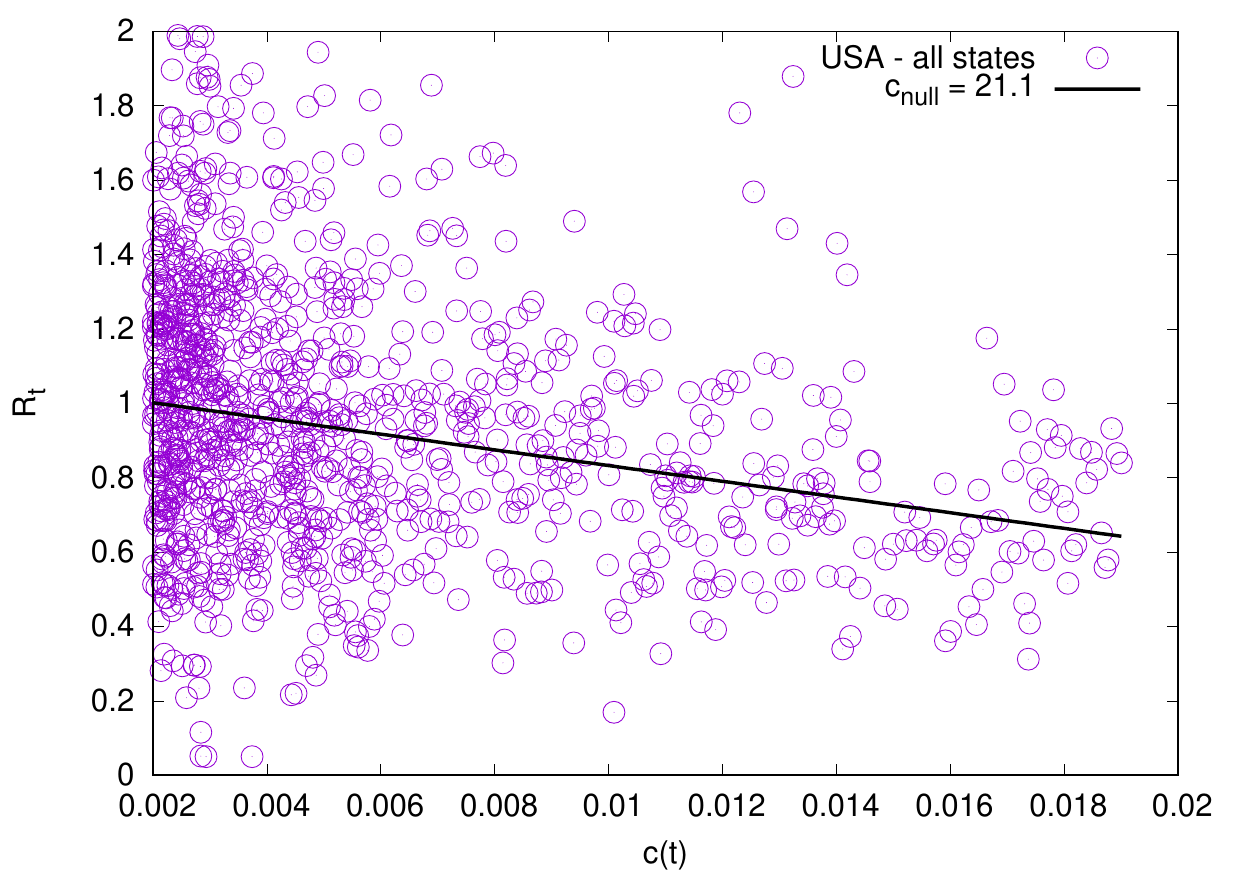}
		\caption{\label{miaseconda} $R$ calculated via convolution as a function of the fraction of officially infected population}
	\end{subfigure}%
	\caption{Estimation of asymptomatic, mildly symptomatic, and undetected individuals in some key Italian regions (a) and the USA (b). The errors in the estimation are about $20\%$ of the fitted value of the slope.}
	\label{fig_luigi}
\end{figure}

The changing trends of the reproduction number may be due to several interrelated reasons apart from physical distancing policies.  These reasons can be collected into two groups. The first has to do with the virus itself and its capacity to spread.  An increase in temperature or the development of less dangerous strains can decrease the effective infectiousness of the 
contagion.  The other group of reasons is connected to the decrease in the number of susceptible
individuals.  Supposing that the latter is the actual reason, we made a linear fit of $R$ as a function of the fraction of the total population infected.  Shown in Fig. \ref{fig_luigi} using 
a very simple ansatz:
\begin{equation}
R(t) = \tilde{R} \left (1 - \dfrac{c}{\lambda} \right ).
\end{equation}
In this way, when the number of susceptibles is zero $R(t) \rightarrow 0$. So, the value of
the officially detected fraction of the population leading to $R(t) = 0$, $c_{null} = \lambda$, gives
the ratio, $\lambda$, between the number of officially detected and the number of actual cases (supposing that this
ratio is approximately constant in time). 
We perform the fit only using points 20 days after the lockdown (day 35 in the figures)
for Italian regions and US states. 
We find that the critical official fraction of infected persons is $c_{null} = 4.7\%$, both for Italian regions and US states. The data and the best fit are shown in Fig. \ref{fig_luigi}.
This means that the actual infected population fraction should be obtained by multiplying the officially detected cases by a factor of $\Lambda = 1/c_{null} \simeq 20$. 
These results are a-posteriori partially confirmed by some preliminary results from antibody testing
performed in Italy.  They are also in line with the estimated test fraction, $\Lambda$, found through models  used by officials to decide on policies like shelter-in-place orders, such as the \citet{coviactnow} initiative and \citet{flaxman2020report}.  		

\begin{figure}[!ht]
	\centering
	\begin{subfigure}[c]{0.48\textwidth}
		\centering
		\includegraphics[angle=0,origin=c,width=0.9\linewidth]{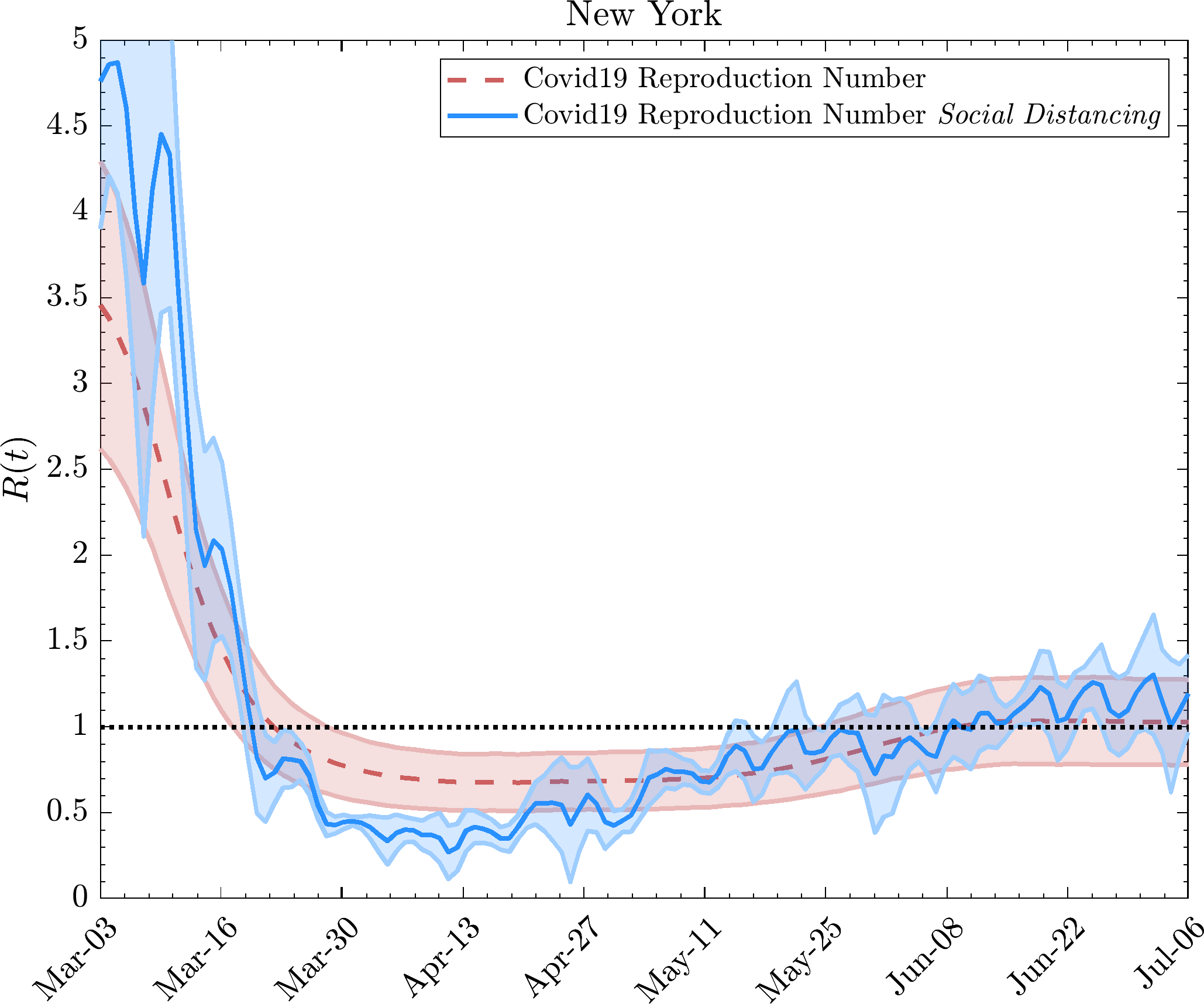}
		\caption{}
	\end{subfigure}%
	\begin{subfigure}[c]{0.48\textwidth}
		\centering
		\includegraphics[angle=0,origin=c,width=0.9\linewidth]{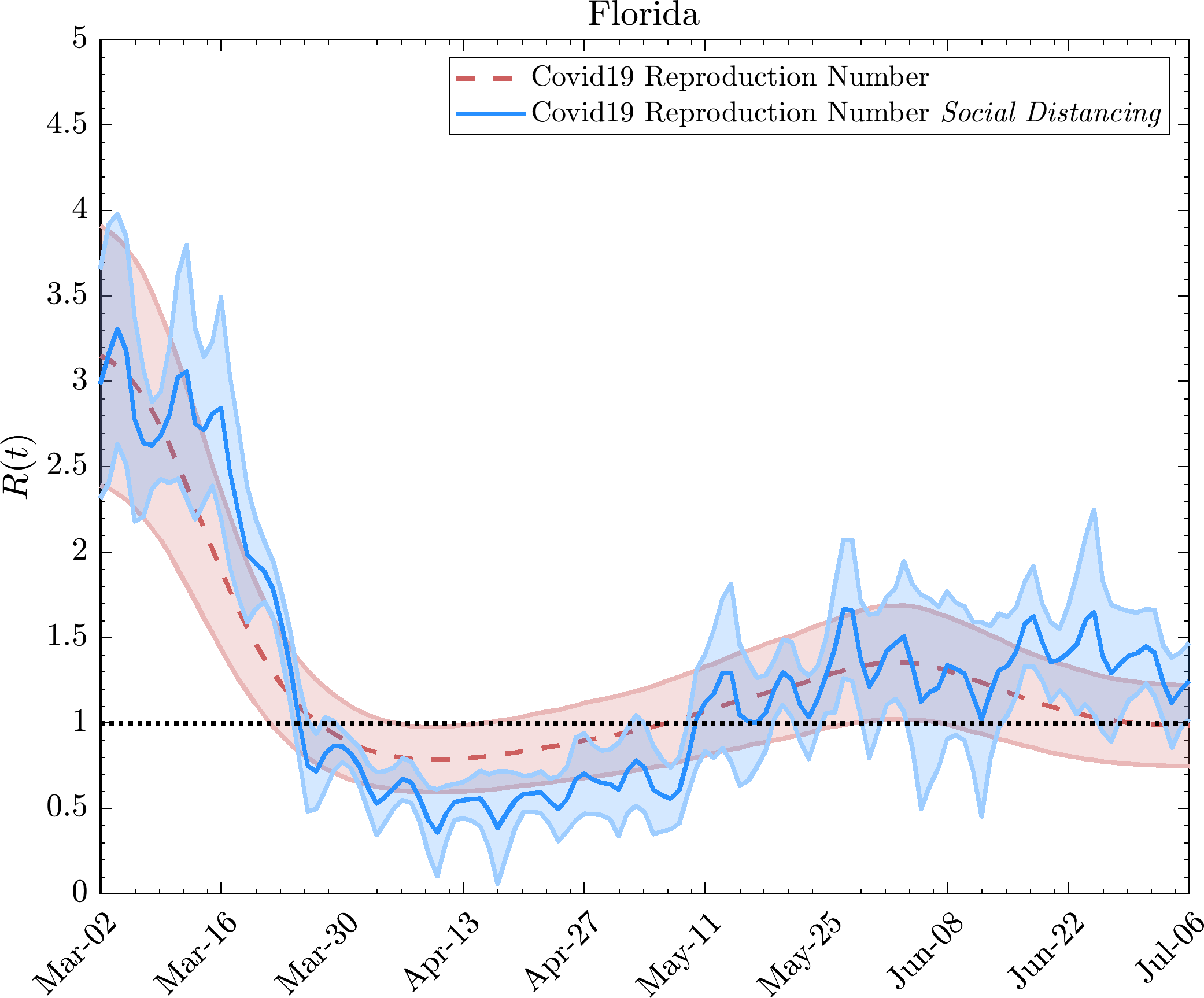}
		\caption{}
	\end{subfigure}%
\\ \vspace{1cm}
	\centering
	\begin{subfigure}[c]{0.48\textwidth}
		\centering
		\includegraphics[angle=0,origin=c,width=0.9\linewidth]{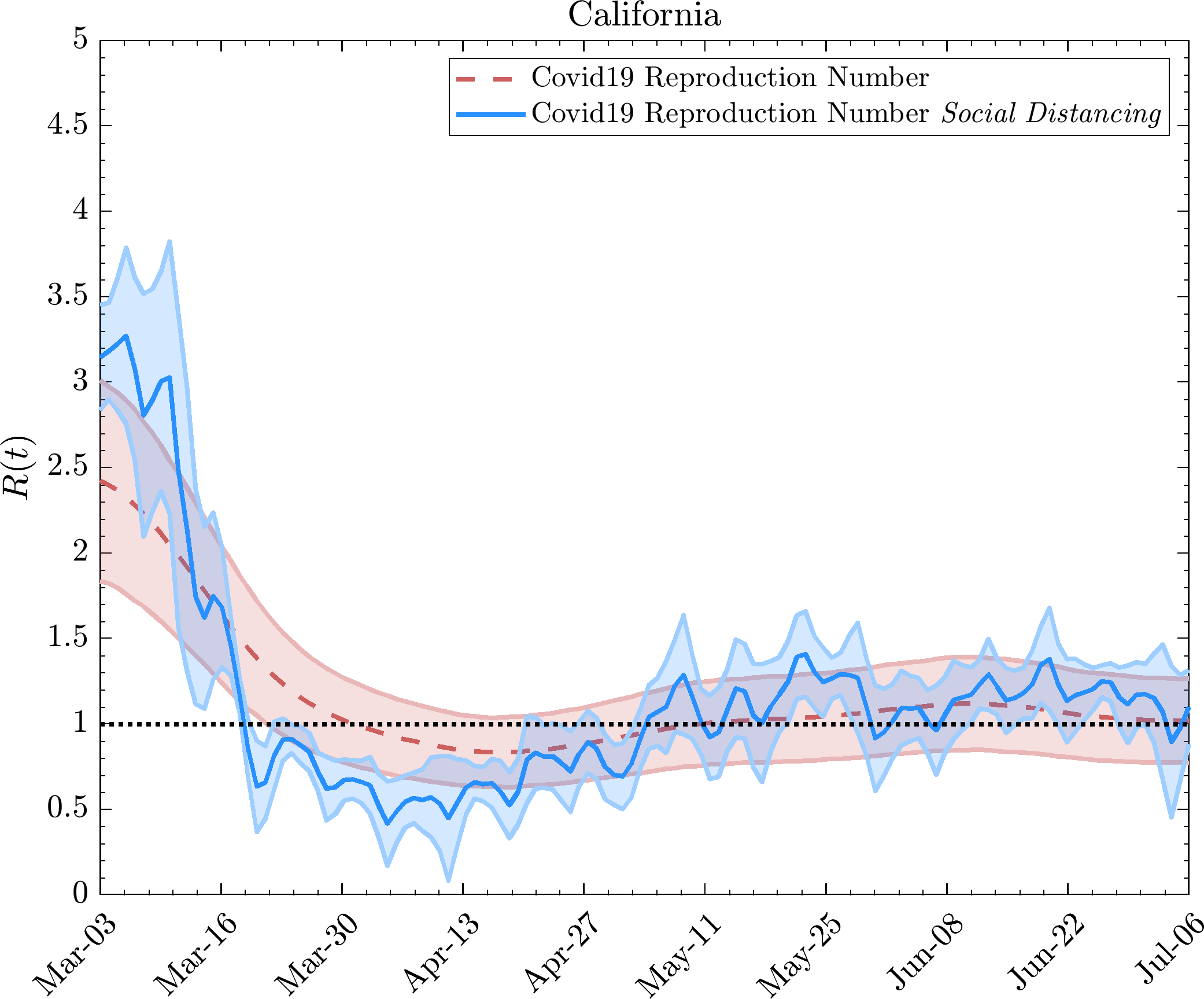}
		\caption{}
	\end{subfigure}%
	\begin{subfigure}[c]{0.48\textwidth}
		\centering
		\includegraphics[angle=0,origin=c,width=0.9\linewidth]{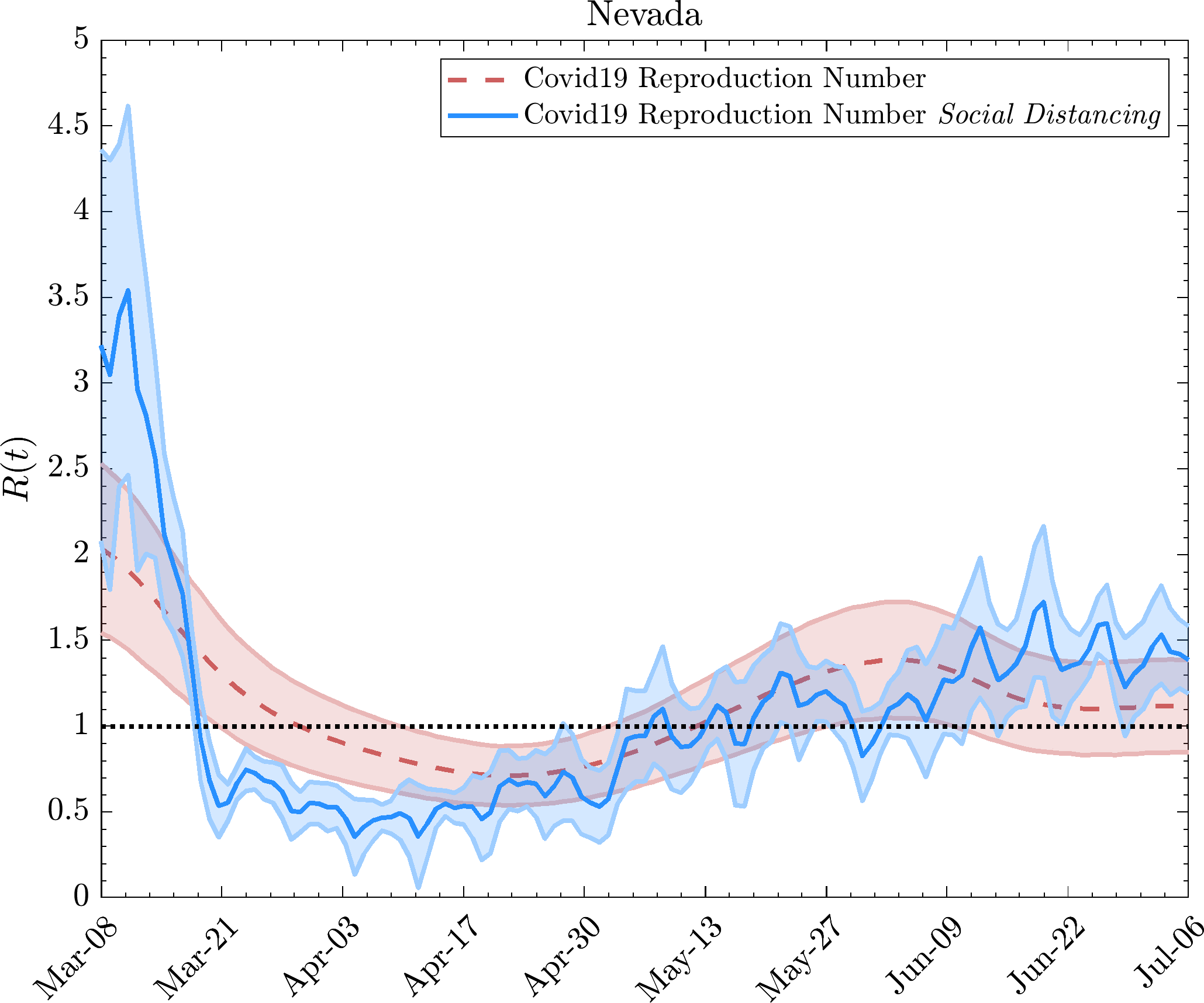}
		\caption{}
	\end{subfigure}%
	\caption{Effective reproduction number data courtesy of Rtlive \cite{rtlive} and social distancing data courtesy of Unacast \cite{unacast}.}	
	\label{fig_USuna1}  
\end{figure}

\begin{figure}[!ht]
	\centering
	\begin{subfigure}[c]{0.48\textwidth}
		\centering
		\includegraphics[angle=0,origin=c,width=0.9\linewidth]{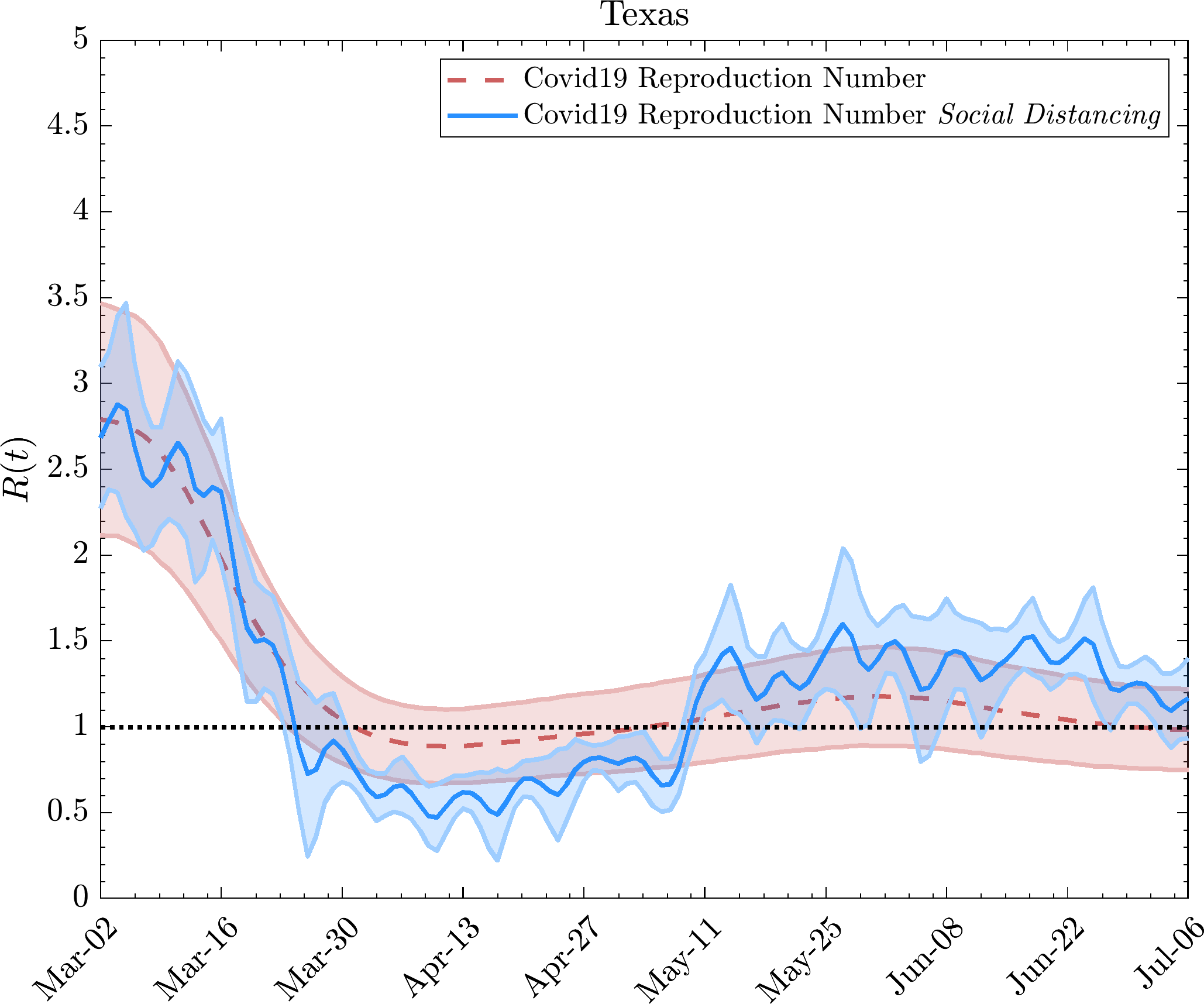}
		\caption{}
	\end{subfigure}%
	\begin{subfigure}[c]{0.48\textwidth}
		\centering
		\includegraphics[angle=0,origin=c,width=0.9\linewidth]{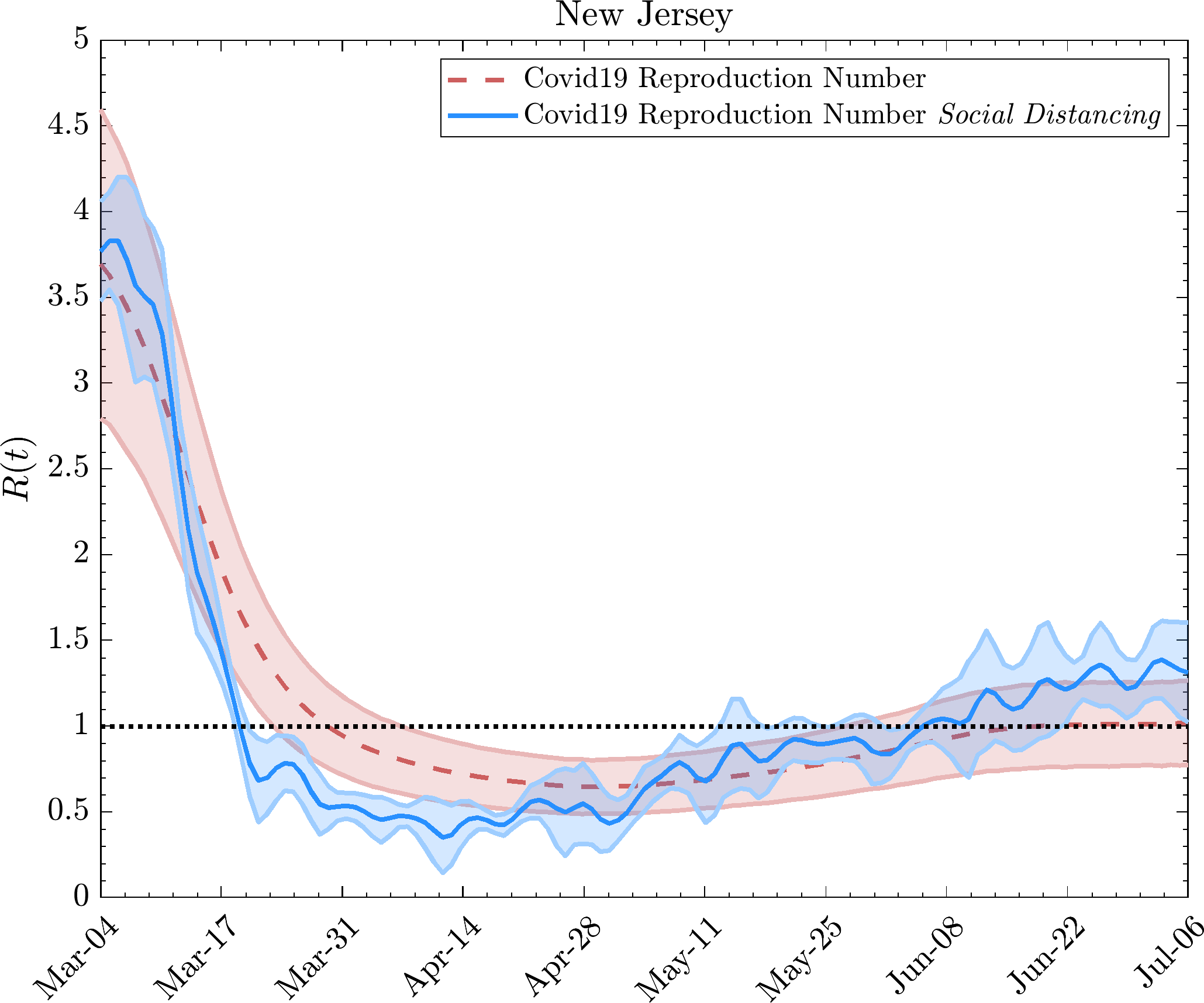}
		\caption{}
	\end{subfigure}%
	\\ \vspace{1cm}
	\centering
	\begin{subfigure}[c]{0.48\textwidth}
		\centering
		\includegraphics[angle=0,origin=c,width=0.9\linewidth]{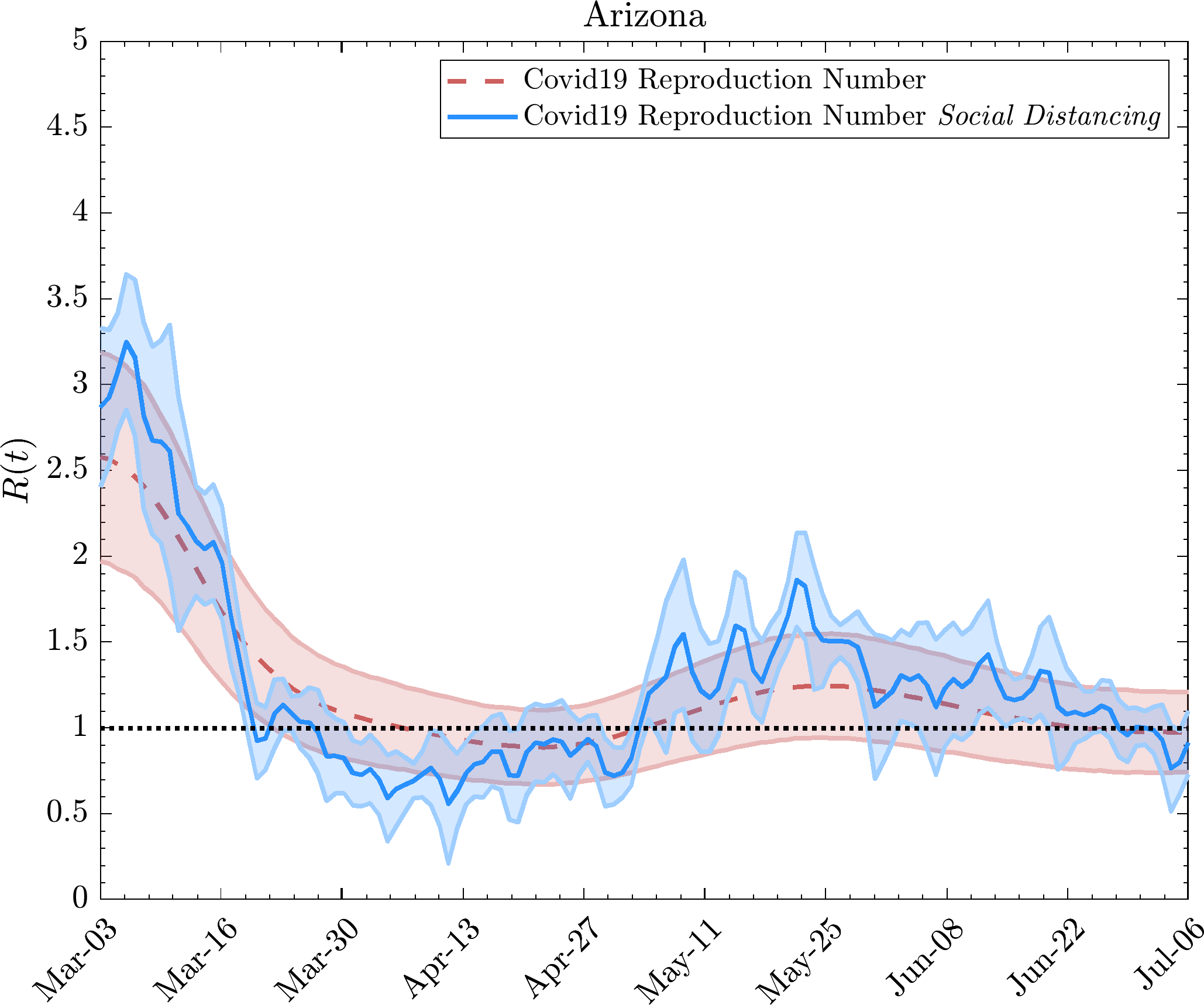}
		\caption{}
	\end{subfigure}%
	\begin{subfigure}[c]{0.48\textwidth}
		\centering
		\includegraphics[angle=0,origin=c,width=0.9\linewidth]{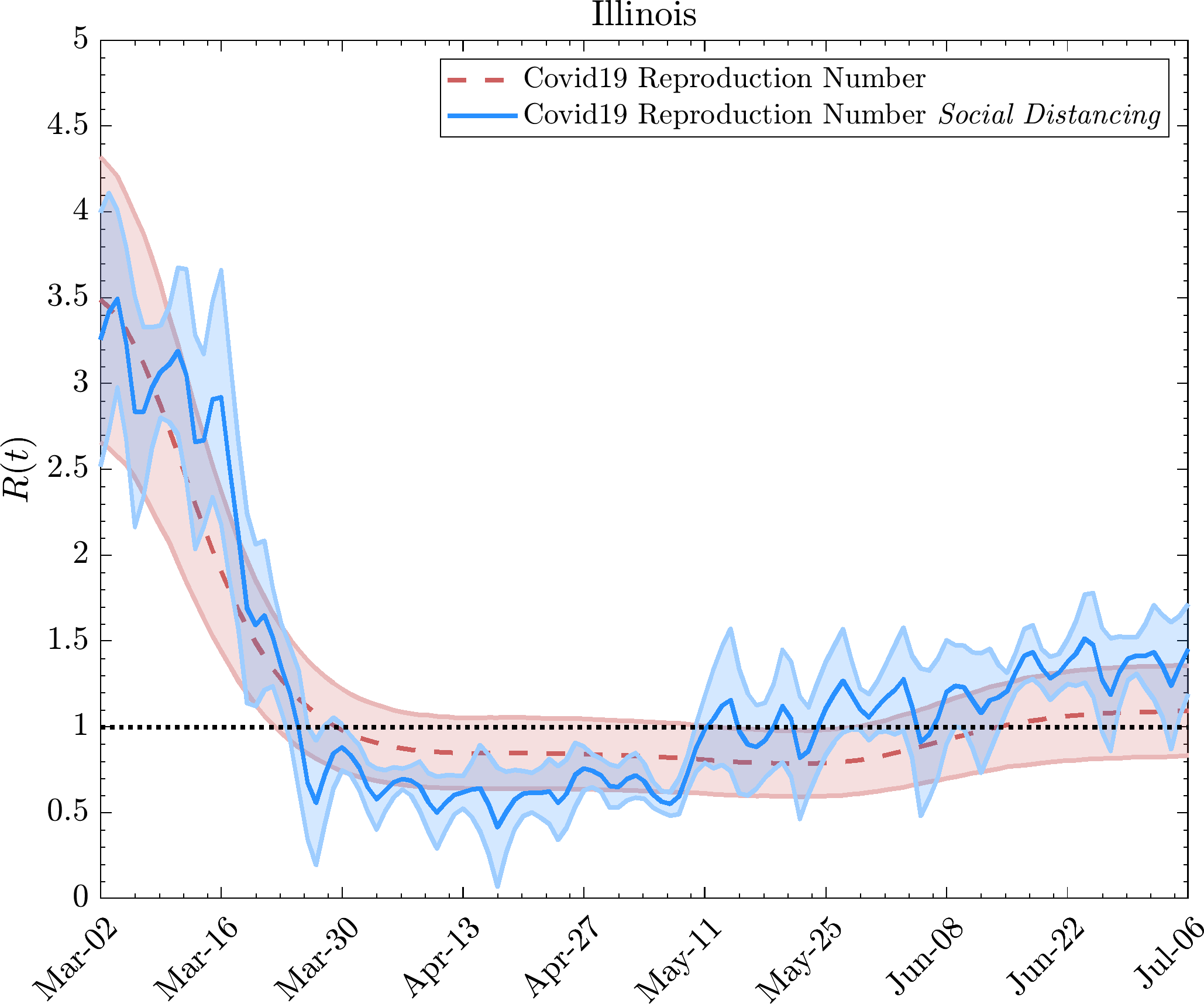}
		\caption{}
	\end{subfigure}%
	\caption{Effective reproduction number data courtesy of Rtlive \cite{rtlive} and social distancing data courtesy of Unacast \cite{unacast}.}	
	\label{fig_USuna2}  
\end{figure}

\begin{figure}[!ht]
	\centering
	\begin{subfigure}[c]{0.48\textwidth}
		\centering
		\includegraphics[angle=0,origin=c,width=0.9\linewidth]{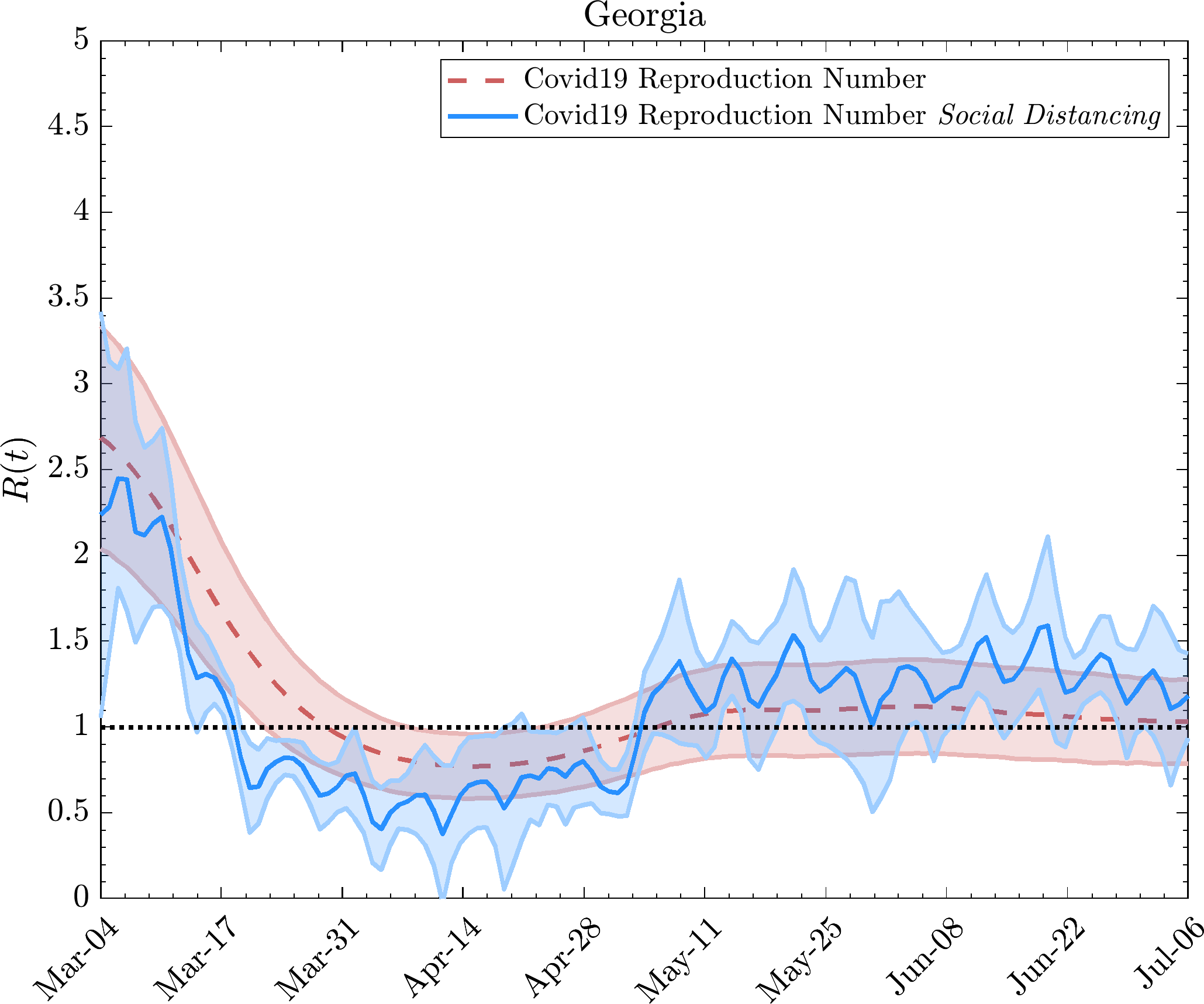}
		\caption{}
	\end{subfigure}%
	\begin{subfigure}[c]{0.48\textwidth}
		\centering
		\includegraphics[angle=0,origin=c,width=0.9\linewidth]{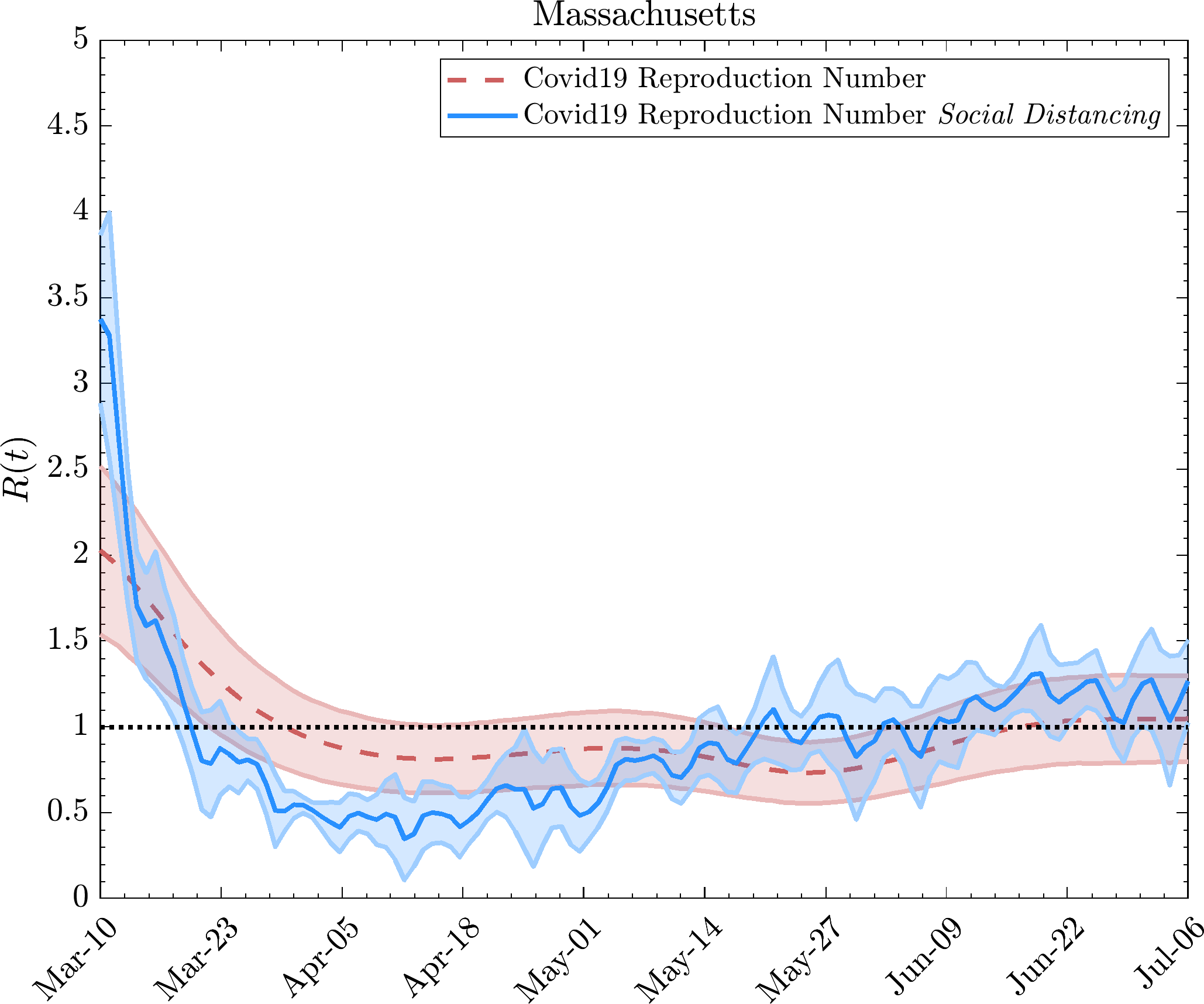}
		\caption{}
	\end{subfigure}%
	\\ \vspace{1cm}
	\centering
	\begin{subfigure}[c]{0.48\textwidth}
		\centering
		\includegraphics[angle=0,origin=c,width=0.9\linewidth]{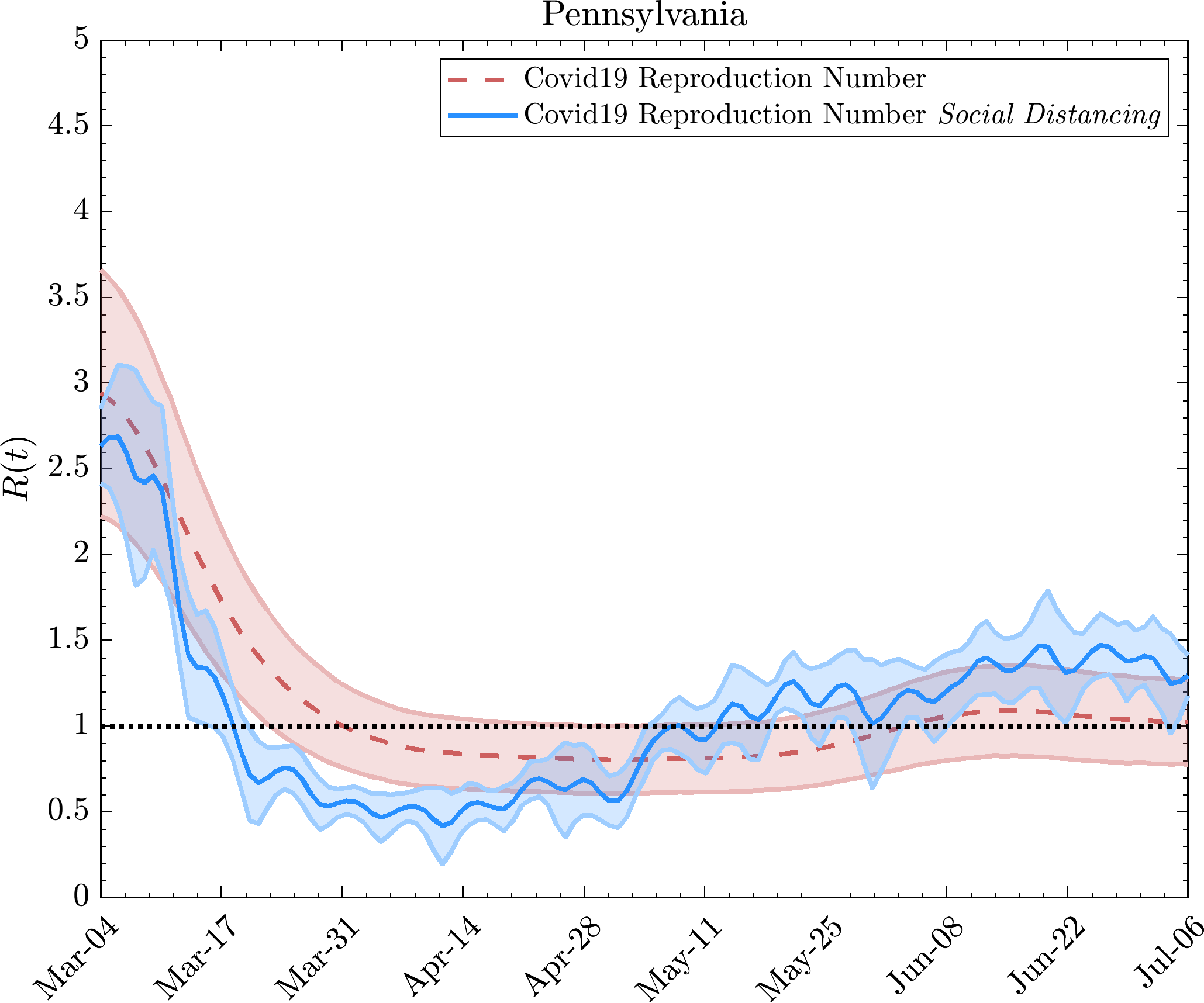}
		\caption{}
	\end{subfigure}%
	\begin{subfigure}[c]{0.48\textwidth}
		\centering
		\includegraphics[angle=0,origin=c,width=0.9\linewidth]{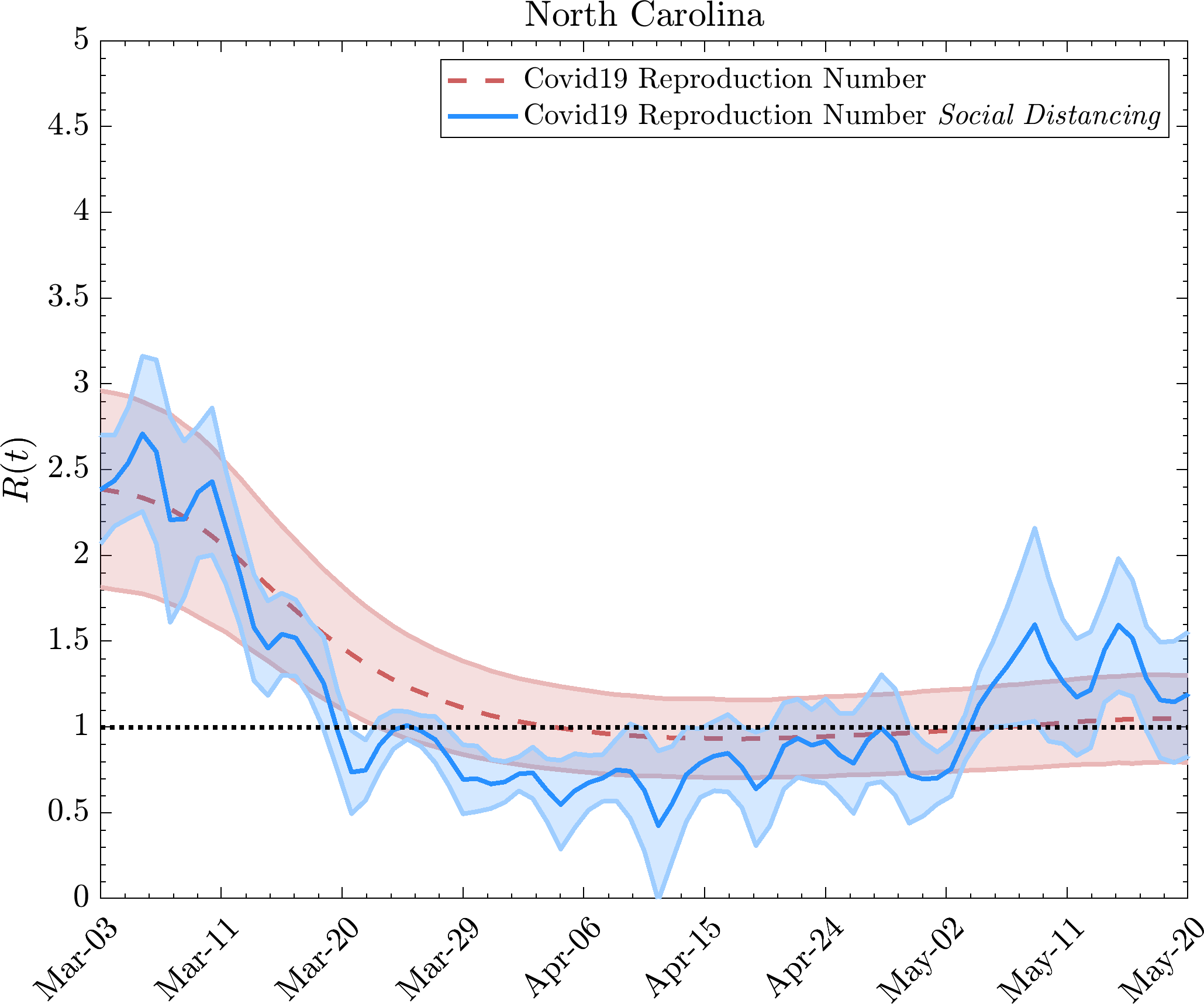}
		\caption{}
	\end{subfigure}%
	\caption{Effective reproduction number data courtesy of Rtlive \cite{rtlive} and social distancing data courtesy of Unacast \cite{unacast}.}	
	\label{fig_USuna3}  
\end{figure}

\begin{figure}[!ht]
	\centering
	\begin{subfigure}[c]{0.33\textwidth}
		\centering
		\includegraphics[angle=0,origin=c,width=0.9\linewidth]{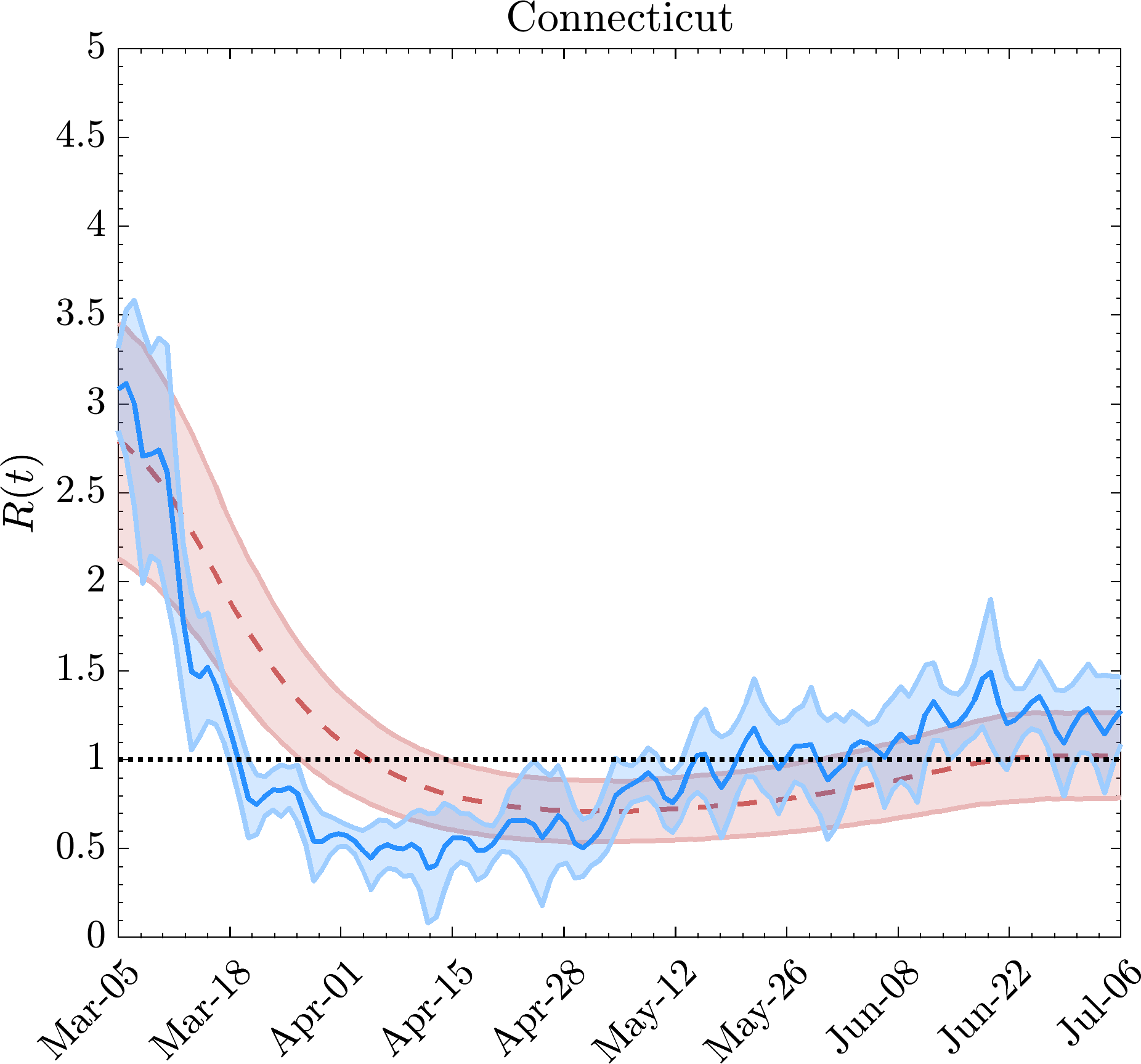}
		\caption{}
	\end{subfigure}%
	\begin{subfigure}[c]{0.33\textwidth}
		\centering
		\includegraphics[angle=0,origin=c,width=0.9\linewidth]{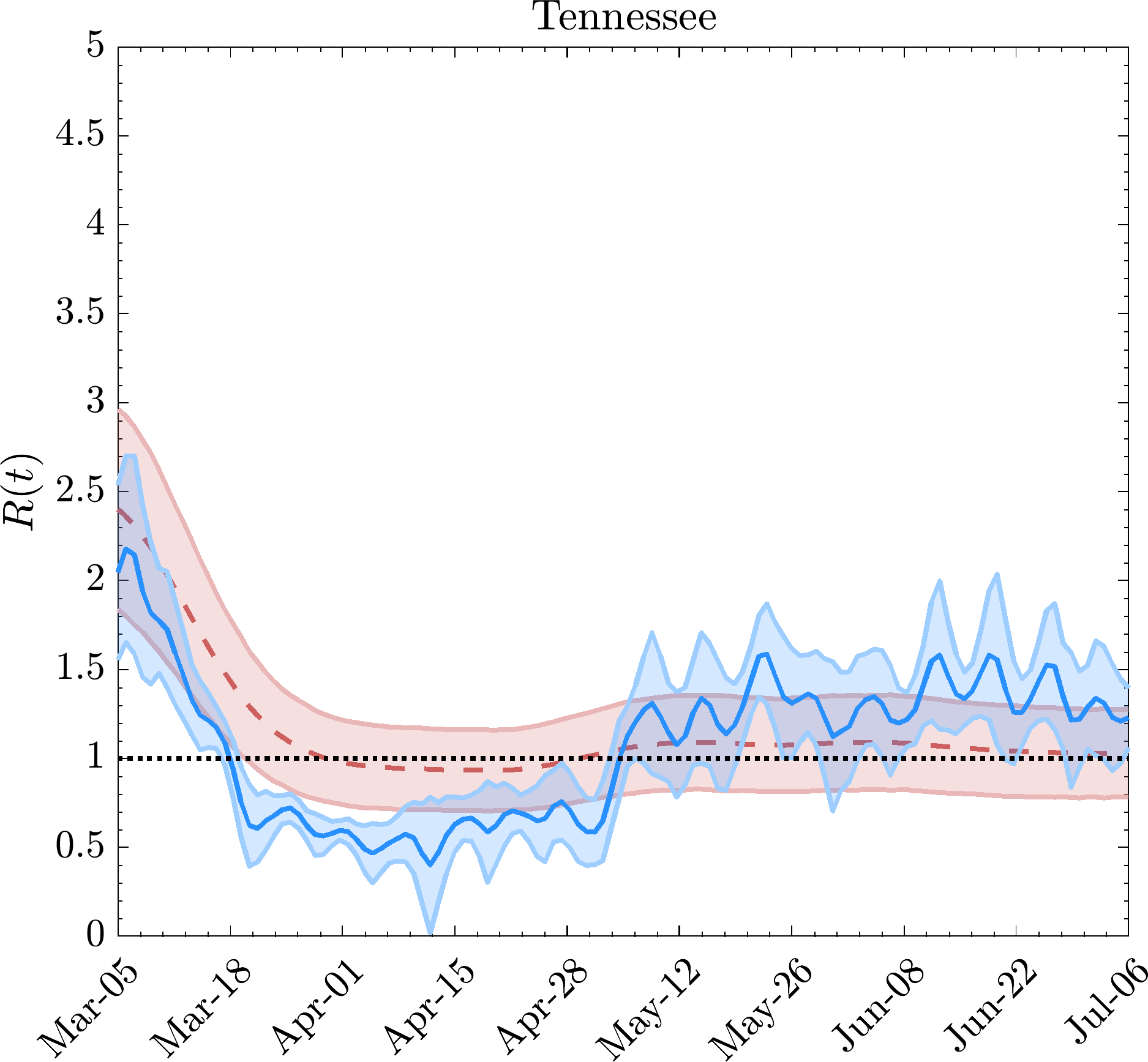}
		\caption{}
	\end{subfigure}%
	\begin{subfigure}[c]{0.33\textwidth}
	\centering
	\includegraphics[angle=0,origin=c,width=0.9\linewidth]{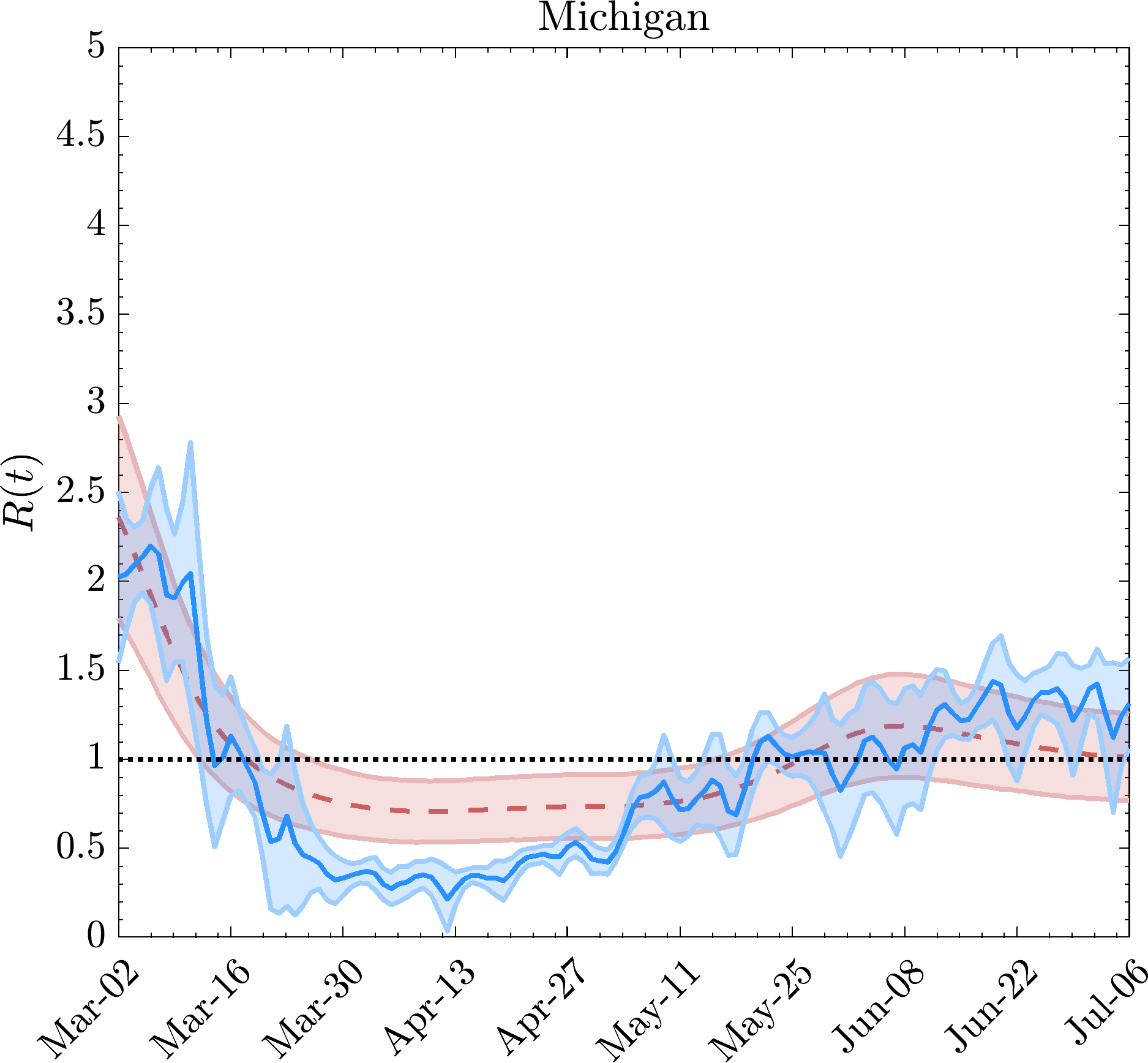}
	\caption{}
\end{subfigure}%
	\\ \vspace{1cm}
	\centering
	\begin{subfigure}[c]{0.33\textwidth}
		\centering
		\includegraphics[angle=0,origin=c,width=0.9\linewidth]{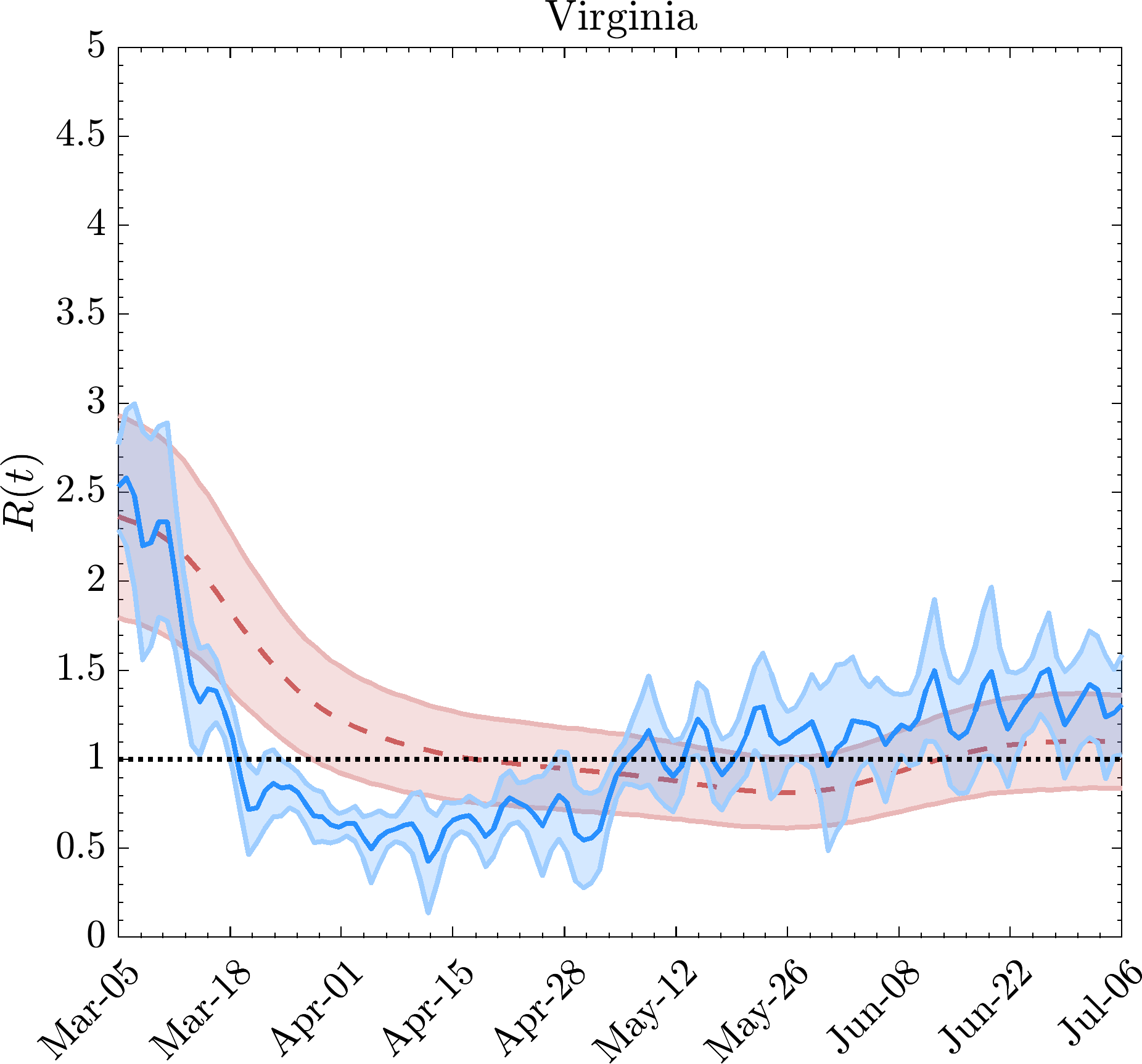}
		\caption{}
	\end{subfigure}%
	\begin{subfigure}[c]{0.33\textwidth}
		\centering
		\includegraphics[angle=0,origin=c,width=0.9\linewidth]{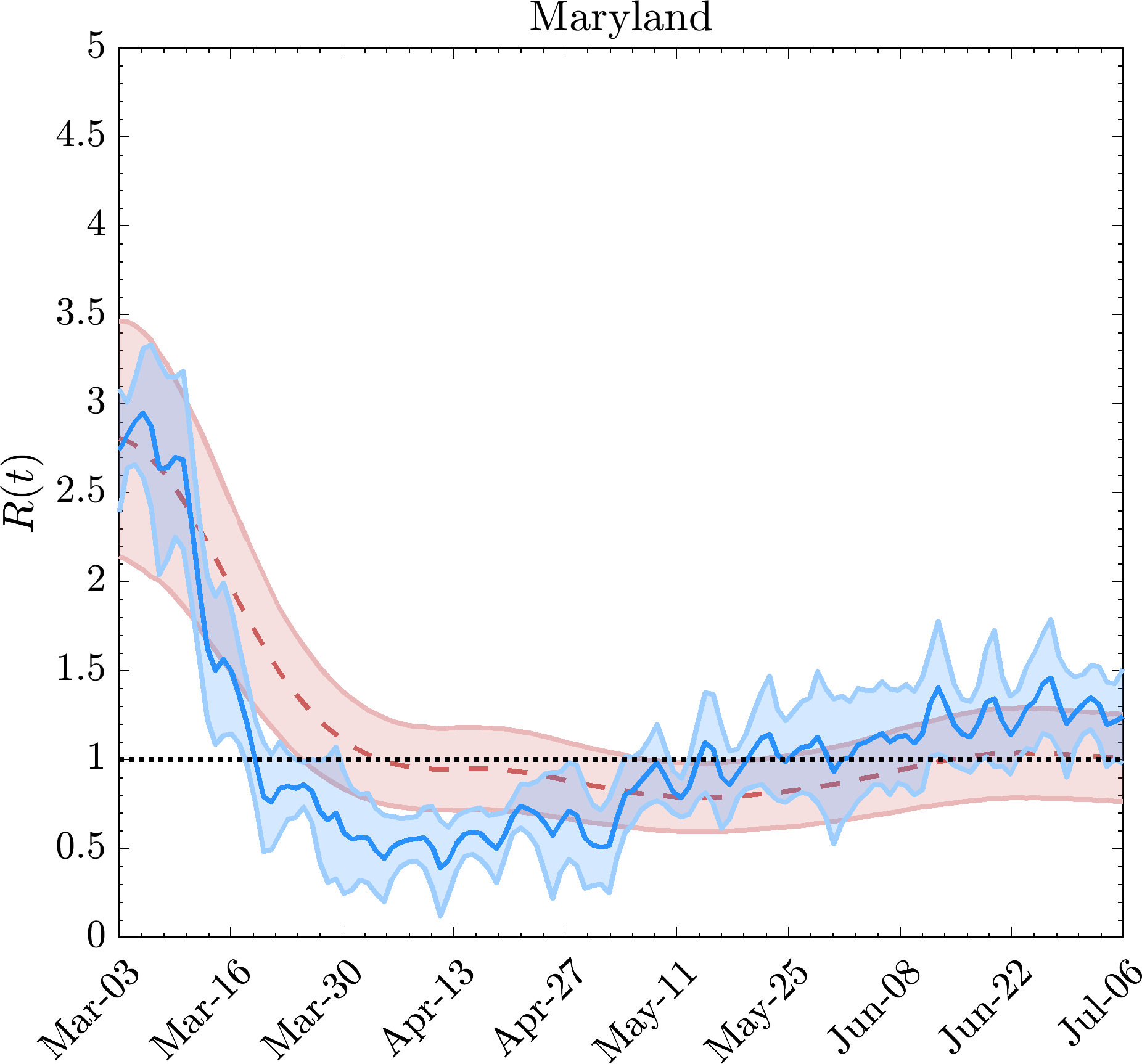}
		\caption{}
	\end{subfigure}%
	\begin{subfigure}[c]{0.33\textwidth}
	\centering
	\includegraphics[angle=0,origin=c,width=0.9\linewidth]{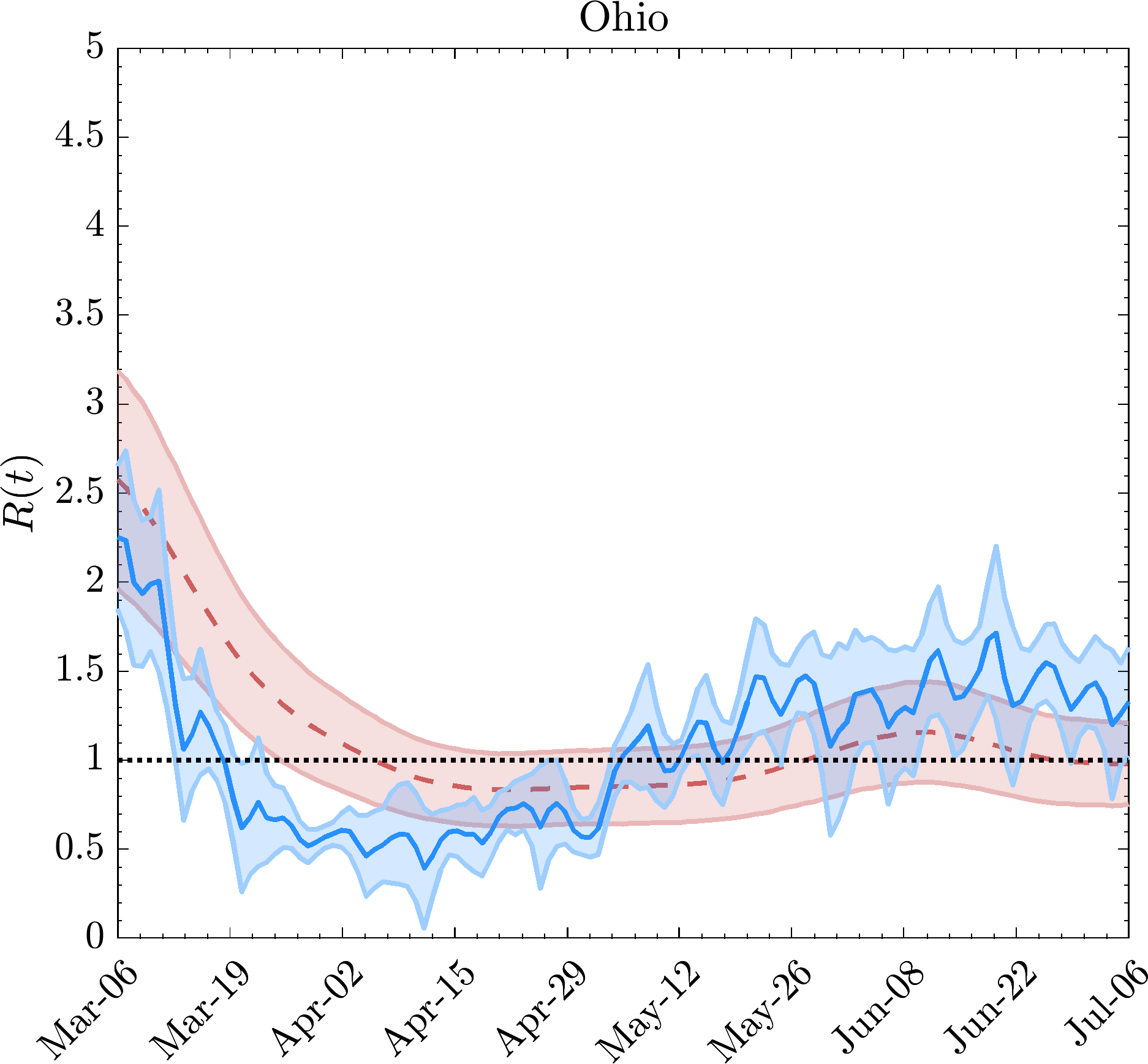}
	\caption{}
\end{subfigure}%
	\\ \vspace{1cm}
\centering
	\begin{subfigure}[c]{0.33\textwidth}
	\centering
	\includegraphics[angle=0,origin=c,width=0.9\linewidth]{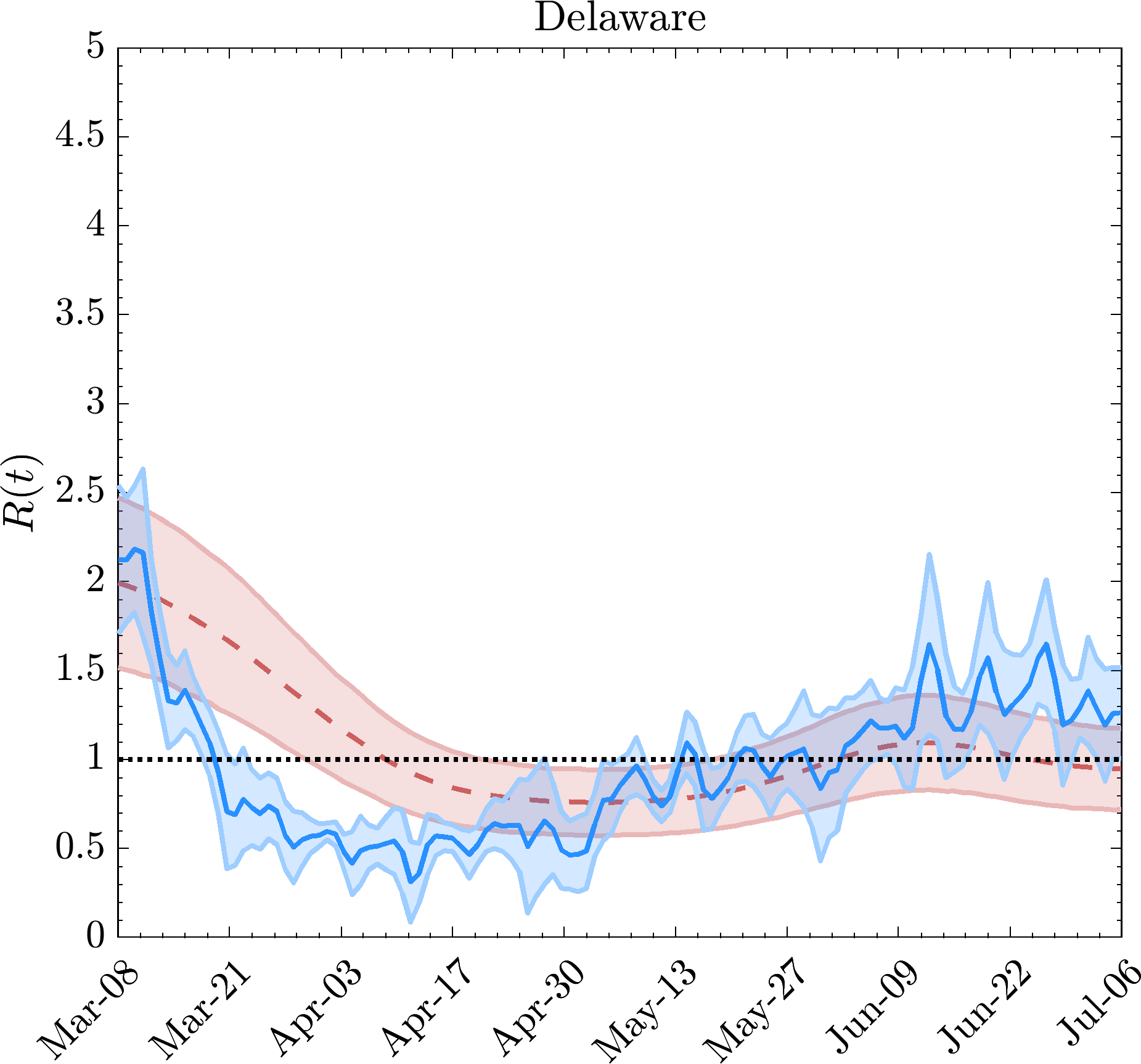}
	\caption{}
\end{subfigure}%
\begin{subfigure}[c]{0.33\textwidth}
	\centering
	\includegraphics[angle=0,origin=c,width=0.9\linewidth]{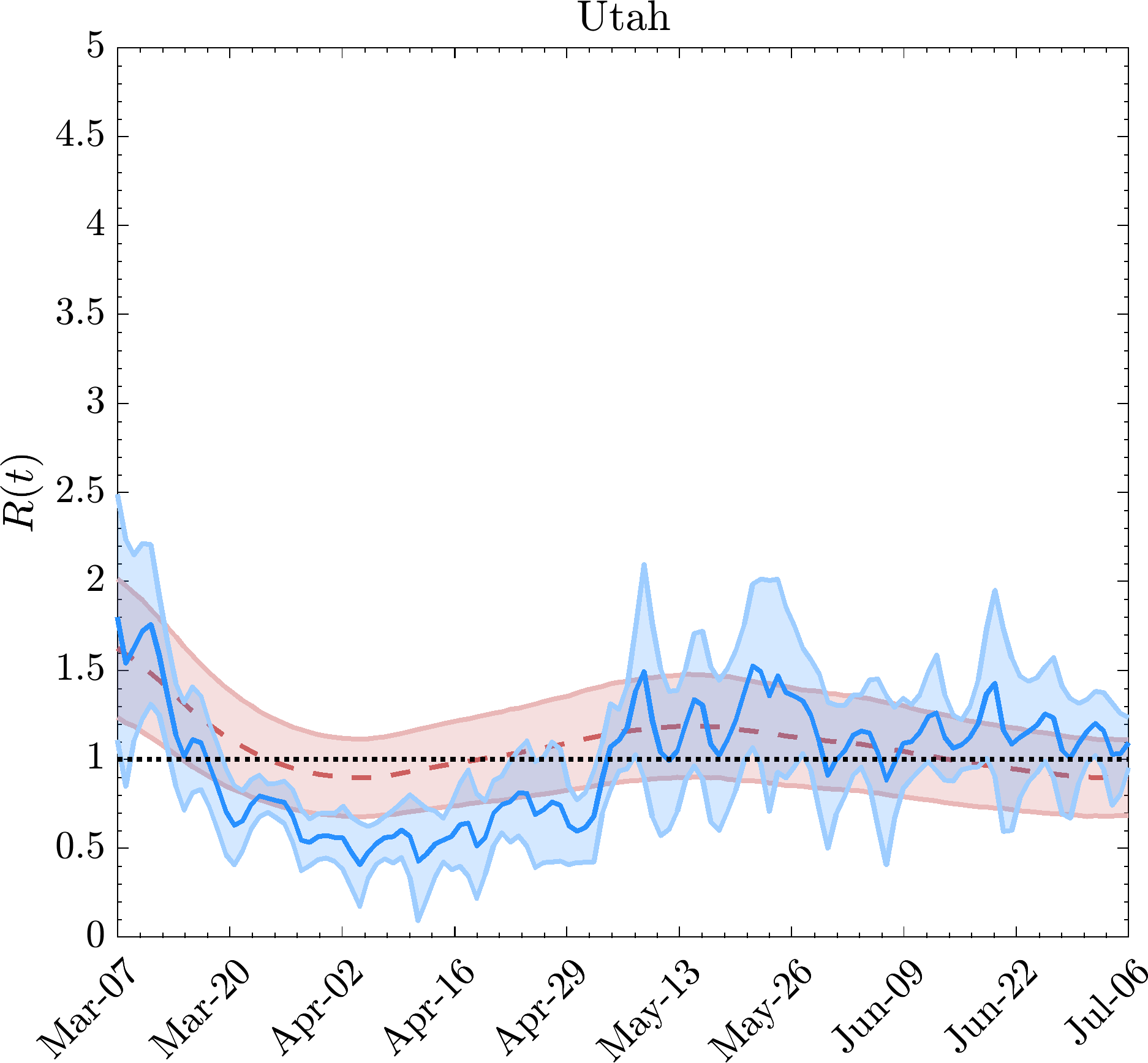}
	\caption{}
\end{subfigure}%
\begin{subfigure}[c]{0.33\textwidth}
	\centering
	\includegraphics[angle=0,origin=c,width=0.9\linewidth]{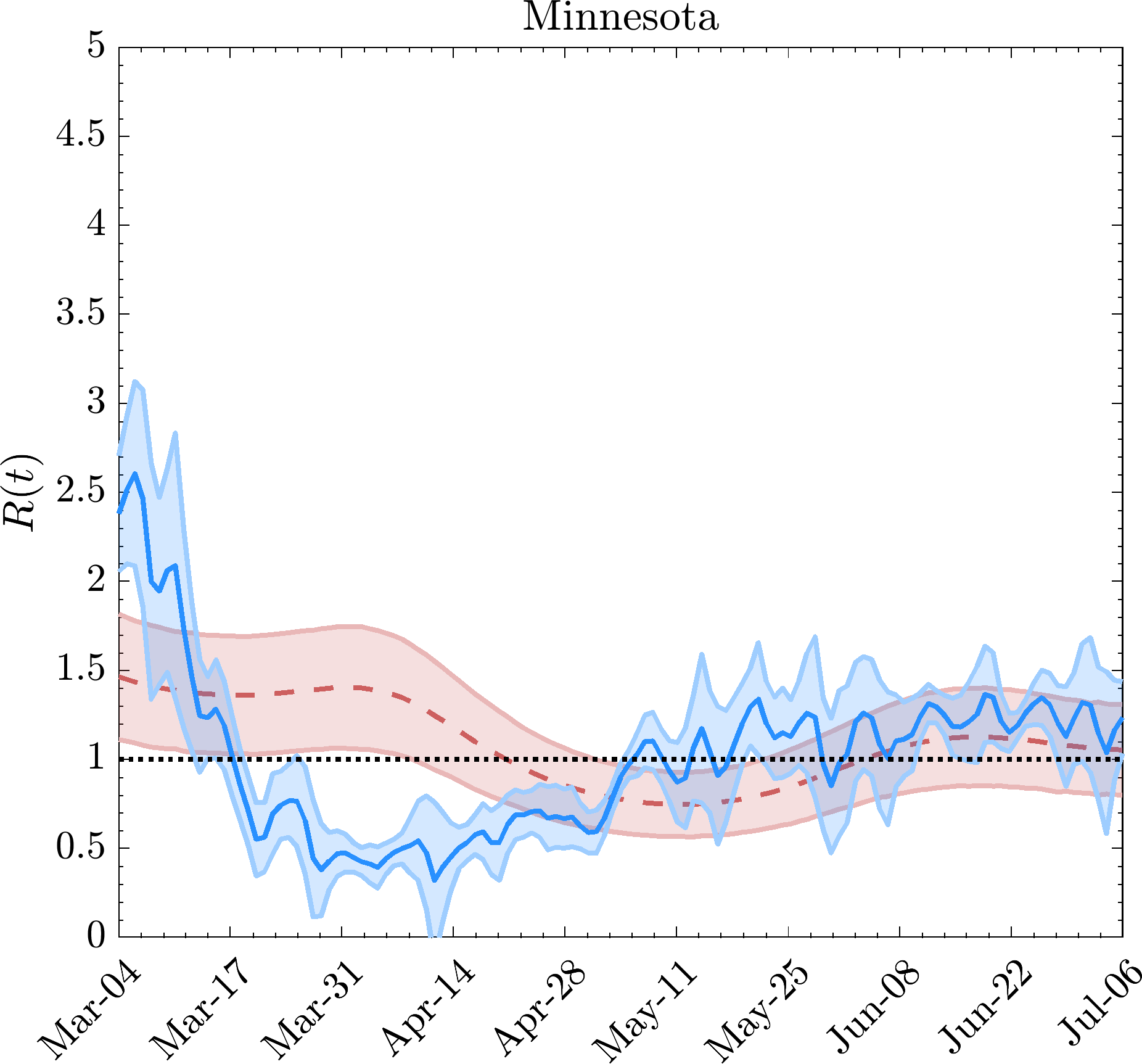}
	\caption{}
\end{subfigure}%
	\caption{Effective reproduction number data courtesy of Rtlive \cite{rtlive} and social distancing data courtesy of Unacast \cite{unacast}.}	
	\label{fig_USuna00}  
\end{figure}

As a further step to check the robustness of our findings, we have analyzed an independent dataset provided by \citet{unacast}.  These data offer  metrics for physical distancing based on GPS devices.  They provide proxies for social mobility (interpreted as average distance traveled) and interpersonal proximity (human encounters), calculated with respect to the 4 weeks before the CoViD-19 outbreak. The former is the percent reduction in the total distance traveled per device, averaged across all devices located in given US state. The latter is an estimate of close encounters between two devices per square kilometer, expressed as a fraction of the baseline. This estimate counts two users from the same province as having come in contact if they were within a circle of radius of
50 meters of each other within a 1-hour period.
This data provides a check on the accuracy of our previous results and allows us to extend them to all 50 US states.  A selection of these results is shown in Figs.\ref{fig_USuna1}-\ref{fig_USuna00}. These figures indicate that the results using Unacast data are consistent with the other databases for social mobility and human proximity.


\section{Epidemic risk per categories and remote working}
\begin{figure}[!ht]
			\centering
			\begin{subfigure}[c]{0.50\textwidth}		
				\centering
				\includegraphics[width=1\linewidth]{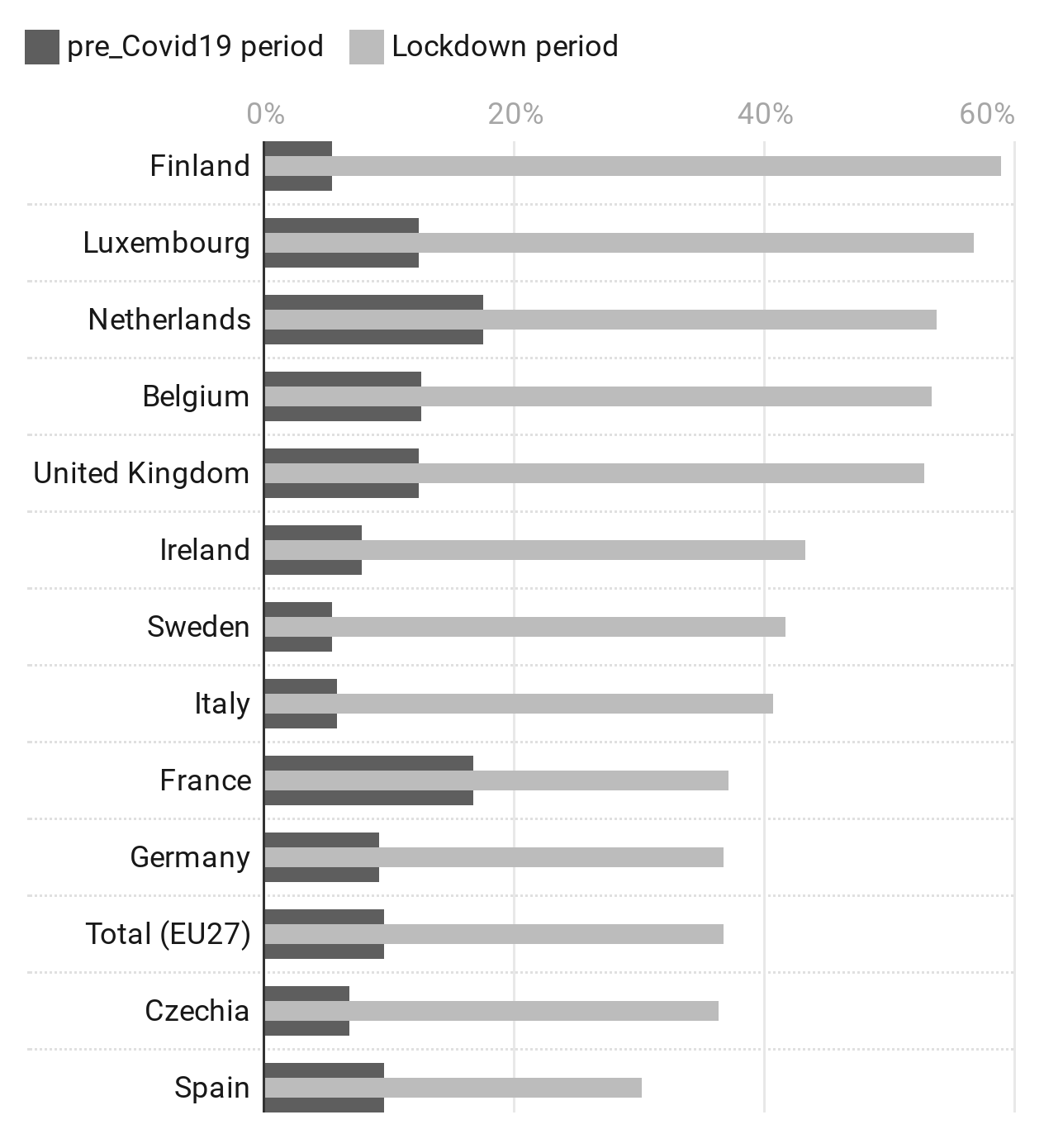}
				\label{fig_teleworka}
			\end{subfigure}%
			\begin{subfigure}[c]{0.45\textwidth}
				\centering
				\includegraphics[width=1\linewidth]{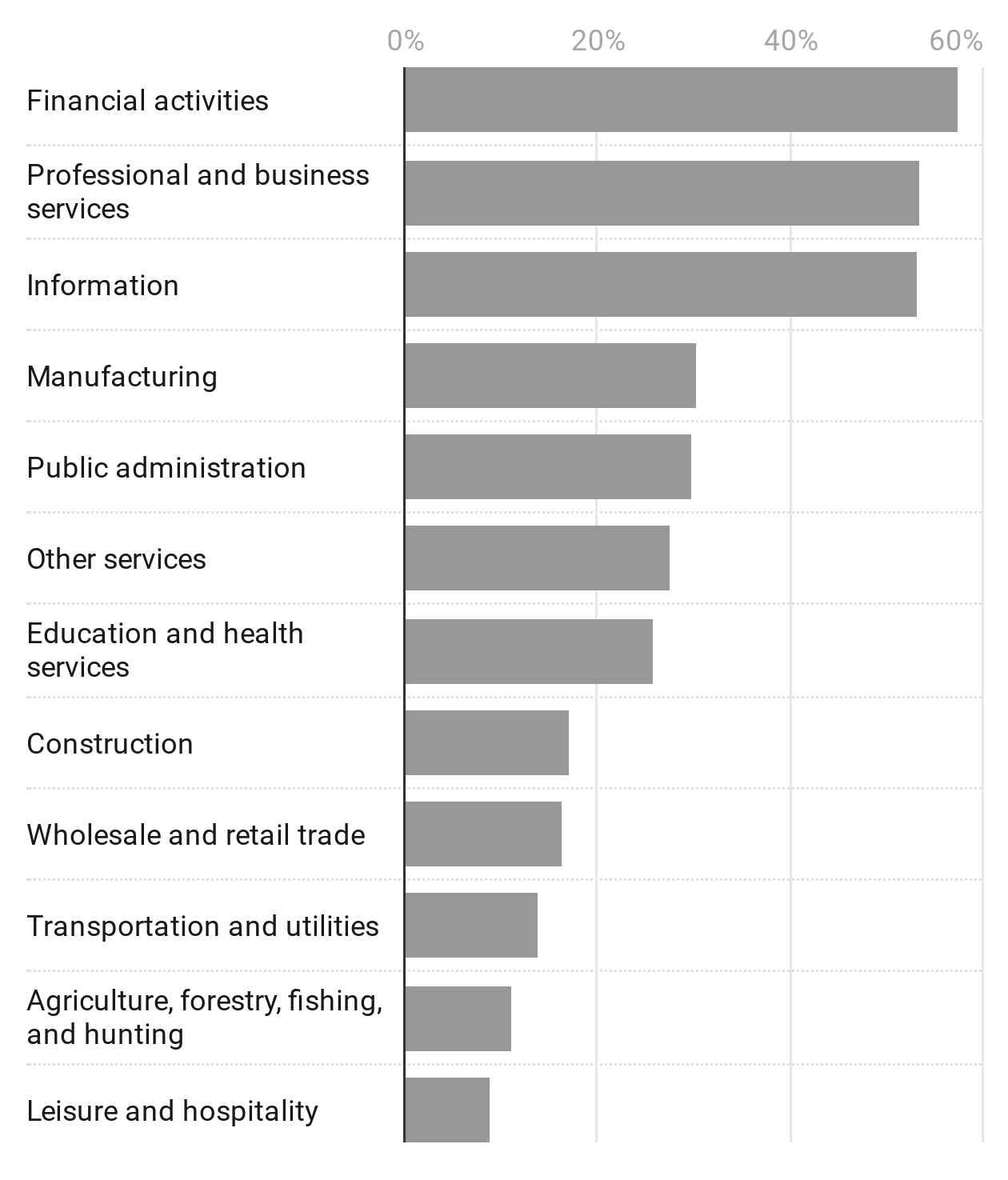}
			\end{subfigure}%
			\caption{Remote-working from home in the EU and US. (a) Teleworking with daily frequency  in some European countries before (dark gray tick bars) and after (light gray thin bars) the outbreak of CoViD-19. Source: \citet{telew}. (b) Share of US workers who can telework, by industry, source: \citet{BureauL}.}
			\label{fig_telework}  
		\end{figure}

	The outbreak of the CoViD-19 pandemic has pushed many countries towards a response that envisages social distancing policy the implementation of which has an important social and economic impact on the organization of production and of the work process.  Working from home has been a necessary practice for many firms and workers during the lockdown period of the CoViD-19 crisis, so it represents a crucial measure for sustaining production during the crisis, even if its effects on productivity are unclear, as discussed by \citet{oecd20}.
	Recent analyses reveal that in the range of $30\%-40\%$ of all the job occupations in western countries could be performed from home (on average if one accounts for different teleworkability of specific activities) as studied in different countries: for Italy in \citet{cetrulo2020privilege,irpet},  for UK in \citet{office2020coronavirus}, for the US by \citet{gould2020not,BureauL}. As a consequence, lockdown measures should take into account such limitations and allow for other activities so as to adopt all the possible social distancing measures that can be effective for each workplace (air ventilation, face masks and coverings, etc.), see \citet{chu2020physical}.  On top of that, a more complex task consists of also considering the demand side of the lockdown, with many people avoiding spending money in activities that are considered to potentially increase risk of infection. 
	It is beyond the scope of this paper to explore the economic loss and social damage due to pandemics.  However, one might account for the heterogeneity of types of contacts by writing the reproduction number in eq.\eqref{eq_Rtmob} as a sum of different reproduction numbers across categories of individuals evolving in time according to the state of the infection:
	\begin{equation}\label{eq_multiRt}
	R(t)=\sum_{\alpha=1}^{W} R_{\alpha}(t) = \sum_{\alpha=1}^{W}R_{\alpha}(t_0) \,\frac{ n_{\alpha}(t-\tau_g)\eta_{\alpha}(t-\tau_g) \mu_{\alpha}(t-\tau_g)  }{\rho_{\alpha}(t-\tau_g)}c_{\alpha}(t_0),
	\end{equation}
	where the normalization constant (set at $t=t_0$, the initial time) is $c_{\alpha}(t_0)=\tfrac{\rho_{\alpha}(t_0-\tau_g)}{n_{\alpha}(t_0-\tau_g)\eta_{\alpha}(t_0-\tau_g) \mu_{\alpha}(t_0-\tau_g)  }$.  Note that in eq.\eqref{eq_multiRt}, the subscript $\alpha$ indexes the types of contact, which are assumed  to be mutually exclusive and exhaustive. In general, these assumptions are not necessarily fulfilled\footnote{One person  might go to work as well as to a restaurant, to a bar, to the shopping center, then on vacation, etc.} , but they are plausible if we consider the two categories to be working activities, for example.
	Since different categories of contacts can have, e.g., different average proximity, mobility and duration of contact (or even environmental conditions), the number of susceptibles corresponds to the number of people at risk in that particular category\footnote{The effective reproduction number may vary as well, because the communities in different locations may differ in their level of immunity. Similarly, the basic reproduction number, which is  the reproduction number when there is no immunity from past exposures, may vary across locations because contact rates among people may differ due to differences in population density and cultural differences.}. 	
	For example, one could consider the impact of lockdown restrictions on the occupational structure of a country, and quantify the jobs that can be done at home, defining the composition of the underlying labour force in terms of occupational, wage and contractual distributions.
		So, in principle, we could split $R(t)$ into two categories according the teleworkability of the activity, and consequently estimate the risk of an epidemic outbreak according to lockdown policies.  As shown in Fig.\ref{fig_telework}a, during the period before the outbreak of CoViD-19, the  diffusion of teleworking and smart working in Europe was very unequal\footnote{We use the terms teleworking and smartworking interchangeably. However, teleworking is a way to work independently of the geographic location of the office or business.  On the other hand, smartworking is a new version of telework, with an innovative workflow based on a strong element of flexibility in terms of hours and location of job, which applies to companies with flexible organizational models.}. As reported by \citet{telew}, the average is about $17\%$ (sum of regular and occasional teleworkers), but in some countries there are peaks over $30\%$ and in others -including Italy -teleworkers were less than $10\%$ of the workforce.  Turning to the smart working activities during the period of lockdown, European countries have responded with a strong increase in teleworking in differing degrees, ranging from near $60\%$ for Finland to $30\%$ for Spain.  This is because each country has a different share of workers who can telework, depending on the activity and the organization of labour, as discussed in Fig.\ref{fig_telework}b for the US. 
	For simplicity, we imagine two general categories of interactions outside the home.  We classify these interpersonal contacts into occasional contacts and structured ones as described in Appendix \ref{app_occasional}.  The first represent the erratic movement of an individual used in our kinetic framework, the latter is the situation of allocation of individuals to bounded areas such as schools, workplaces, or hospitals. Let us define a unique sector of the economy where $\alpha$ is represented as teleworking activity which mainly follow a structured contact pattern, while the remaining $1-\alpha$ of the productive activity cannot be performed remotely.  Then the reproductive number can be expressed as:
\begin{equation}
R(t)= R_{\alpha}(t)+R_{1-\alpha}(t)\sim R(t_0)\tfrac{\eta(t-\tau_g)}{\eta(t_0-\tau_g)}\Big( \alpha m\tfrac{\, n_{\alpha}(t-\tau_g)}{ n_{\alpha}(t_0-\tau_g)} + (1-\alpha)p\tfrac{\, n_{1-\alpha}(t-\tau_g)}{n_{1-\alpha}(t_0-\tau_g)}  \Big).
\end{equation}
	Here $m\approx m_0\alpha$ is the residual mobility for people working remotely, but still moving for ordinary needs ($m_0\approx 0.3$ is the residential movement trend from Google's mobility data, which is multiplied by $\alpha$, i.e., the share of people working remotely\footnote{Recall that the Google dataset  measures visitor numbers to specific categories of locations every day and compares the change in this number relative to an average baseline day before the pandemic outbreak.}).  The variable $p$ is the average distance between individuals in workplaces, taking into account all types of individual protection measures (masks or other devices that together reduce the probability to be infected)\footnote{If one follows the research from \citet{chu2020physical} and considers all the individual protection procedures, one can obtain a minimal value of $p\approx 0.2$, as best-case scenario.}. For example, in an workplace or a school, where structured contacts occur,  one might approach the problem of reducing the risk of disease transmission using individual position allocation according to physical distancing prescriptions. 
	
	A crude estimate can be made by considering a constant biological-environmental state, a number of population at risk equally distributed among the two categories, and at the beginning of an epidemic, so that $R_t\sim R_0\big( m_0\alpha ^2  + p(1-\alpha) \big)$. The share of teleworking activity $\alpha$ in order to minimized the epidemic risk is given by $\alpha=p/2m_0$.  This value of $\alpha$ yields a reproduction number equal to $R_t^{min}=2m_0\alpha(1-\alpha/2)R_0$. 
	Now consider a realistic estimated teleworking percentage of $\alpha= 30\%$, and  a best-case scenario where physical distancing protocols allow one to reduce the contact risk to $p=0.2$.  In this scenario, the minimum value of the reproduction number is  $R_t\sim R_0/6\approx 0.5$, where we consider the basic reproduction number at the beginning of the Lockdown to be $R_0=3$, as also discussed in \citet{park2020systematic,zhuang2020preliminary,d2020assessment}. The value of $R_t\approx 0.5$ we have estimated, incidentally, is the minimal value reached during the lockdown in Italy.  This was acheived with a percentage of remote working of about $40\%$, compared to a capacity of $30\%$ allowed by remote systems organization. Moreover, the percentage of those with training in remote working constituted a share of only about $15\%$ of the workforce.  This excessive amount of remote working may have impacted the productivity of the Italian economy.  Note that the estimations above are considered in the best-case scenario where individual safe position allocation and both the occasional and structured contacts are assessed in the most precise possible way and under favorable environmental conditions.  As a consequence, such reasoning represents only a theoretical attempt pointing towards future directions for policy making strategies.
\section{Conclusions}
The renewal equation is a powerful tool for analyzing and modelling epidemic
data.  We have found it to be both practically and conceptually useful.
In combining the renewal equation with a kinetic collisional model for
infection propagation, we have been able to derive a set of predictive
equations for the short-to-medium-term behavior of an epidemic.  These 
tools allowed us to disentangle the effects of population mobility,
physical proximity, and depletion of susceptibles.  Knowing the effects
of each of these components of the response of the government and society
to the CoViD-19 epidemic should allow for less costly and more effective
strategies for defeating epidemics.  In particular, the collision model approach to
estimation of infection risk should allow local, regional, and national
governments to better assess the continuing threat of CoViD-19 to the public welfare.

Some future directions for this research are: extension of the model to
be more realistic, extension of the analysis to obtain more useful information
about the propagation of the epidemic, and incorporation of the lessons
learned into more comprehensive methods for combating CoViD-19.
This analysis has focused on the lockdown, but the same theoretical tools
along with additional technology and data resources show promise for the
analysis of the post-lockdown response and further mitigation of this disease.

We have mainly focused our study on the spread of a contagion in a homogeneous population, and at this stage, we do not investigate the dynamics of the severity of the disease.  This is interrelated with the mechanisms of immune response to the SARS-CoV-2 infection. In order to examine these dynamics, we would need to focus our attention on the microscale corresponding to viral particles and immune cells.  Since these agents induce the dynamics of the varying intensities of the disease observed at the macroscopic scale of the human population.  Furthermore, to assess the severity of an epidemic in a population, one should take into account both the reproduction number $R(t)$ and the absolute number of cases. A high $R(t)$ is manageable in the very short run as long as there are not many people sick to begin with. An important aspect of $R(t)$ is that it represents only an average across a region. This average can miss regional clusters of infection. Another subtlety not captured by $R(t)$ is that many people never infect others, but a few 'superspreaders' pass on the disease many more times than average, perhaps because they mingle in crowded, indoor events where the virus spreads more easily. This means that bans on certain crowded indoor activities could have more benefit than blanket restrictions introduced whenever the $R(t)$ value hits one. In conclusion, in addition to $R(t)$ one should look at trends in numbers of new infections, deaths, hospital admissions, and cohort surveys to see how many people in a population currently have the disease, or have already had it. Fatality rates and intensive care hospitalization rates are related to disease severity.  Virulence increases with repeated contacts, since it is related to the number of exposures to the virus and the infectious dose. Our estimation of the rate of collision is equal to $\gamma =4\pi d  \mu r$. In fact, mask wearing, physical distancing and hygiene may also be reduce the infectious dose that people encounter in the population at large.

\section{Acknowledgments}
Fabio Vanni acknowledges support from the European Union's Horizon 2020 research and innovation programme under grant agreement No.822781 GROWINPRO - Growth Welfare Innovation Productivity.
\section{Conflict of interest}
  The authors declare that there are no conflicts of interest regarding the publication of this paper.

\bibliographystyle{plainnat}
\bibliography{referencesE} 
\addcontentsline{toc}{chapter}{Bibliography}

	\clearpage
	\newpage
	\appendix
	\section*{{Appendix}}

	\section{A semi-analytical estimate of $R_t$ and data calibration }\label{sec_appendixR}
The classic approach to the renewal equation for epidemics, cf. \citet{nishiura2010time,champredon2018equivalence,breda2012formulation},
in its common version assumes that the non-linearity of an epidemic is
characterized by the depletion of susceptible individuals alone (i.e.,
contact and recovery rates are independent of calendar time).  Under
such assumptions, the kernel $A$ can be described in terms of
the so called generation-time distribution $g(\tau)$ which measures
{the time between when a person gets infected and when they
  subsequently infect other people}. The generation-time distribution
is made up of two factors: the first is the probability of being
infectious $\tau$ time units after initially becoming infectious.  The
second contribution is given by the ``transmission potential'', that
is, the average number of secondary infections at ``infection age''
$\tau$.

Then, the \textit{instantaneous reproduction number} is defined as:
\begin{equation}
R_i(t):=\int_0^{\infty} A(t,\tau)d\tau,
\end{equation}
which is understood as the expected number of secondary infections transmitted
by a typical infectious individual at calendar time $t$.  For $t=0$, it corresponds to $R_0$,
which is the basic reproduction number. We highlight that the instantaneous reproduction number is different from the effective one, at least in principle, but under certain assumptions they are relatively similar, and such conditions are particularly fulfilled if the lockdown policy are in action, so that the two measures for the reproduction number coincide,  when the number of cases per day is roughly constant, both definitions should give R = 1, or rather if $\beta=\beta _0$. 

\begin{figure}[!h]
 	\centering
 	\begin{subfigure}[c]{0.9\textwidth}
 		\centering
 		\includegraphics[angle=0,origin=c,width=0.9\linewidth]{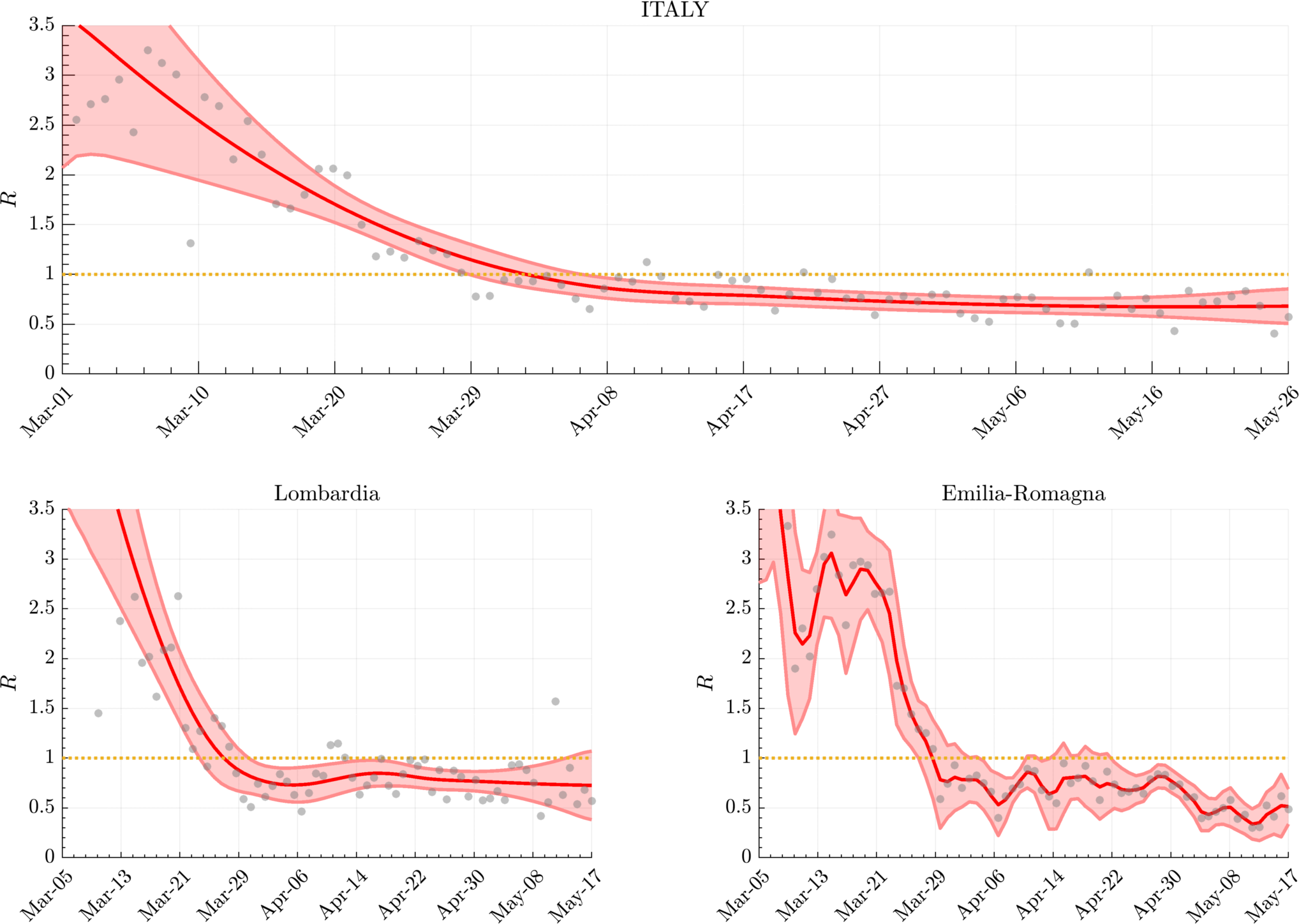}
 	\end{subfigure} 
 	\caption{$R_t$ values obtained from cases reported by date of diagnosis described in the appendix. Source: \citet{ProtezCivileCov}}
 	\label{fig_Rdpc}  
 \end{figure}

\begin{figure}[!h]
 	\centering
 	\begin{subfigure}[c]{0.9\textwidth}
 		\centering
 		\includegraphics[angle=0,origin=c,width=0.9\linewidth]{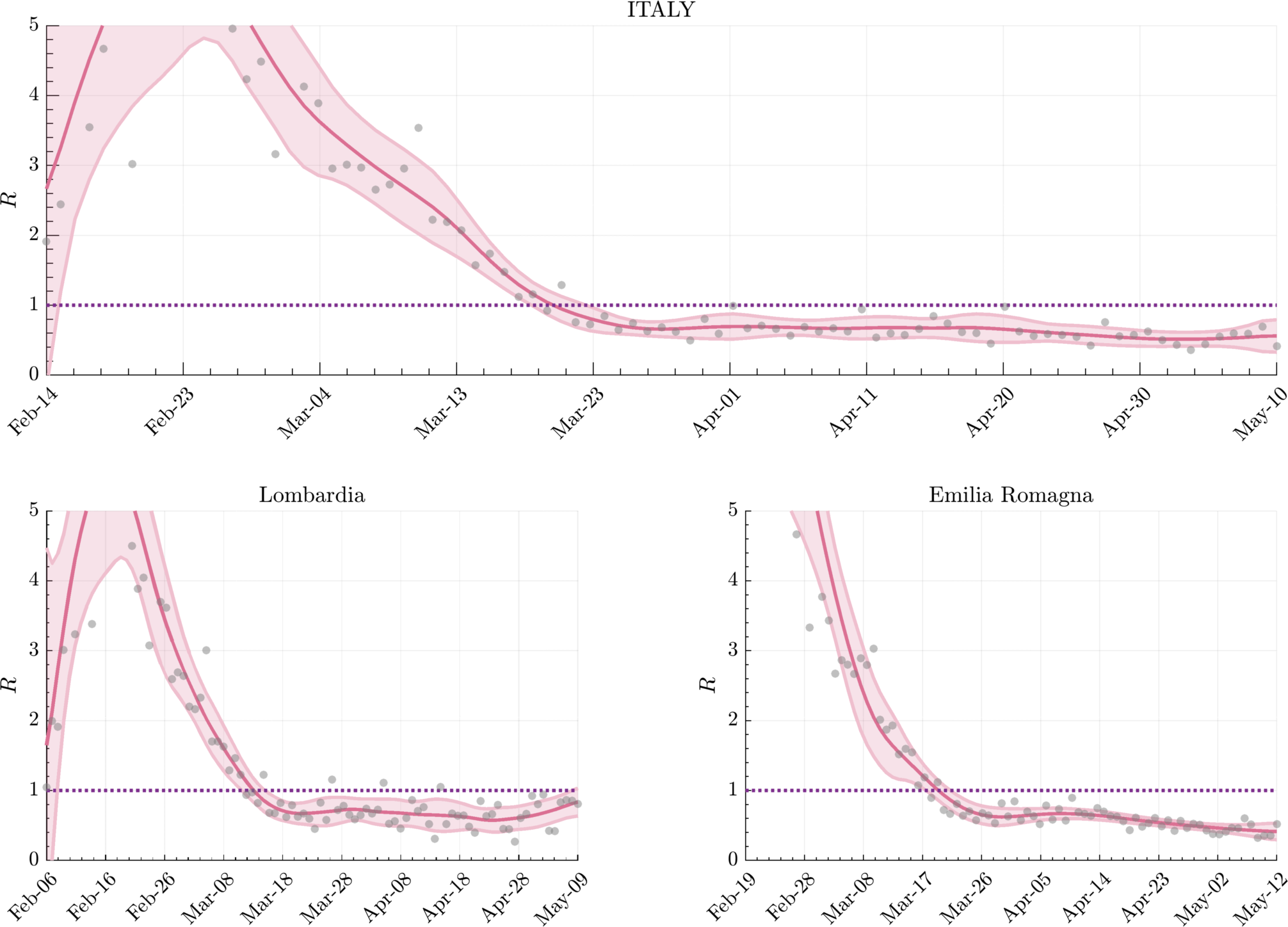}
 	\end{subfigure} 
 	\caption{$R_t$ values computed for cases reported by date of onset of symptoms. Source: \citet{epicentro}. The time shift with respect to Fig.\ref{fig_Rdpc} is due to the fact that the onset of symptoms is typically
 		one week before the official detection of the case.}
 	\label{fig_Repi}  
 \end{figure}

In general, the infectivity kernel $A \geq 0$ is an arbitrary function of the infection age $\tau$, but as shown in \citet{meehan2019global,magal2010lyapunov,fraser2007estimating} it can be interpreted as the probability to be infectious (capable of transmitting the disease) with age of infection $\tau$.  This probability is variable as the disease progresses within an infected individual. 
Following the renewal equation approach to estimation of the reproduction number as in \citet[Ch.8]{yan2019quantitative} and used in \citet{nouvellet2018simple,flaxman2020report,aleta2020modeling}, let us call $\tau _A$ and $\tau_g$ the maximum and the minimum infection age at which an average individual can contribute to the force of infection, respectively. 
A general approach to estimation of $R(t)$, widely adopted in epidemiological modelling, inverts the general convolution equation eq.\eqref{eq_full} supposing the kernel distribution to be known. We, instead, obtain information about the infection distribution, $\Gamma(\tau)$, from the response of the epidemic's evolution to a decrease in $\beta$ during lockdown.  We do this by seeking the form of $\Gamma(t)$ that leads to a stable and step-like behavior of $\beta(t)$ resembling the behavior of the population's mobility (see, e.g., Fig.\ref{fig_Rdpc}). Notice that eq.\eqref{eq_full} is linear in $j(t)$ and the nonlinearity due to saturation phenomena is hidden inside $n_s(t)$. This means that the results we obtain are also valid if we rescale $j(t)$ by a constant, $\lambda$ (due, e.g., to the asymptomatic and other non-detected cases).
Under the assumption that $\beta$ is independent of the infection-age, we invert eq.\eqref{eq_full}, the discrete anti-convolution equation, obtaining:
\begin{equation}
\beta_i n_{s,i} = j_i \left ( \sum\limits_{k=1}^i \Gamma_k j_{t-k} \right )^{-1}.
\end{equation}
Trying various forms of $\Gamma_i$, we observe that the key ingredient to obtain a flatter behavior of $\beta_i n_{s,i}$, is an abrupt transition of $\Gamma_i$ to zero after a time $\tau_A$.  In particular, after the date corresponding to 15 days after the lockdown in Italian regions, yielding $\tau_A = 15$ days. The time evolution of $R$ estimated from our approach, supposing $n_s$ equals $1$ (total population susceptible) is:
\begin{equation}
R_i = \beta_i   \sum\limits_{k=1}^{\infty} \Gamma_k.
\end{equation}
The simplest choice of the infection-age probability is:
\begin{equation}
\Gamma(t) = \left \{
\begin{array}{l}
1 \quad \mathrm{if} \: \tau_g\leq t < \tau_A\\
0 \quad \mathrm{} \: \text{otherwise}
\end{array}
\right.
\end{equation}
This function produces a regular and intuitive behavior of the reproduction number $R$ of CoViD-19 that resembles very nearly the mobility pattern (see Fig.\ref{fig_Rdpc} and \ref{fig_Repi}) with $\tau_A=15$ days and where $\tau_g=4$.\footnote{We similarly evaluated a gamma distributed $\Gamma_k$ with mean $6.5$ and scale factor of $0.65$, obtaining equivalent results for the time period of the lockdown.}, see \citet{lauer2020incubation,ganyani2020estimating,du2020serial}. Let us notice that, the $R_t$  does depend on $\Gamma(\tau)$, but only through its integral over all possible values of tau. As a consequence, the most changing $\Gamma$ can do is change R(t) by a re-scaling.  If the mean infectious period is the same for two generation-time distributions, then R will be the same. However, effectively, infectious age distribution  depends on $t$.  Since contact tracing, testing, and isolation (as well as treatments) will tend to reduce the active infectious period (and their use depends on t). However, the scale of $R_t$ is important, since the value of 1 is a fixed point that one would like to be below.

In the renewal equation approach, $A(t, \tau)$ is usually decomposed as
$A(t, \tau) = R_t g(\tau)$, where $g(\tau)$ is the generating-time
distribution, so that
\begin{equation}
j(t)=R_t\int_{0}^{\infty}g (\tau) j(t-\tau)d\tau.
\end{equation}
 Typically, the generation distribution is unknown, though it can be
 approximated by assuming it is the same as the \textit{serial-interval
 distribution,} which refers to the time between successive
 cases in a chain of transmission (the time interval between infection and
 subsequent transmission).\footnote{ The serial interval may be the
   same as the generation time if the onset of symptoms is the same as
   the onset of infectiousness and the latent period is constant. This is not the case when the
   incubation period of the primary case depends on the time from
   onset to secondary transmission}.

 If we look at the data regarding $j(t)$, we see that, after that the effect of the
 lockdown becomes apparent, $j(t)$ shows an exponential decay.  This is
 especially clear in the case of the data referred to the date of symptom
 onset.  
 
 If we follow this approach, the value of $R_t$ can be evaluated
 analytically. Indeed we have that, if $j(t) = j_l \exp(- \eta t)$
 \begin{equation}
 j_l \exp(-\eta t) = R_t \int_{0}^{\infty}g(\tau) j_l \exp(-\eta t) \exp(\eta \tau) \mathrm{d} \tau.
 \end{equation}
 Cancelling $j_l\exp(-\eta t)$, and multiplying both sides by $\int_0^{\infty} \Gamma(\tau) \mathrm{d}\tau$,
 we have
 \begin{equation}
 R_t = \dfrac{\hat{\Gamma}(0)}{ \hat{\Gamma(-\eta)}}.
 \end{equation}
 Here $\hat{\Gamma}(s) $ is the Laplace transform of $\Gamma(t)$. In the case of a window-like
 $\Gamma(t)$ between $\tau_g$ and $\tau_A$ we have
 \begin{equation}
 R_t = \dfrac{\eta (\tau_A - \tau_g)}{ \exp(\tau_A \eta) - \exp(\tau_g \eta)}.
 \end{equation}
 Setting $\eta = 1/22.14$ (days)$^{-1}$ here yields $R_t \simeq 0.66$.
 
 Now, let us use instead the gamma distribution $\Gamma^{\prime}(t)$.  Recall that 
 its Laplace transform is 
 \begin{equation}
 \hat{\Gamma^{\prime}}(s)= \dfrac{\beta^{\alpha}}{(s + \beta)^{\alpha}}.
 \end{equation}
 Using the values of the parameters found in the literature \cite{flaxman2020report}, namely
 $\alpha =  1.87$, $\beta = 0.28$, and $\eta = 1/22.14$, we obtain
 $R_t \simeq 0.72$.  This value is slightly larger than that found with the window-like
 approach we followed in the main part of the paper.

 The renewal approach can be connected to deterministic, compartmental models
such as SIR models.  For the SIR model, $\beta$ is considered constant
with respect to infection age and is called the transmission rate.  The 
infectious survival probability is $\Gamma(t)=e^{-\gamma t}$, with $\gamma = D^{-1}$,
defined to be the recovery rate, which is the same as the inverse of the mean infectious time $D$. 
Considering the renewal collision equation eq.\eqref{eq_incidence}, we substitute the collisions of individuals moving in a region with a rate of contacts $\beta(t,\tau)$ (an average value representative of the whole region), so that we can neglect geographic factors.
Thus, we have a correspondence with compartmental deterministic models considered in terms of  the integral kernel:
\begin{align}
A(t,\tau)&=\beta e^{-\gamma \tau} \quad \text{SIR}\\
A(t,\tau)&=\beta \frac{\sigma }{\gamma -\sigma}(e^{-\sigma \tau}-e^{-\gamma \tau}) \quad \text{SEIR},
\end{align}
both for the SIR and for the corresponding SEIR model, with average duration of latency equal to $1/\sigma$.
Examining the basic reproduction number at the beginning of the epidemic ($t\to 0$) we find the popular basic reproduction number $R_0$,
\begin{equation}
R_0:=\int A(t=0,\tau)d\tau = \frac{\beta}{\gamma}.
\end{equation}
This is equivalent to the reproduction number obtained in the SIR model, where $R_0$ is the ratio between effective contact rate and the removal rate (i.e., the inverse of the expected duration of infection).

\begin{figure}[!ht]
	\centering
	\begin{subfigure}[c]{0.9\textwidth}
		\centering
		\includegraphics[width=0.8\linewidth]{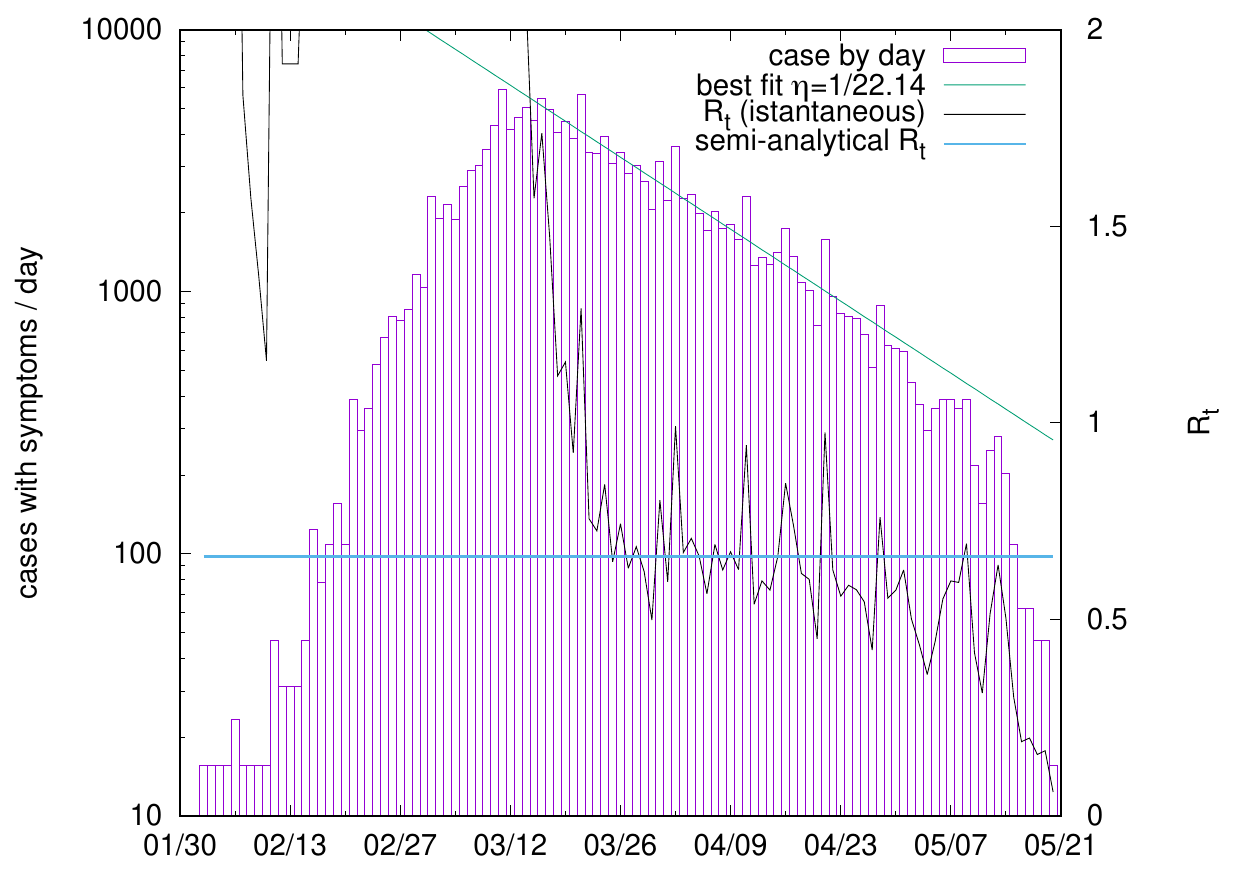}
	\end{subfigure}%
	\caption{Daily cases reported to the date of the onset of symptoms (source ISS \cite{epicentro}) in lin-log scale. The green line is the best exponential fit  to the data between March 13th  and May 5th. The black line is the instantaneous $R_t$ while the horizontal line refers to the semi-analytical estimate corresponding to $\eta = 22.14$, $\tau_g = 4$, and $\tau_A = 14$.}\label{fig_epi_log}
\end{figure}

So, as regards the  \textit{effective reproduction number} we can write for the SIR model:
\begin{align}
R(t)&=\text{IPR}_t\cdot D\\
&=\frac{\beta}{\gamma}n_s(t).
\end{align}
For the SIR model, the reproduction number is evaluated as
\begin{equation}
R_{\text{{\tiny SIR}}}(t) = \dfrac{j_o(t)}{I(t)}.
\end{equation}
In several regions and countries we observe that after the lockdown the behavior of the number of daily infections has responded to abrupt changes with a delay of approximately 15 days.  After that, the time evolution of $j(t)$ is very similar to an exponential decay, to a new steady value. See, for example, the growth and decrease exponential decays fitted to the Italian national data supplied by the \citet{epicentro}, referred to the date of the onset of symptoms in Fig.\ref{fig_epi_log}.
The same data, if interpreted as the output of an SIR or SEIR-like model, leads to a value of of the reproduction number $R_{\text{{\tiny SIR}}} = \dfrac{j_o(t)}{I(t)}$ (where $I(t)$ is the number of active infected persons reported at time $t$) which decays approximately exponentially, beginning just after the abrupt decrease in mobility due to the lockdown effect.  Note that in our renewal description, the active cases are the total number of infected people which are not isolated, and consequently have the same mobility as the rest of the population.  We call the ratio of the number of non-isolated infected people to the total population the prevalence $p(t)\neq I(t)$ as used in the SIR description.
This behavior is quite surprising because, as shown in Fig.\ref{fig_mob}, the mobility remains quite constant after the lockdown, so it is not clear why $\beta_{SIR}$ decays until the end of May 2020. For this reason we think that, due to the obvious incompleteness of data, it is better to use our approach in evaluating the epidemiologically significant value of $R_t$. 
%

\section{Temporal intervals in epidemics}\label{app_epitimes}
\begin{figure}[!ht]
	\centering
	\begin{subfigure}[c]{0.8\textwidth}
		\includegraphics[width=0.9\linewidth]{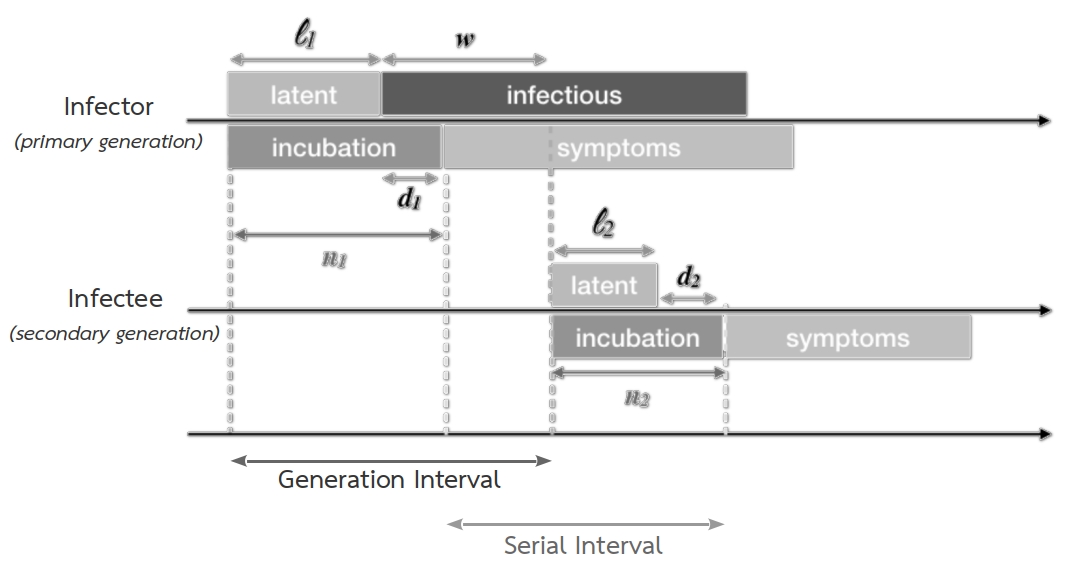}
	\end{subfigure}%
	\caption{ Clinical and epidemiological periods and parameters.  Figure modified from \citet{champredon2018equivalence}.}
	\label{fig_infectiontime}  
\end{figure}

Understanding the time intervals between successive generations	of infected individuals is crucial to appropriately quantifying the transmission dynamics of infectious diseases.  As discussed in  \citet{champredon2018equivalence,fine2003interval,svensson2007note,britton2019estimation}, there are three fundamental time periods that determine transmission from one individual to another for directly-transmitted infectious diseases: the latent ($l$), incubation ($n$), and infectiousness periods, as summarized in \ref{tab_times}.
		\begin{table}[!ht]
	\centering\captionsetup{justification = centering}
	\caption{typical time periods in infectious disease evolution}
	\label{tab_times}
	{
		\setlength\arrayrulewidth{.001pt}
		\begin{tabular}{Sc|Sc}
			\textbf{Term} & \makecell{ \textbf{Description}} \\ \toprule 
			\rowcolor{Gainsboro!20}
			\makecell{\textsl{ in clinical analysis  } }     & {\footnotesize  \makecell{ } }   \\ \hline
			{ \makecell{{Incubation period} } }    & {\footnotesize  \makecell{time from  infection to first clinical symptoms \\ (one can be infectious in this period) }} \\ \hline
			{ \makecell{{ Latent period } } }    & {\footnotesize   \makecell{ time from infection to onset  of infectiousness\\ (one cannot transmit yet) }}  \\ \hline
			{ {Infectious period}    } &  {\footnotesize \makecell{time during which  an infected individual  \\ can transmit a pathogen to others \\(An infectious individual may not show symptoms) }}  \\ \midrule
			\rowcolor{Gainsboro!20}
			\makecell{\textsl{in epidemic modeling } }     & {\footnotesize  \makecell{ } }   \\ \hline
			\makecell{{Generation time} }     & {\footnotesize  \makecell{ the time duration from the onset of infectiousness\\ in a primary case to the onset of infectiousness in a secondary case\\ infected by the primary case. (not easily observable)} }   \\ \hline
			\makecell{ {Serial interval}}     & {\footnotesize  \makecell{ duration of time between the onset of symptoms\\ in a primary case and the onset of symptoms in a secondary case\\ infected by the primary case (a readily observable period).}  } \\ \bottomrule
			
		\end{tabular} 
	}
\end{table}
Calling $w$ the interval of time between the end of the infector's latent period and the time of disease transmission to an infectee,  the difference between the latent and incubation periods is
noted $d = l - n$. In particular, as in figure \ref{fig_infectiontime}, we define the generation time as $g=l_1+w$ and the serial interval as $s=g-n_1+n_2=(l_2+w)+(d_1-d_2)$. Assuming $l$ and $d$ to be identically distributed, the generation-interval distribution  and serial-interval distribution  have the same mean.  If all the time periods are also uncorrelated in addition to being identically distributed: $\text{var}(s)=\text{var}(g)+2\text{var}(d)$.  Thus, if the variance of the difference between the latent and incubation periods is small, the variance of the serial and generation intervals are similar.
Recent studies on CoViD-19,as in \citet{ferretti2020quantifying,park2020time}, have highlighted that there are significant contributions of different transmission routes to the distribution of generation times (time from infection to onward transmission), and consequently the basic reproduction number $R_0$ and the disease dynamics.  The different routes hinge mainly on the role of non-symptomatic carriers in transmission. In particular, if the generation-interval distribution of asymptomatic transmission differs from that of symptomatic transmission, then estimates of the basic reproduction number which do not explicitly account for asymptomatic cases may be systematically biased.

\section{Packing occasional and structured contacts}\label{app_occasional}
Let us briefly discuss a naive way to minimize the transmission of  airborne diseases occurring via human contacts, which can be split into two categories: structured and occasional contacts. 
In our collisional kinetic framework we have considered contacts among individuals to be random or, in other words, occasional. In addition to these erratic contacts (happening, for example, in the streets), one can consider structured contacts occurring at home, in hospitals, workplaces, and schools, just to mention a few of the possibilities. For structured contacts, we should consider the use of a different approach than collision theory.

To address the case of structured contacts, we consider a geometrical viewpoint known as circle packing theory.  Circle packing theory studies the arrangement of individual circular zones (infectious zones in our case) on a given surface such that no overlap occurs and so that no circle can be enlarged without creating an overlap.
In order to minimize exposure to the virus, and so reduce the viral load in closed and clustered spaces like workplaces and hospitals, it should preferable to arrange individuals inside a given boundary such that no two infectious zones overlap and some (or all) of them are mutually tangent. Using the most efficient hexagonal packing one can obtain a packing density of $\pi/12$ so that $90.6\%$ of the working area will be covered by workers. The actual packing density will be less than this value because of boundaries which will force one to either use a sub-optimal packing method or leave gaps at the edges.  If each person should be at least $r$ meters away from others to be a safe distance of infection,in an environment box with side length $L$, the number of workers (or students in a class) that could be allocated is:
$$ n=\frac{1}{12}\left(\frac{L}{r}\right)^2.$$ 
However, using circle packing theory to arrange people in bounded area is only a part of the strategy to reduce the infectious dose.  Other effects are also important, for example, the same arrangement of people can have different effects depending on whether it occurs indoors or outdoors. However, a better packing can help to use the same spaces and buildings in a more efficient way, which is especially relevant for schools and workplaces.

Now we turn to the occasional contacts among individuals which makes the path of each individual erratic. Such random movements could also be organized in order to reduce the relative velocity of individuals.  For example, when workers have to move it would be beneficial choose repelling paths so as to avoid contact or collisions. In our kinetic framework,  particle interactions, whether repulsive or attractive, are so weak that they are also negligible. Nevertheless, such repulsive behavior can result in very erratic walks.  Thus, increasing the relative velocity of active particles and consequently increasing the unpredictability of the trajectories. Nevertheless,  one can imagine persons to move in a coordinated way so as to minimize their relative speeds. Again, according to kinetic theory, the particles of a gas are in state of continuous random motion. The particles move in different directions with different speeds and we use the mean relative speed between particles. In any case, we violate the hypothesis that the velocities of particles are uncorrelated. The best way to reduce the relative velocity while allowing the particles to keep on moving is an ordered motion. For example, in a box one can move in parallel on overlapping lanes with identical seeds. Apart from some remaining fluctuations the mobility as relative velocity is minimized with an optimal packing allocation of space. In this way we could reach the maximal reduction of collisions without blocking the overall social movements.

\begin{figure}[!ht]
	\centering
	\begin{minipage}{.45\linewidth}
		\begin{subfigure}[t]{.9\linewidth}
		\centering
\includegraphics[angle=0,origin=c,width=0.9\linewidth]{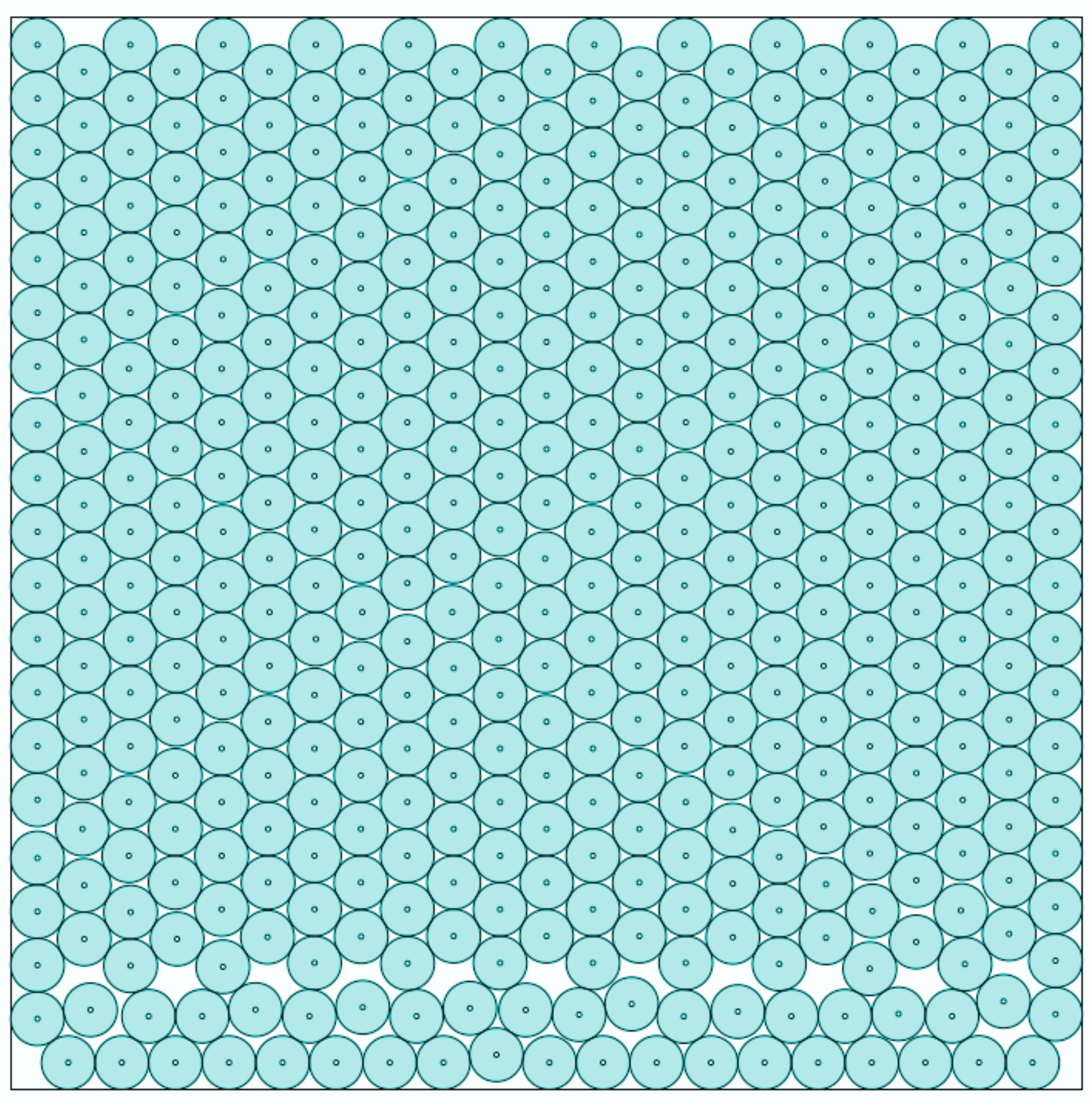}
			\caption{Circle Packing}
			\label{fig:weather_activity}
		\end{subfigure}
	\end{minipage}
	\begin{minipage}{.45\linewidth}
		\begin{subfigure}[t]{1\linewidth}
		\centering
\includegraphics[angle=0,origin=c,width=0.8\linewidth]{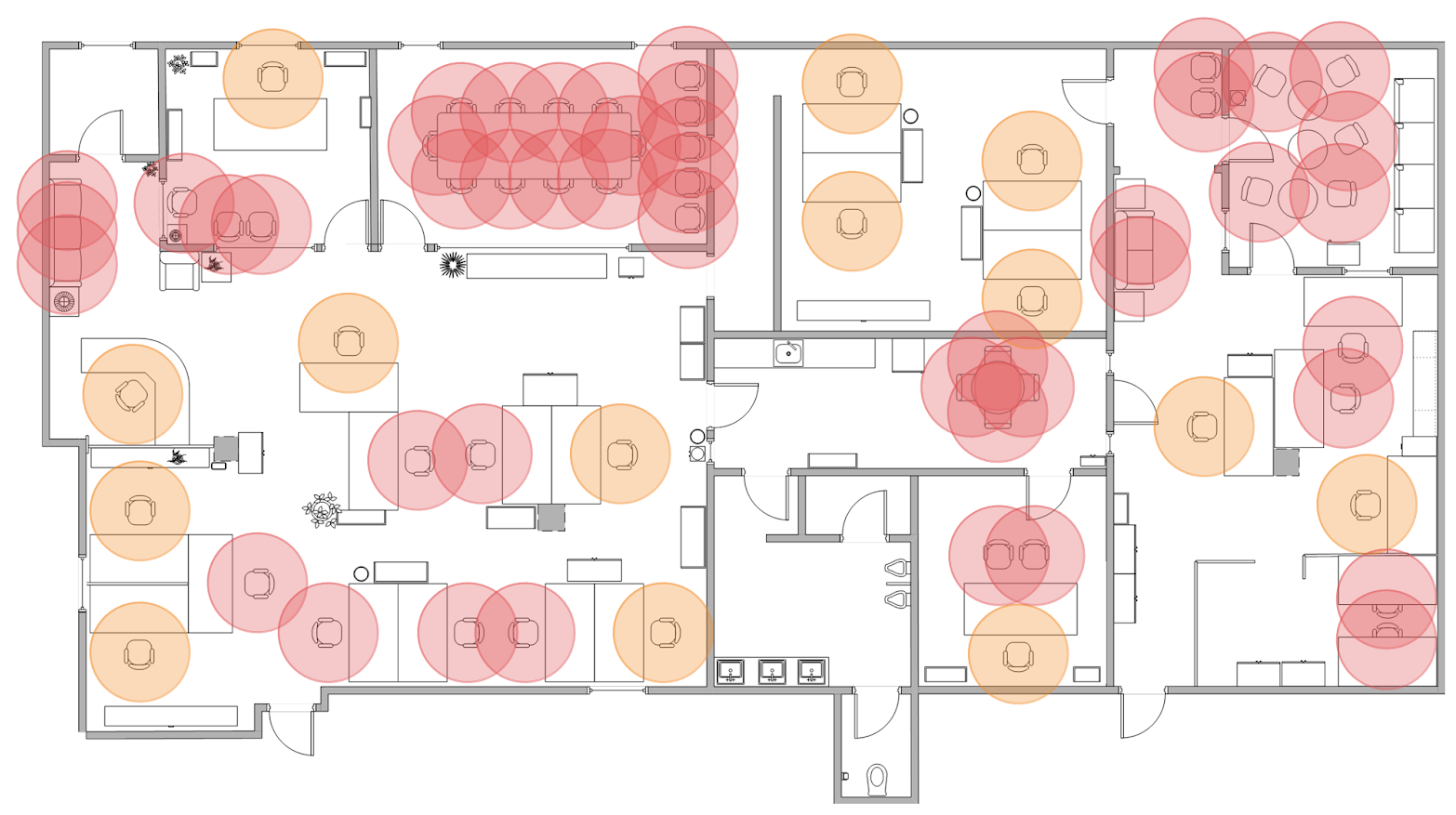}
			\caption{Workplace office}
			\label{fig:weather_filter1}
		\end{subfigure} 
	\\
		\begin{subfigure}[b]{1\linewidth}
		\centering
\includegraphics[angle=0,origin=c,width=0.6\linewidth]{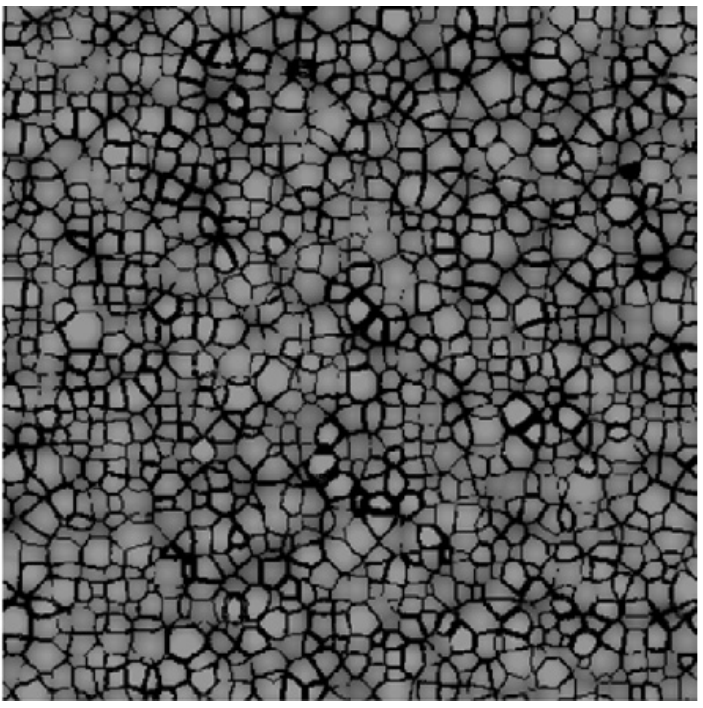}
			\caption{Crown-Shyness}
			\label{fig:weather_filter2}
		\end{subfigure} 
	\end{minipage}
	\caption{Physical distance in structured environments. (a) Packing optimum: side length of square is $L=40$, infectious zone for individuals has radius $r=2$. Thus we can fit $n=442$ persons in the region, using the hexagonal packing arrangement.  This wastes only $13.2\%$ of the available space. (b) Example of a office workplace where each location is surrounded by a circle $2$ meters in diameter, source: \citet{floorC}. However, in addition to physical distance, one should consider other factors such as air movement and ventilation, shared spaces and face coverings, to produce safe working environments. (c) crown delineation, which distinguishes crowns and identifies the species of each crown, source: \citet{goudie2009empirical}. }
\label{fig_packing} 
\end{figure}

This sort of coordinated spatial allocation might be inspired by crown shyness in trees, see \citet{franco1986influence,hastings2020tree,goudie2009empirical}. Networks of treetop chasms have been documented in forests around the world in which canopies maintain gaps, in a phenomenon called crown shyness, that may help trees share resources and stay healthy. This phenomenon consists of limitations on the growth of the canopy the trees, in such a way that the leaves and branches of adjacent trees do not touch each other. This effect allows a greater penetration of light into the forest and permits forest plants to perform photosynthesis more efficiently. Moreover, it avoids damaging the branches and leaves in case of storm or gusts of wind and prevents diseases, larvae, and insects that feed on leaves from spreading easily from one tree to another.
This example of a natural social distancing strategy makes us suggest the possibility of building models of mobility inspired by this and similar phenomena with the addition of a transport component to accomodate movement needs.

\section{Predicting ahead one generation time}\label{app_growth}
We examine the growth rate of epidemics considered as a nonlinear system given by:
\begin{equation}
X(t+1)=F\left(X(t),\mathbf{p}\right), \quad t\in \mathbb{Z},
\end{equation}
where $X(t)\in \mathcal{R}$ is the state vector representing the total number of cases of infected individuals, $\mathbf{p}\in \mathbb{R}^l$ is a parameter vector, and $F:\mathbb{R}^n\times \mathbb{R}^l \to \mathbb{R}^n$ is a continuously differentiable function.
The growth factor is defined as:
\begin{equation}
\mathcal{G}(t)=\frac{\Delta X(t+1)}{\Delta X(t)}=|f'(X(t),\mathbf{p})|.
\end{equation} 
Which indicates the tendency of the epidemic to increase ($\mathcal{G}(t)>1$) or decrease ($\mathcal{G}(t)<1$).
Geometrically, this critical value of the growth factor is an inflection point of the cumulative number of cases.  The growth factor is also a proxy of the basic reproduction number, because it is approximately monotonically related to the true reproduction number and crosses $1$ when the true reproduction number crosses $1$, as explained in \citet{wallinga2007generation}.

Since $\Delta X(t)=J(t)$, the renewal equation allows us to estimate the generational growth rate to be:
\begin{equation}\label{eq_growthR}
\mathcal{G}(t+\tau_g) = \frac{J(t+\tau_g)}{J(t)}\approx \frac{\eta(t)N_s(t)}{\eta(t-\tau_g)N_s(t-\tau_g)}\frac{r(t)}{r(t-\tau_g)}\frac{\mu(t)}{\mu(t-\tau_g)}\dfrac{\sum \limits_{\tau=0}^{\tau_A-\tau_g}J_o(t-\tau)}{\sum \limits_{\tau=\tau_g}^{\tau_A}J_o(t-\tau)}.
\end{equation}
By definition, $\mathcal{G}(t)$ calculates the multiplicative increase in the number of incident cases over approximately one serial interval, but without requiring one to specify the serial interval distribution.  We also notice that for CoViD-19, the detection time (estimated using the serial interval) averages about $\tau_g=5$ days, while the maximum infection age is estimated to be $\tau_A=14$ days, cf.~\citet{ganyani2020estimating,li2020serial,nishiura2020serial}.

We see that the mobility is an essential piece of information when social and governmental forces, like the onset of lock-downs or other social and economical actions, modify the spread of an epidemic. 
Of current interest are the effects when lockdown policies are abandoned or weakened, so that mobility begins to increase.  We have supposed that the number of susceptible individuals has minimally changed.  However, it is possible to estimate from data the individuals which are at risk of infection by considering the number of undetected individuals, as shown below.
We used mobility data from \citet{Google19} and \citet{Apple19} as well as from forthcoming tracking systems.
  Finally, in order to evaluate the impact of different components of social distancing in the Growth rate dynamics \eqref{eq_growthR} we use Dynamic Time Warping (DTW). DTW is used to quantify the similarity or calculate the distance between two time series with different lengths. In time series analysis, DTW is one of the algorithms for measuring similarity between two temporal sequences, which may vary in speed. DTW indicates that even by introducing only the mobility data we could improve our forecast by about 10 percent.

\begin{figure}[!ht]
	\centering
	\begin{subfigure}[c]{0.99\textwidth}
		\centering
		\includegraphics[angle=0,origin=c,width=0.7\linewidth]{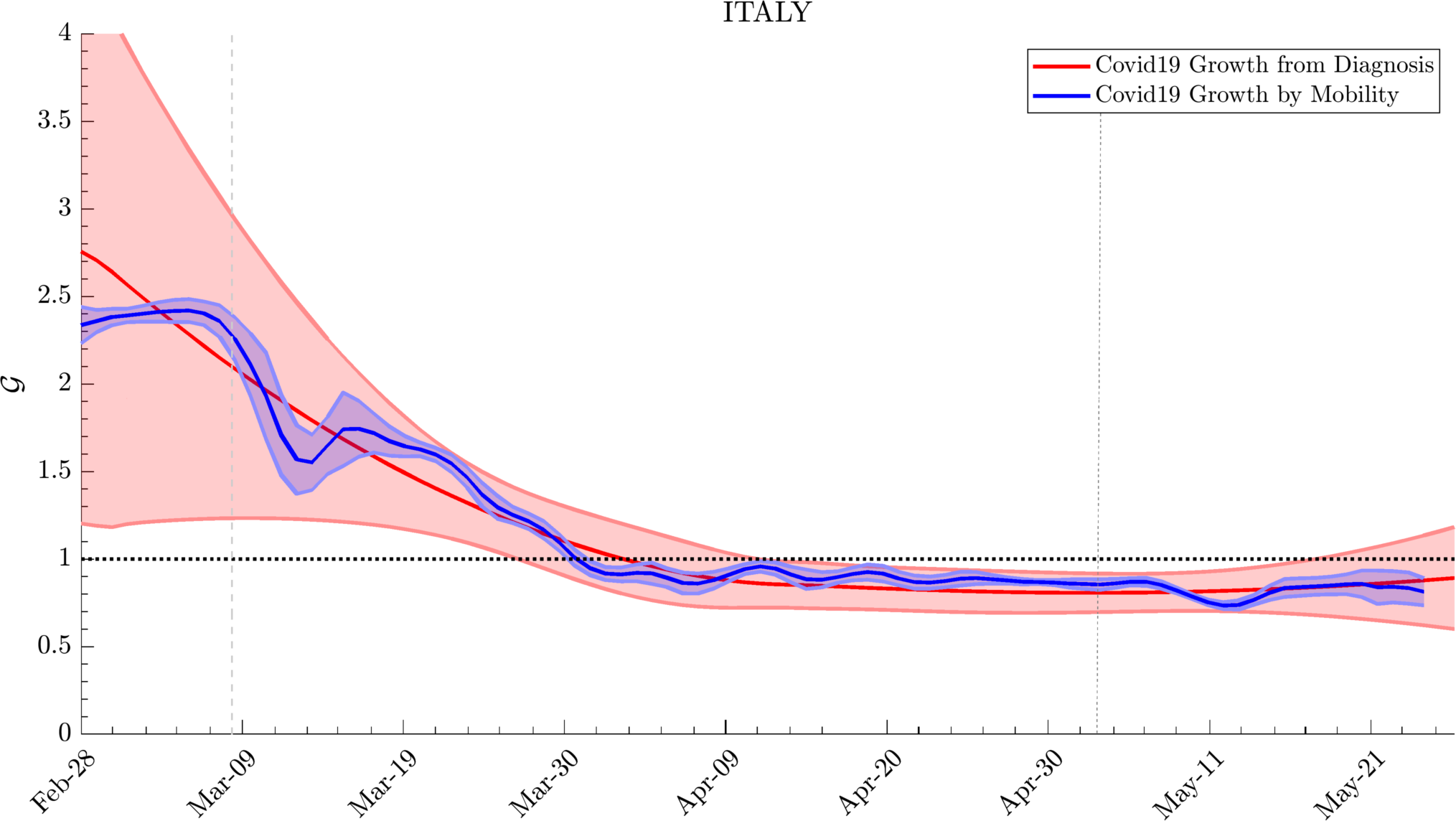}
		\caption{}
	\end{subfigure}%
	\\  \vspace{0.7cm}
	\begin{subfigure}[c]{0.45\textwidth}
		\centering
		\includegraphics[angle=0,origin=c,width=0.9\linewidth]{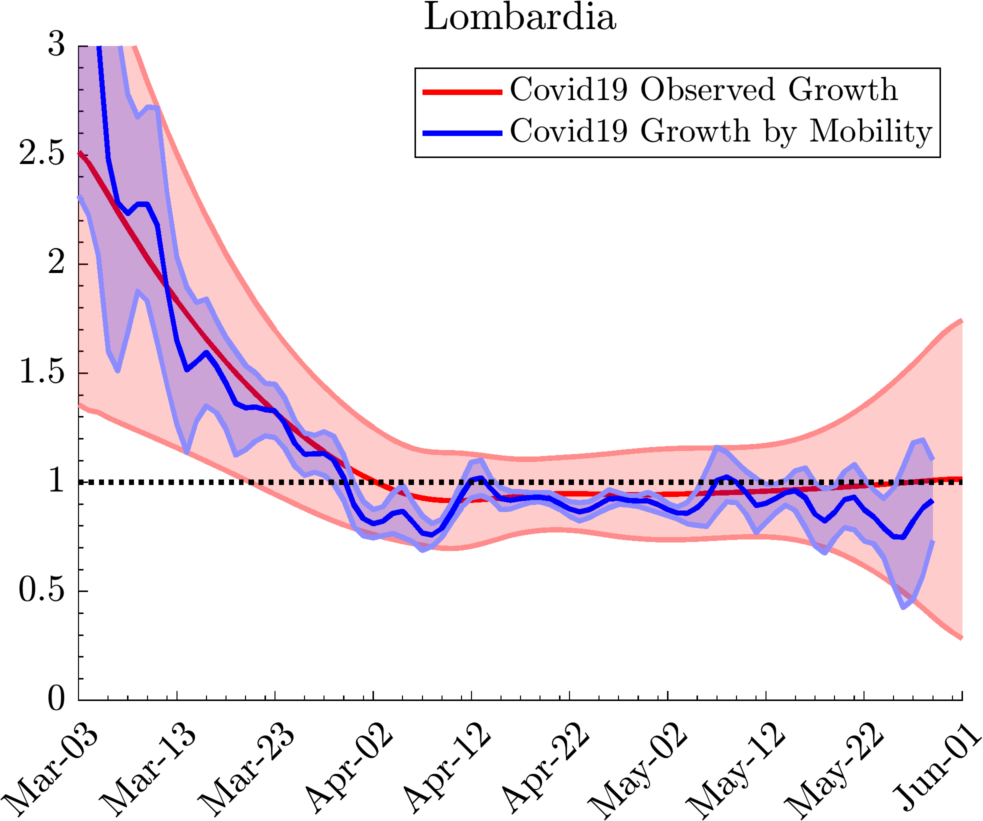}
		\caption{}
	\end{subfigure}%
	\begin{subfigure}[c]{0.45\textwidth}
		\centering
		\includegraphics[angle=0,origin=c,width=0.9\linewidth]{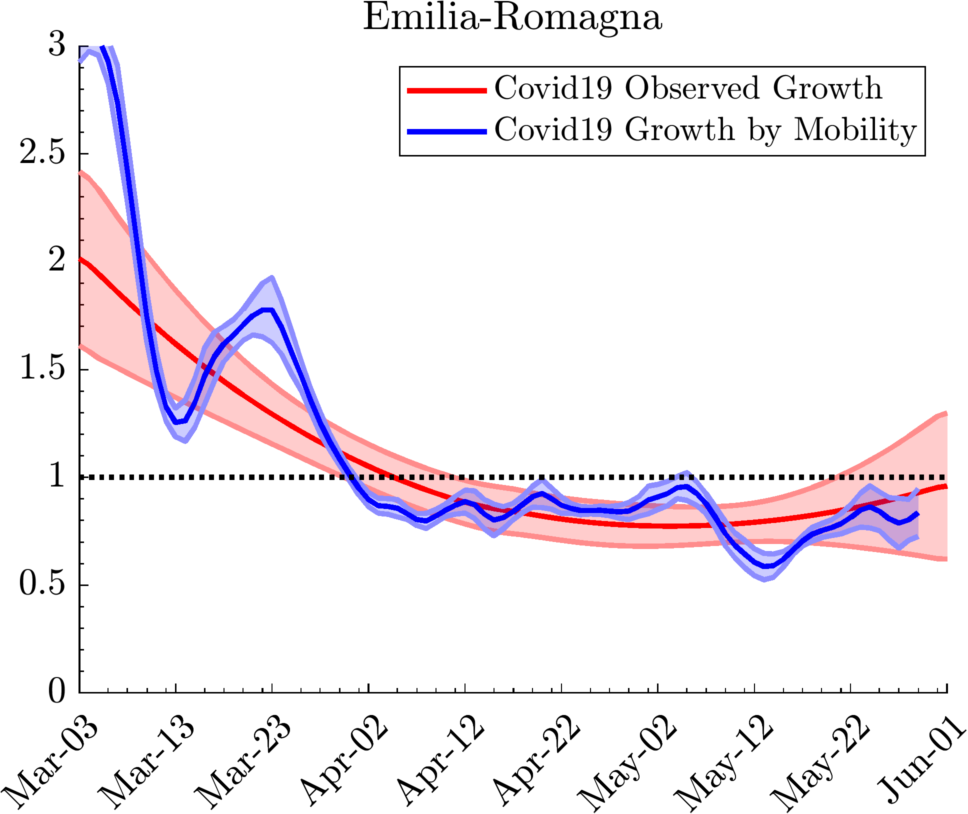}
		\caption{}
	\end{subfigure}%
	\caption{Infected individuals reported by date of laboratory diagnosis: Italy}
	\label{fig_itaGR}  
\end{figure}

\begin{figure}[!ht]
	\centering
	\begin{subfigure}[c]{0.99\textwidth}
		\centering
		\includegraphics[angle=0,origin=c,width=0.7\linewidth]{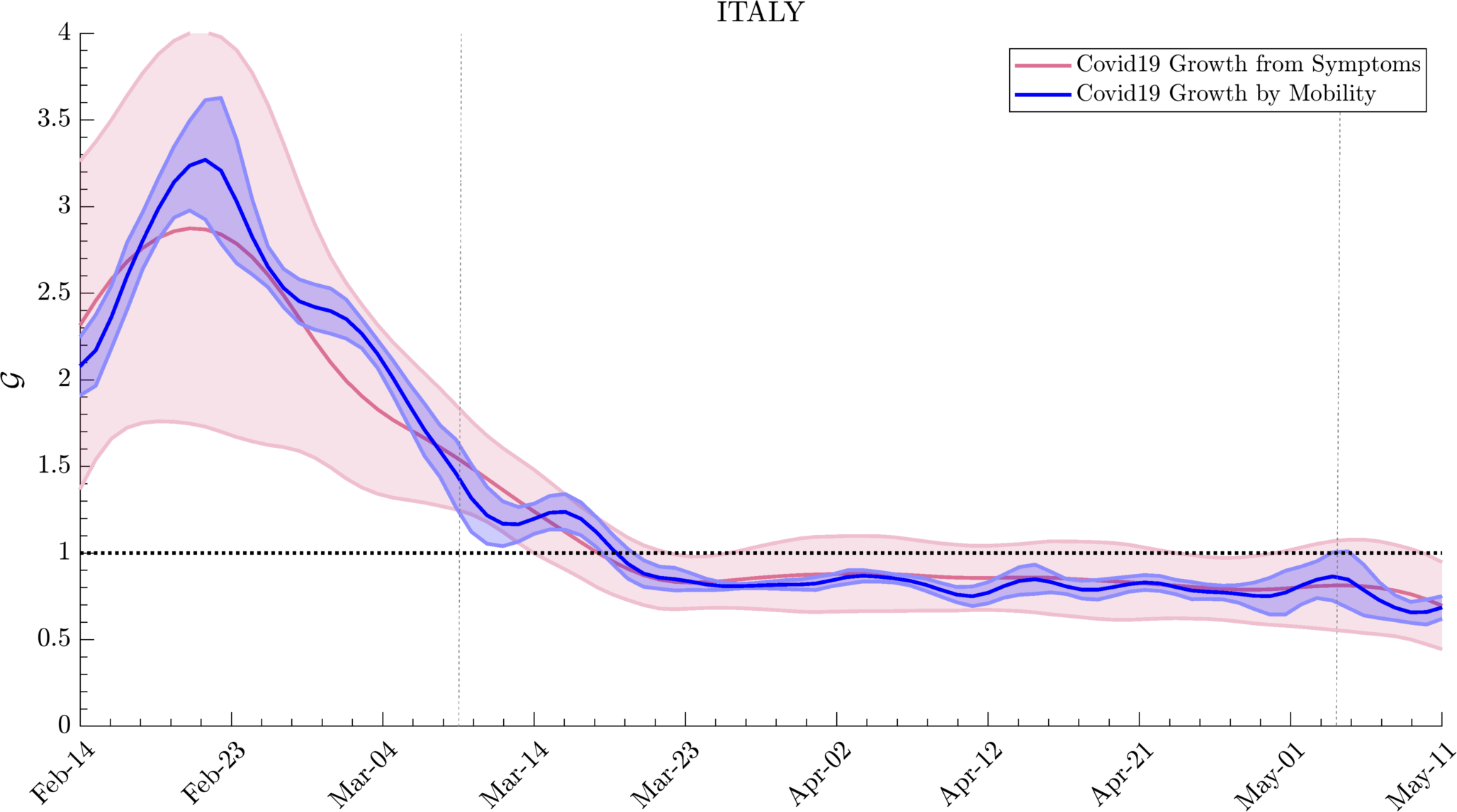}
		\caption{}
	\end{subfigure}%
	\\  \vspace{0.7cm}
	\begin{subfigure}[c]{0.45\textwidth}
		\centering
		\includegraphics[angle=0,origin=c,width=0.9\linewidth]{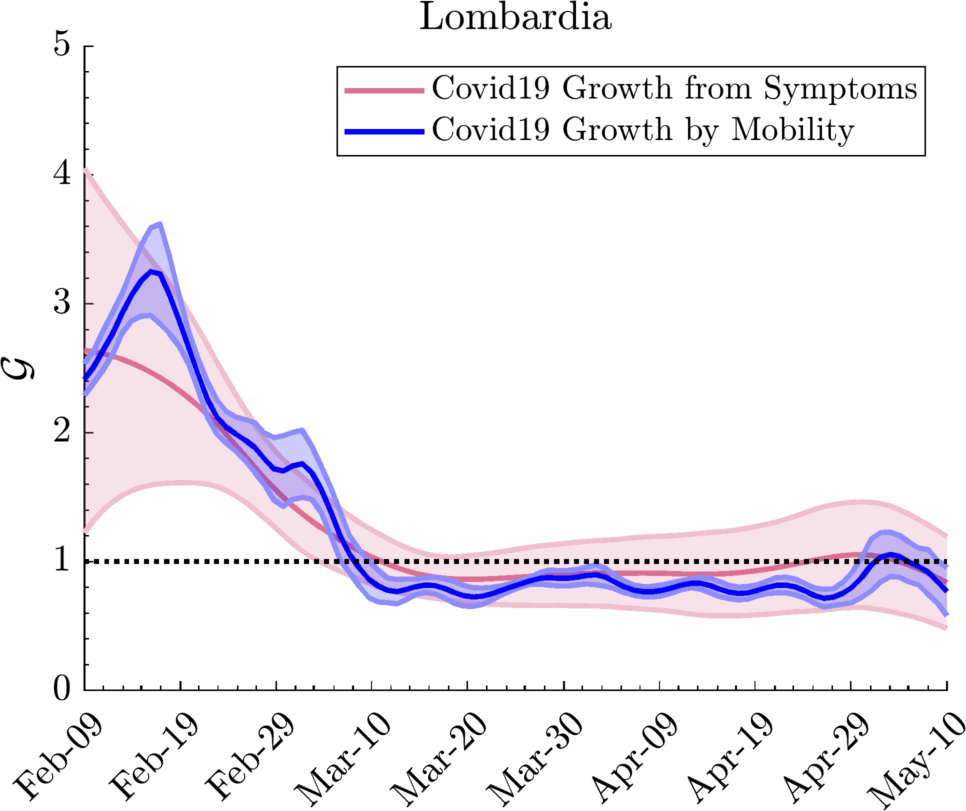}
		\caption{}
	\end{subfigure}%
	\begin{subfigure}[c]{0.45\textwidth}
		\centering
		\includegraphics[angle=0,origin=c,width=0.9\linewidth]{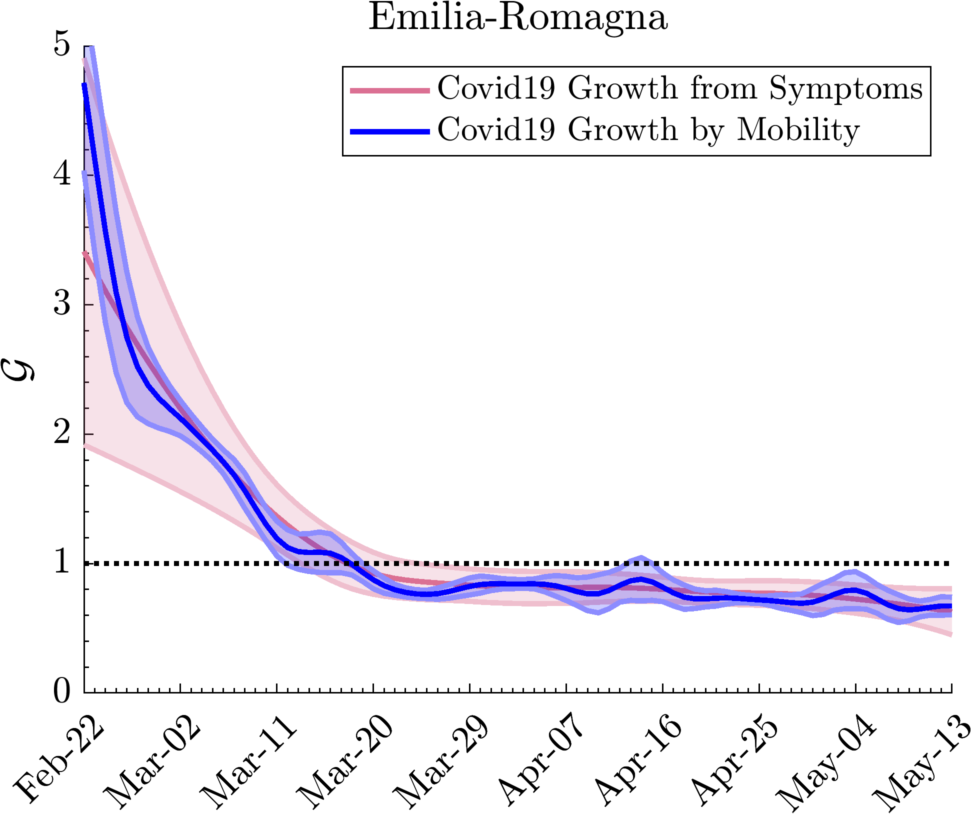}
		\caption{}
	\end{subfigure}%
	\caption{Infected individuals reported by date of onset of symptoms: Italy}
	\label{fig_itaepi}  
\end{figure}

\begin{figure}[!ht]
	\centering
	\begin{subfigure}[c]{0.99\textwidth}
		\centering
		\includegraphics[angle=0,origin=c,width=0.7\linewidth]{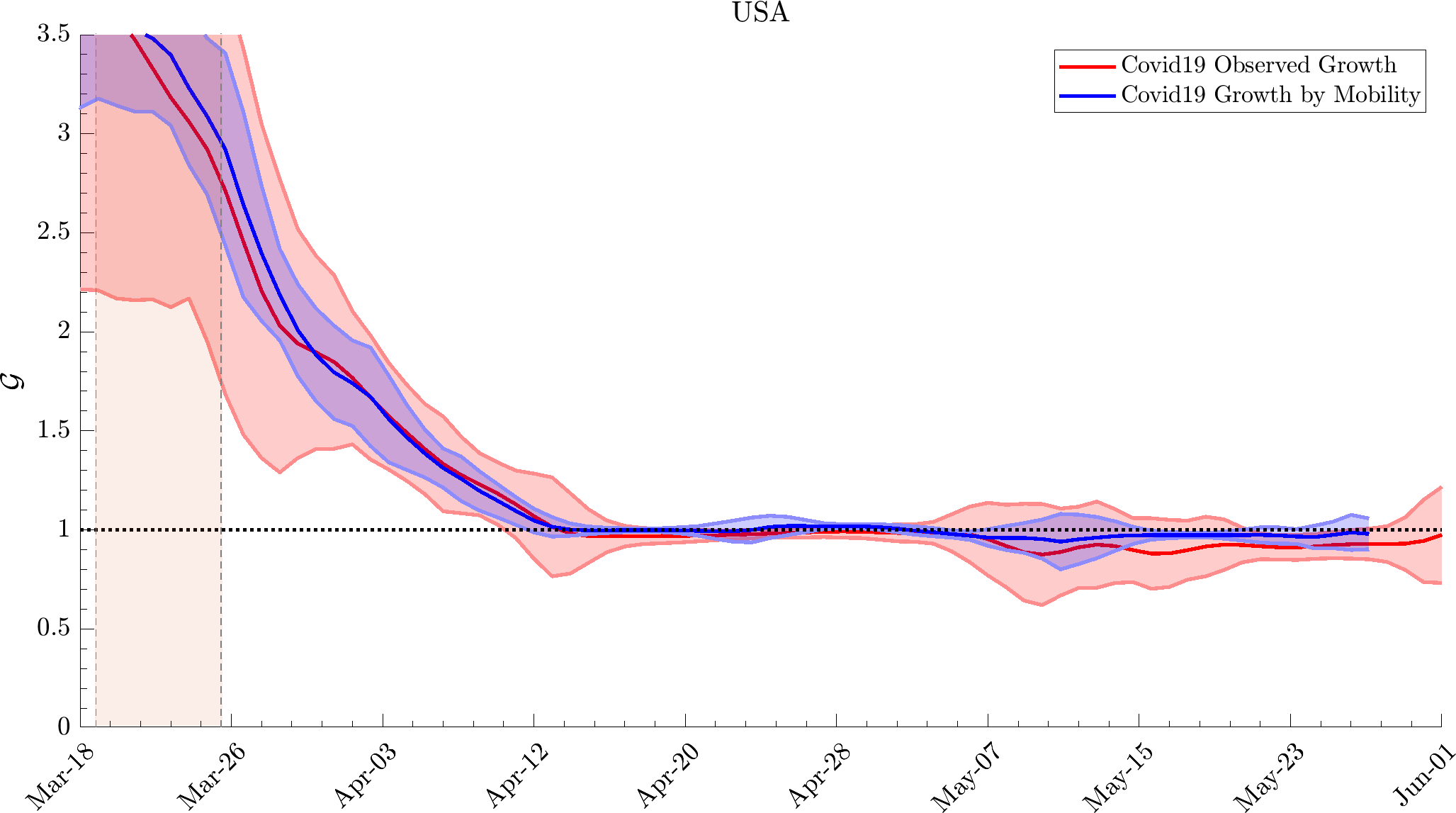}
		\caption{}
	\end{subfigure}%
	\\  \vspace{0.7cm}
	\begin{subfigure}[c]{0.45\textwidth}
		\centering
		\includegraphics[angle=0,origin=c,width=0.9\linewidth]{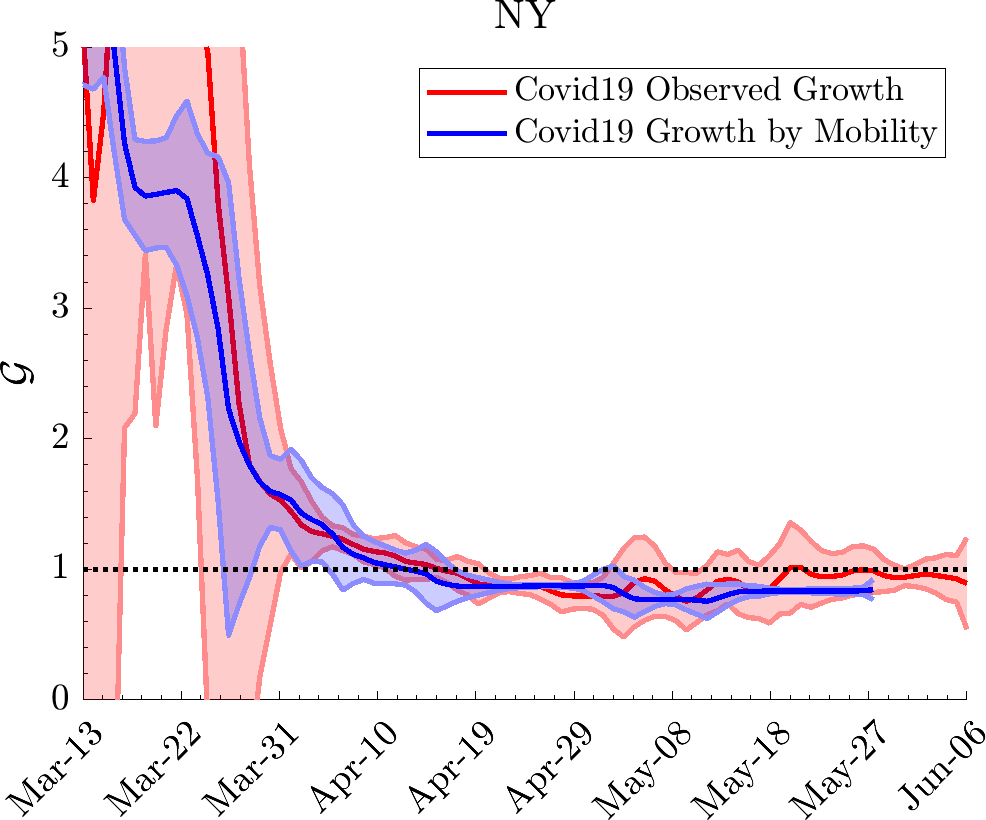}
		\caption{}
	\end{subfigure}%
	\begin{subfigure}[c]{0.45\textwidth}
		\centering
		\includegraphics[angle=0,origin=c,width=0.9\linewidth]{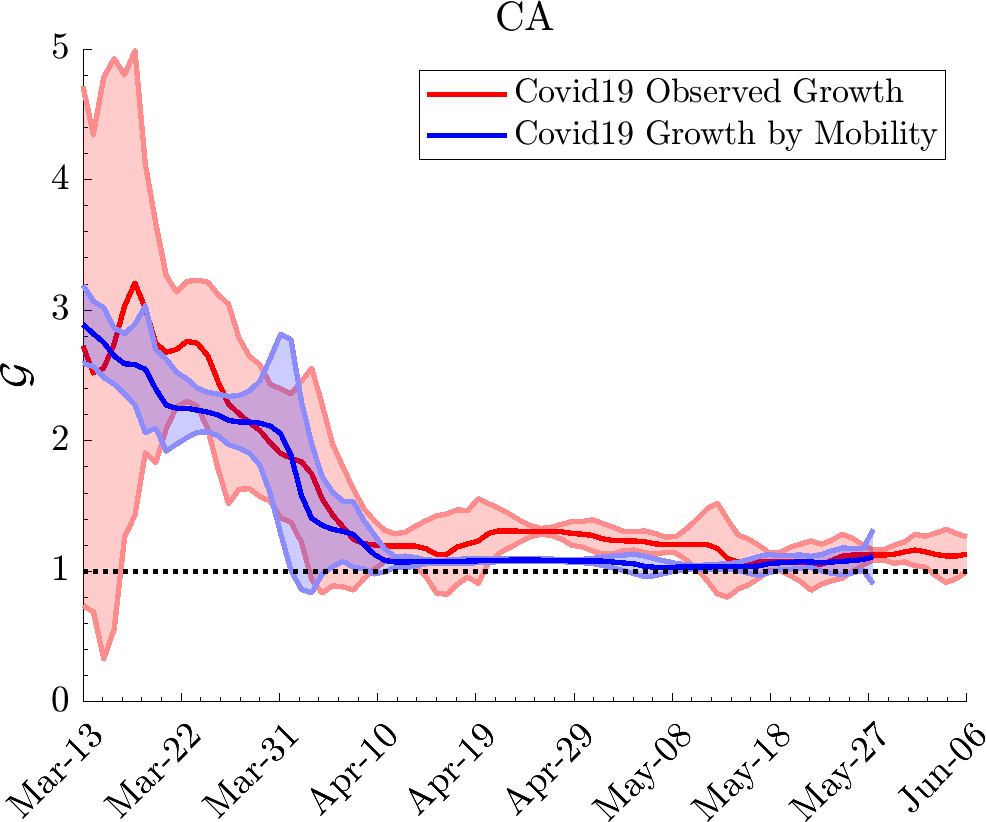}
		\caption{}
	\end{subfigure}%
	\caption{Infected individuals by date of laboratory diagnosis: USA.  Loess regression with $95\%$ confidence interval has been used to highlight the growth trend.}
	\label{fig_USAGR}  
\end{figure}
		\clearpage
	\epigraph{{\footnotesize\textit{There are \textsl{seeds} of many things that fly about that are both sources of life and sources of death, and when, \textsl{by some chance}, the latter are gathered together and disturb the sky, the air below becomes diseased.}}}{ \textbf{Lucretius}, De Rerum Natura, book VI }

	\end{document}